\documentclass[30pt,a4paper]{article}
\usepackage{graphicx,epsfig}
\usepackage{fancyhdr}
\usepackage{graphicx}
\usepackage{amsmath, mathrsfs}
\usepackage{multirow}
\usepackage{slashed}
\usepackage{graphicx,rotate,multicol}
\usepackage{wrapfig}
\usepackage{dcolumn} 
\usepackage{slashed,color,amsmath,amssymb}
\usepackage{jheppub}
\usepackage{blindtext}
\usepackage{enumitem}
\usepackage{epsfig}
\usepackage{latexsym, amssymb} 
\usepackage{enumitem}
\usepackage{slashbox}
\usepackage{caption}
\usepackage{multirow}
\usepackage{bigcenter}
\usepackage{verbatim}
\usepackage{footnote}
\usepackage{float}
\usepackage[caption=false]{subfig}
\newcommand{\be}{\begin{equation}}
\newcommand{\ee}{\end{equation}}
\newcommand{\bea}{\begin{eqnarray}}
\newcommand{\eea}{\end{eqnarray}}
\newcommand{\jet}{\mathbf{J}}
\newcommand{\vp}{\varphi}
\newcommand{\tvp}{\widetilde{\varphi}}

\def\ckin{C_{H,\text{kin}}}
\usepackage{bm}
\preprint{}

\title{Exploring Higgs-Photon Production at the LHC}

\author{\large{Tisa Biswas$^1$, Anindya Datta$^1$}}

\affiliation{\vspace*{0.1in}$^1$\normalsize Department of Physics, University of Calcutta,  
	92 Acharya Prafulla Chandra Road, \\ Kolkata 700009, India}

\emailAdd{tibphy\_rs@caluniv.ac.in}
\emailAdd{adphys@caluniv.ac.in}

\abstract{
	We have investigated the signal for physics beyond the Standard Model via Higgs plus photon final state, hitherto unobserved at the LHC, in the framework of Standard Model Effective Field Theory. Using the relevant dimension-6 operators, we probe two distinct classes of interactions, based on the Lorentz structure of Higgs couplings to gauge bosons and fermions. To begin with, constraints on the Wilson coefficients of these operators have been derived from existing experimental data. We then  focus on the hadronic decay of the Higgs boson to two bottom quarks in the boosted regime, leading to a high $p_T$ fat-jet  recoiling against a hard photon. Following a  CMS Run II search for a heavy resonance decaying into a Higgs  boson and a photon, a detailed signal and background analysis for this channel has been done and limits on the relevant Wilson coefficients have been obtained. By performing a cut-based analysis, we identify some kinematic observables that distinguish between  signal and SM background. Minimum values of the Wilson coefficients that will yield  $3\sigma$ signal significance at the 14 TeV run of the LHC with $3000 ~\rm fb^{-1}$ data, have been obtained. A multivariate analysis using a boosted decision tree and exploiting the  jet substructure techniques further help to  isolate the regions of phase space where the contribution of SMEFT driven signal is significantly enhanced. Finally, we briefly discuss the parameter dependencies and interpretations of the allowed values of the coefficients on a particularly interesting UV complete model namely, the Minimal Supersymmetric Standard Model.
}
\begin{document}
	\maketitle
	\section{Introduction}
	\label{sec:intro}
	
	With the Large Hadron Collider (LHC) in operation, our venture to shed more light on the fundamental interactions of elementary particles has seen impressive advances. After a few years that the first beam circulated at the LHC, the discovery of the 125 GeV scalar  boson~\cite{Aad:2012tfa,Chatrchyan:2012xdj} has paved the way towards the Era of Higgs Physics and intense efforts have been put in the decade thereafter, studying its properties to a high degree of accuracy. So far, properties of this 125 GeV scalar have been found to be in good agreement with those of the Standard Model (SM) Higgs boson. However, the exact mechanism for electroweak symmetry breaking remains to be verified. It is also possible that so far elusive signal of new physics beyond the SM may arise from the interactions of the Higgs boson itself. In this endeavour, precise measurements of  Higgs  couplings are very important. The continuously increasing amount of data recorded at the LHC opens the possibilities to explore properties of Higgs boson in a multitude of ways. Improved analysis techniques, especially in the context of High Luminosity-LHC (HL-LHC), will  be extremely useful in the pursuit of physics beyond the SM.
	
	There are several approaches to search for the   new particles and their interactions. One way is to look for the signature of  new particles that can be produced at resonance at high energies. Yet, after years of intense collective efforts searching for new particles, no clear evidence of such new resonances  have been found. Another fully consistent and theoretically-sound framework is  Effective Field Theory (EFT). This framework relies on the idea that new physics affecting the LHC measurements is too heavy to be produced on-shell.  However, such heavy states leave their footprints via their contributions when they act as propagators either at tree level or higher order  processes involving the SM particles  leading to  deviations  in their couplings. So, it is of utmost importance to measure the interactions involving SM particles at LHC energies as an indirect probe for such new heavy particles and their properties~\cite{Weinberg:1978kz,Buchmuller,Leung:1984ni,Dawson:2018dcd}. We will follow this second avenue and stick to a framework where the SM is augmented by a set of higher dimensional operators respecting all the symmetries of the SM, namely, the Standard Model Effective Field Theory (SMEFT),
	
	\begin{equation}
	\mathcal{L}_{\rm SMEFT}=
	\mathcal{L}_{\rm SM}+\sum_{d=5}^{\infty} \sum_{i=1}^{n}{C_i^{(d)}\over\Lambda^{d-4}}\mathcal{O}_i^{d}
	\label{eq:smeft}
	\end{equation}
	Here, $\mathcal{O}_i^{d}$ are the mass dimension-$d$ operators, constructed out of the SM fields only. Any non-SM physics effects are encapsulated in the Wilson coefficients $C_i^{(d)}$. The energy scale $\Lambda$ can be connected to the masses of the heavy particles which manifest themselves via virtual effects. With these assumptions, there are 59 linearly independent dimension-6 operators in the Warsaw basis \cite{Grzadkowski:2010es} which we will use to probe any new physics that they contain in their structure.
	
	In this work, we will investigate the Higgs boson production in association with a photon. The Higgs boson production in association with a vector boson, $(W^\pm,Z)$ has been extensively studied in the SM~\cite{Gupta:2014rxa,Cohen:2016bsd}. In the limit of vanishing quark masses, Higgs production in association with a photon from quark$-$anti-quark annihilation, is absent at the tree level.  This process can also take place via internal particle loops, rate for which is also not very promising. Therefore, any deviation from the SM predictions and observation of $h\gamma$ signal at the LHC will point to the presence of beyond-SM (BSM) physics. At hadron colliders, it is very challenging to observe this particular Higgs signal due to the large background and its small cross-section. 
	
	Thus, it is important to understand how the  new physics can be manifested via this channel  and accordingly optimise the search strategy. In this work, we will identify the SMEFT operators that give leading contributions to the production of the Higgs and a photon. The Higgs decays into a pair of bottom quarks, resulting in a final state consisting of a Higgs-like fat-jet and a hard photon. The CMS collaboration has looked for this signal using $35.9~\rm fb^{-1}$ data~\cite{CMS:2018cno} at 13 TeV run in search of a heavy resonance decaying into a Higgs and a photon. By comparing our estimated rates, with those published in Ref.~\cite{CMS:2018cno}, we put model independent constraints on the Wilson coefficients of the relevant dimension-6 operators. These Wilson coefficients of SMEFT operators relevant in  Higgs  physics  have  been  constrained  through  various channels, for example in Refs.~\cite{Sanz,Cepeda:2019klc,Biswas:2021qaf}.	We compare the bounds obtained from our analysis with the existent limits on such couplings.
	
	With these constraints on the operator coefficients, we study the impact of possible BSM physics in this channel at the HL-LHC. Our study  proposes more than one kinematic variables which are very instrumental in separating the new physics effects from the SM background.
	
	In the following, we highlight the novel features of our study:
	\begin{itemize}
		\item   To the best of our knowledge, this is the first comprehensive study of $h \gamma$ production in the framework of SMEFT at the LHC in the boosted kinematic region of the phase space. The earlier investigations \cite{Dobrescu:2017sue,Khanpour:2017inb,Aguilar-Saavedra:2020rgo} in this channel, dealt with chosen few dimension-6 operators involving either only bosonic fields or fermionic fields without  contrasting the results with available experimental data. 
		
		\item In previous works on this particular channel \cite{Abbasabadi:1997zr,Aguilar-Saavedra:2020rgo,Gabrielli:2016mdd,Khanpour:2017inb}, only the SM contribution arising from tree-level $q \bar{q} \to h \gamma$ process has been taken into account. In this work, we have included the SM effects  arising from both the tree-level heavy quark-antiquark annihilation $\to h \gamma$ and one-loop contribution to $q \bar{q} \to h \gamma$ and report our findings including the non-zero  interference  effects between  the  dimension-6 operator-induced  amplitudes  and  the  SM amplitudes (see Fig. \ref{tab:Feynmandiag}).
		
		\item We focus in the high energy regime where the Higgs is boosted and will manifest itself as a fat-jet containing two b-tagged subjets.  We investigate topologies of the final state with such a fat-jet in association with a hard photon for possible signature of  new interactions with Lorentz structures different from the SM. 
		
		\item To better discriminate between signal and background, a multivariate analysis is performed using Boosted Decision Tree (BDT) techniques with jet substructure observables such as N-subjettiness, jet mass and energy correlators as input variables. Going beyond the standard cut-based approach and using the BDT technique shows a clear improvement in terms of signal significance as explicitly demonstrated in the following.
		
		\item We have segregated two types of interactions based on the operator structures. We propose to correlate these distinct types of interactions in producing $h\gamma$ by using the ratio of rates in the $\gamma \gamma$ and $Z \gamma$ decay modes of Higgs and ratio of rates of $Vh$ and $VBF$ Higgs production. The effect of various dimension-6 operators considered in this paper, can also be observed in the process $pp \to h \to b \bar{b} \gamma$. We briefly discuss the effect of such operators in the final state kinematics for this process.
		
		\item We briefly tried to discuss the possible sources of such dimension-6 operators arising in a UV complete theory like the MSSM.
	\end{itemize}
	
	The plan of the paper is the following. In section~\ref{sec:eft_framework}, we enlist  the  relevant  operators  in  the Warsaw basis that contribute to the $p p \to h \gamma$ process. This has been followed by a discussion on various constraints on these operators in the light of  experimental data. In section~\ref{sec:collider_analysis}, a detailed collider analysis studying the kinematical features of the signal and background processes is performed. Firstly, we present these features with a cut-based analysis. Thereafter, we discuss the projected sensitivities to the EFT couplings.  The improvement over the cut-based results using boosted decision trees has been explored. In section~\ref{sec:further_interpretation}, we comment on the numerical results and interpretations and discuss some parameter dependencies of the cross-section in detail. Finally, we draw our conclusions in section~\ref{sec:discussions}.
	\section{Higher Dimensional Operators Parametrisation}
	\label{sec:eft_framework}
	We will follow the effective field theory approach to probe for any possible deviations from the SM interactions in the Higgs sector. The Higgs observed at the LHC is assumed to be a part of an $SU(2)_L$ doublet and that the $SU(2)_L \times U(1)_Y$ symmetry is linearly realized in the effective theory. Starting with the experimental inputs in terms of the measured values of $\alpha , M_Z$ and $G_F$, we consider the following general set of dimension-6 gauge invariant operators which give rise to such anomalous $hV\gamma$ interactions and the $h\gamma ff$ contact interactions.
	\begin{itemize}
		\item Operators containing the Higgs doublet $\vp$ and its derivatives:
		\begin{equation}
		\mathcal{O}_{H\Box}  =  (\vp^\dag \vp)\raisebox{-.5mm}{$\Box$}(\vp^\dag \vp);~~~
		\mathcal{O}_{H D} =  \left(\vp^\dag D^\mu\vp\right)^\star \left(\vp^\dag D_\mu\vp\right);~~
		\mathcal{O}_{H}  =  (\vp^\dag\vp)^3  
		\label{eq:op1}
		\end{equation}
		Here, $D_\mu$ has the usual meaning referring to the covariant derivative and contain $SU(2)_L$, $U(1)_Y$  gauge couplings and bosons. The operators $\mathcal{O}_{H\Box}$ and $\mathcal{O}_{HD}$ contribute to kinetic energy of the Higgs field, $h$. So, we require a finite wave function renormalisation in order to obtain the Higgs kinetic term to the canonical form. Thus, the Higgs doublet is expanded as

\begin{align}
\vp &= \frac{1}{\sqrt 2} \left(\begin{array}{c}
0 \\
\left[ 1+ \ckin \right]  h + v_T
\end{array}\right), & \quad  \ckin &\equiv \frac{1}{\Lambda^2}\left( C_{H\Box}-\frac14  C_{HD}\right),
\label{Hvev}
\end{align}
Here $\langle \vp^\dagger \vp \rangle $ has been defined to include corrections due to dimension-6 $\mathcal{L}$ so that $v_T \equiv ( 1+ 3 C_H v^2/8 \lambda\Lambda^2 ) v$ where $\sqrt{2 \langle \vp^\dagger \vp \rangle_{SM}} \equiv v$.
		\item These set of  operators can induce $h\gamma qq$  contact interactions. Due to the gauge field strengths, there is a presence of explicit dependence on the momentum of the gauge bosons in such operators. This will play a significant role in kinematics in which we are interested in this analysis.
		\begin{equation}
		\mathcal{O}_{uW} = (\bar q_p \sigma^{\mu\nu} u_r) \tau^I \tvp\, W_{\mu\nu}^I;~~~
		\mathcal{O}_{dW} = (\bar q_p \sigma^{\mu\nu} d_r) \tau^I \vp\, W_{\mu\nu}^I \nonumber
		\end{equation}
		\begin{equation} 
		\mathcal{O}_{uB} = (\bar q_p \sigma^{\mu\nu} u_r) \tvp\, B_{\mu\nu};~~~
		\mathcal{O}_{dB} = (\bar q_p \sigma^{\mu\nu} d_r) \vp\, B_{\mu\nu} 
		\label{eq:op2}
		\end{equation}
		The field content consists of $q$, the quark doublet under $SU(2)_L$ and $u$, $d$ are the $SU(2)_L$ singlet quarks, with $p$
		and $r$ denoting the generation indices and $SU(2)_L$ and $U(1)_Y$ gauge fields are denoted by $W$ and $B$, respectively. Neglecting fermion masses, the magnetic dipole type operators connect fermions of different helicities. 
		\item The anomalous $hV \gamma$ interactions arise from the following set of dimension-6 operators \cite{Buchmuller,Grzadkowski:2010es}:
		\begin{equation}
		\mathcal{O}_{HW} = \vp^\dag \vp\, W^I_{\mu\nu} W^{I\mu\nu};~~~
		\mathcal{O}_{HB} = \vp^\dag \vp\, B_{\mu\nu} B^{\mu\nu};~~~
		\mathcal{O}_{HWB} = \vp^\dag \tau^I \vp\, W^I_{\mu\nu} B^{\mu\nu}
		\label{eq:op3} 
		\end{equation}
		\vspace{1mm}
		These operators induce $hV \gamma$ couplings which are SM-like in Lorentz structure, while modify the $hVV$ couplings by introducing new Lorentz structure in the Lagrangian. 	
	\end{itemize}
	The couplings involving the Higgs boson and the photon are generated by the combinations of the aforementioned operators. We note down the additional new interactions in the effective Lagrangian:
	\begin{eqnarray}
	\mathcal{L}_{eff}&= &g_{h\gamma \gamma} h A_{\mu \nu}A^{\mu \nu}+ g_{hZ\gamma} h A_{\mu \nu} Z^{\mu \nu}\nonumber \\
	&&+\left(g_{uuh\gamma} (\overline{u}\sigma^{\mu\nu} P_Ru)hA_{\mu\nu} + g_{ddh\gamma} (\overline{d}\sigma^{\mu\nu} P_Rd)hA_{\mu\nu} + h.c.\right) \, ,
	\label{eq:L_eff}
	\end{eqnarray}
	
	\vspace{1mm}
	The effective couplings contributing to $h \gamma$ production in $pp$ collisions in Eqn. \ref{eq:L_eff} are related to the coefficients of the operators appearing in Eqns.~\ref{eq:op1}-\ref{eq:op3} as follows:
	
	\begin{table}[h!]
		\centering
		\renewcommand{\arraystretch}{1.8}
		\begin{tabular}{|c||c|}
			\hline
			& Warsaw  Basis 
			\\
			\hline\hline
			$g_{h\gamma \gamma}$	& $\frac{2 m_W}{\Lambda^2}\frac{1}{g}\left(s_W^2 C_{HW} + c_W^2 C_{HB}-s_W c_W C_{HWB}\right)$ \\ \hline
			$g_{hZ\gamma}$ 			& $\frac{4 m_W}{\Lambda^2}\frac{1}{g}s_W c_W \left( C_{HW} -  C_{HB}\right) + \frac{2 m_W}{\Lambda^2}\frac{1}{g} (s_W^2 - c_W^2 ) C_{HWB}$ \\ \hline
			$g_{uuh\gamma}$ 		& $\frac{-1}{\sqrt{2}\Lambda^2} \left(c_W C_{uB} + s_W C_{uW}\right)$  \\ \hline
			$g_{ddh\gamma}$ 		& $\frac{-1}{\sqrt{2}\Lambda^2} \left(c_W C_{dB} - s_W C_{dW}\right)$ \\ \hline
			\hline
		\end{tabular}
		\caption{Anomalous Higgs couplings in the Warsaw basis.}
		\label{tab:rgb}
	\end{table}
	In the SMEFT, the input parameters are shifted from their SM values at the tree level, which creates a dependence on $C_{HD}, C_{HWB}, C_{Hl}^{(3)}$ and $C_{ll}$ \cite{Brivio:2017bnu,Brivio:2018bnu}. Additionally, the effective couplings of the fermions to the Z and W gauge bosons undergo a shift from their SM values. The momentum structures of the vertices remain unchanged by these effects. We shall focus on operators that give dominant effect in the selection efficiencies and kinematic observables to the $h \gamma$ process at the LHC, given its small cross-section in the SM and large QCD backgrounds. Moreover, the method developed here, is of general utility in studying all possible higher-dimensional operators.
	\subsection{Model-independent constraints}
	After parametrising new physics in terms of dimension-6 operators, we now locate the region of parameter space favoured by the existing experimental data.
	
	 The operator coefficients $C_{HWB}$ (that drives $Z - \gamma$ mixing) and $C_{HD}$ are constrained from the electroweak precision data (EWPD).  At the tree level, $C_{HWB}$ and $C_{HD}$ contribute to the oblique precision EW parameters $S$ and $T$, respectively. Taking into account the constraints on S and T parameters from the datasets from $Z$-pole, $W$ mass and LEP-2 $WW$ scattering measurements as well as LHC Run 1 and Run 2 Higgs results, the $2$-$\sigma$ marginalised ranges in a two-parameter fit of  $C_{HWB}$ and $C_{HD}$ are $[-0.00452,0.0053]$ and $[-0.0119,0.0299]$ in ${\rm TeV}^{-2} $, respectively \cite{ellis}. In case all other Wilson coefficients are profiled, the constraints become looser with $C_{HWB} \in [-0.36,0.73]~{\rm TeV}^{-2}$ and  $C_{HD} \in [-0.36,0.73]~{\rm TeV}^{-2} $, at $2$-$\sigma$ level, respectively \cite{Ellisglobalfit2021}. The operators $\mathcal{O}_{HW}$ and $\mathcal{O}_{HB}$ also contribute to the so-called Peskin-Takeuchi $S,T$ and $U$ parameters~\cite{Peskin,Altarelli} and are therefore constrained by electroweak precision observables~\cite{Hagiwara,Garcia} (EWPO). Following Ref.~\cite{Garcia}, the bounds at $95 \%$ C.L., taking one operator at a time, are given by $-1.5~{\rm TeV}^{-2}<\frac{C_{HW}}{\Lambda^2}<2.5~{\rm TeV}^{-2} $,$-1.5~{\rm TeV}^{-2}<\frac{C_{HB}}{\Lambda^2}<2.5~{\rm TeV}^{-2}$\footnote{Ref.~\cite{Garcia} uses a different normalisation, $viz$, ${\cal O}_{HW}= \vp^\dag \widehat  W^{\mu\nu} \widehat W_{\mu\nu} \vp = -\frac{g^2}{4}(\vp^\dag \vp) W^{a\,\mu\nu}W^a_{\mu\nu}$ and ${\cal O}_{HB} = (\vp^\dag \vp) \
				\widehat   B^{\mu\nu} \widehat  B_{\mu\nu}$ .}. We consider the operators $\mathcal{O}_{HW},\mathcal{O}_{HB}$ and $\mathcal{O}_{HWB}$ for analysis in this study. However, Higgs data puts much stronger bounds on these  operators compared to the bounds derived from the oblique parameters. Couplings arising from the bosonic operators affect the tree-level SM Higgs couplings, namely, $hWW$, $hZZ$ and contribute in the loop-induced Higgs decays as well, namely, $h \rightarrow \gamma \gamma$ and $h \rightarrow Z \gamma$. The dipole operators  are  also  subjected  to  the  constraints  imposed  by  the  LHC  data. Taking one operator at a time, bounds on them are obtained by comparing the expected cross-section with experimental data from ATLAS and CMS collaborations (Table~\ref{exptdata}). The results are summarised in Table~\ref{D6constraints}. The operators $\mathcal{O}_{H \Box}$ and  $\mathcal{O}_{HD}$ renormalise the Higgs wavefunction and lead to an overall scaling of the rate of the $q\bar{q} \to h \gamma$ process.  These operators can also affect the Higgs self interactions. When loop corrections are included, $C_{HD}$  affects $HZZ$ interaction and not $HWW$ interaction. We shall briefly analyse the effects of $C_{HD}$ and $C_{H\Box}$ in their contribution to $pp \to h\gamma$ in this study.
			
		The rate of the process $b \to s \gamma$ can be modified in presence of the operator $\mathcal{O}_{bW}$. Stringent bounds on such an interaction has been obtained in~\cite{ParticleDataGroup:2020ssz,Cirigliano:2016nyn}. However, these operators are flavour non-diagonal and we will not use such flavour non-diagonal  interactions further in our discussion. Also, constraints can be put considering the contribution of dipole operators to the anomalous magnetic moment of light-quarks. But for the light quarks, $(g-2)$ is hard to extract in a model independent way, is therefore subjected to large uncertainties~\cite{Giudice:2012ms}.	
	\begin{table}[h]
		\centering
		\begin{tabular}{| c | c | c | c |}
			\hline Channel & Experiment & Energy (GeV)& Luminosity \\ \hline 
			$h\rightarrow \gamma \gamma$ & CMS~\cite{CMS:2018piu} & 13 TeV & 35.9 fb$^{-1}$ \\ \hline$h
			\rightarrow \tau \bar{\tau}$ & CMS~\cite{CMS:2017zyp} & 13 TeV & 35.9 fb$^{-1}$ \\ \hline 
			$h \rightarrow WW$ &CMS~\cite{CMS:2018zzl} & 13 TeV & 35.9 fb$^{-1}$ \\ \hline 
			$h \rightarrow \gamma \gamma$ &ATLAS\cite{ATLAS:2018hxb} & 13 TeV & 36.1 fb$^{-1}$ \\ \hline 
			$h\rightarrow ZZ$ & ATLAS~\cite{ATLAS:2017azn} &13 TeV & 36.1 fb$^{-1}$ \\ \hline
			$h\rightarrow b \bar{b}$ & ATLAS~\cite{ATLAS:2018kot} &13 TeV & 79.8 fb$^{-1}$ \\ \hline
		\end{tabular}
		\caption{Listing of the Higgs signal strength measurements at LHC at  13 TeV  included in our study}
		\label{exptdata}
	\end{table}
	\begin{table}[b!]
		\centering
		\renewcommand{\arraystretch}{1.5}
		\begin{tabular}{|c|c|c|}
			\hline
			Constraints &Unitarity &Higgs data ($95\%$ C.L.)\\ \hline\hline
			$\frac{C_{uW}}{\Lambda^2}$ & $[-7.0,7.0]$&  $[-0.417,0.417] $ \\ \hline
			$\frac{C_{uB}}{\Lambda^2}$ & $[-27.75,27.75]$&  $[-0.745,0.745] $ \\ \hline
			$\frac{C_{dW}}{\Lambda^2}$ & $[-7.0,7.0]$&  $[-0.483,0.483] $ \\ \hline
			$\frac{C_{dB}}{\Lambda^2}$ & $[-27.75,27.75]$&  $[-0.909,0.909] $ \\ \hline
			$\frac{C_{HB}}{\Lambda^2}$ & $[-150.75,150.75]$&  $[-0.062,0.342] $ \\ \hline
			$\frac{C_{HW}}{\Lambda^2}$ & $[-24.75,24.75]$&  $[-0.018,0.201] $ \\ \hline
			$\frac{C_{HWB}}{\Lambda^2}$ & $[-346.5,346.5]$&  $[-0.188,0.341] $ \\ \hline
			$\frac{C_{HD}}{\Lambda^2}$ & $[-37.75,37.75]$&  $[-15.309,18.425] $ \\ \hline
			$\frac{C_{H\Box}}{\Lambda^2}$ & $[-26.25,26.25]$&  $[-6.804,11.138] $ \\ \hline
		\end{tabular}
		\caption{Limits for the fermionic and bosonic operator coefficients in the Warsaw basis. The bounds are given in $\rm TeV^{-2}$ units.}
		\label{D6constraints}
	\end{table}

	On introducing these higher dimensional operators, one may reasonably expect a violation of unitarity in $2 \to 2$ scattering processes ($V_L V_L \rightarrow V_L V_L~{\rm{and}}~f \bar{f} \to VV$, where  $V=W,Z$), depending on the values of the Wilson coefficients at high energies. Bounds on these coefficients may be obtained by demanding that there be no such violation of unitarity below the scale of the effective theory. Unitarity constraints stemming from the Higgs scattering processes on our dimension-6 operators of interest, assuming an incoming parton inside a proton carries typically 1 TeV of energy at the LHC, are summarised in Table~\ref{D6constraints}.
	In the subsequent study, values of the effective coupling strengths consistent with the electroweak data and the Higgs measurements have been used. Within such constraints, we concentrate on the operators $\{\mathcal{O}_{uB},\mathcal{O}_{dB},\mathcal{O}_{uW},\mathcal{O}_{dW}\}$ and $\{\mathcal{O}_{HW}, \mathcal{O}_{HB},, \mathcal{O}_{HWB}\}$ and examine their role in modifying the $h \gamma$ production rates at the LHC.
	
	\subsection{$h \gamma$ production in the SMEFT}	
	We implemented the effective Lagrangian in {\tt{FeynRules}} \cite{Alloul:2013bka} to create the UFO model file for the event generator {\tt{Madgraph5\_aMC@NLO}} \cite{Alwall:2014hca} which is used to generate all signal and background events. These events are generated at the leading order (LO) followed by {\tt Pythia} (v8) \cite{Sjostrand:2006za} for showering and hadronization.  We use {\tt{NNPDFNLO}} parton distribution functions \cite{pdf} for event generation by setting  factorisation and renormalization scales equal to the Higgs mass ($\mu_R = \mu_F = m_H$).  The matching parameter, {\tt QCUT} is appropriately determined for different processes as discussed in Ref \cite{MLMMerging}.  To incorporate detector effects, events are passed through {\tt Delphes-v3.4.1} \cite{deFavereau:2013fsa}. Jets are reconstructed by employing the Cambridge-Achen (CA) \cite{Dokshitzer:1997in} algorithm with radius parameter $R = 0.8$, as implemented in the {\tt Fastjet}-v3.3.2 \cite{Cacciari:2011ma}.
	\vspace{1mm}
	
\begin{figure}[b!]
		\centering
		\includegraphics[width=4.1cm,height=3.1cm]{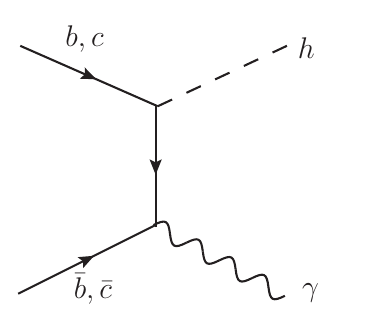}~~
		\includegraphics[width=4.1cm,height=3.1cm]{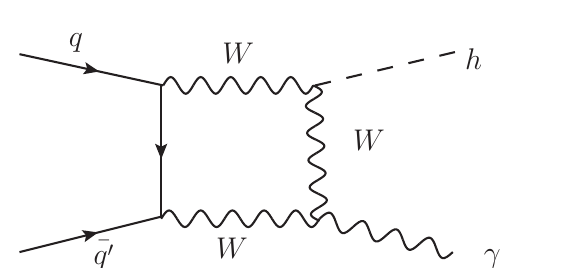}~~
		\includegraphics[width=4.1cm,height=3.1cm]{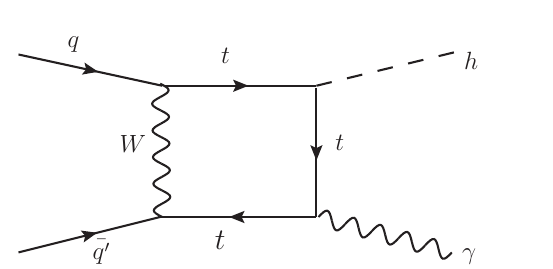}\\
		\vspace{0.6mm}
		(a) \hspace{35mm}(b) \hspace{40mm}(c)  \\
		\vspace{0.6mm}
		\includegraphics[width=4.1cm,height=3.1cm]{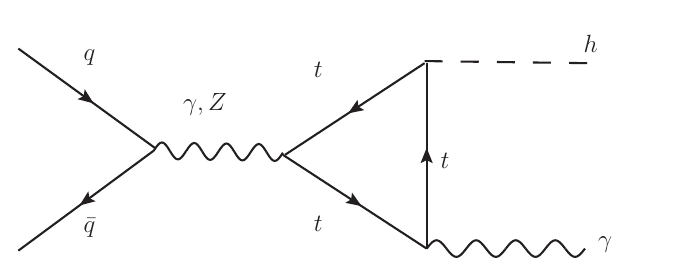}~~
		\includegraphics[width=4.1cm,height=3.1cm]{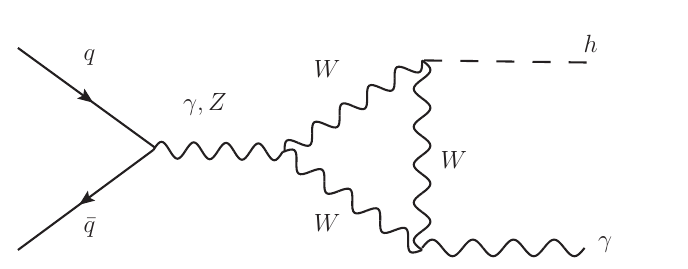}~~
		\includegraphics[width=4.1cm,height=3.1cm]{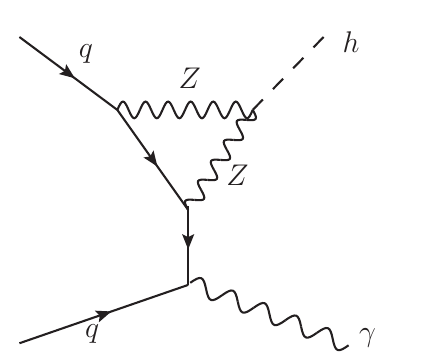}\\
		\vspace{0.6mm}
		(d) \hspace{35mm}(e) \hspace{40mm}(f)   \\
		\includegraphics[width=4.1cm,height=3.1cm]{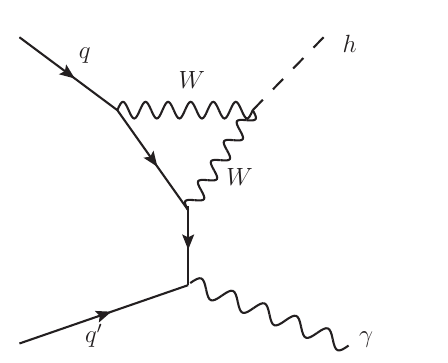}~~
		\includegraphics[width=4.1cm,height=3.1cm]{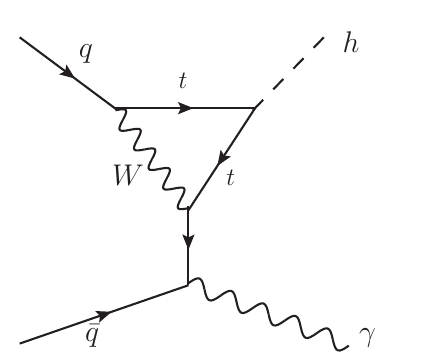}~~
		\includegraphics[width=4.1cm,height=3.1cm]{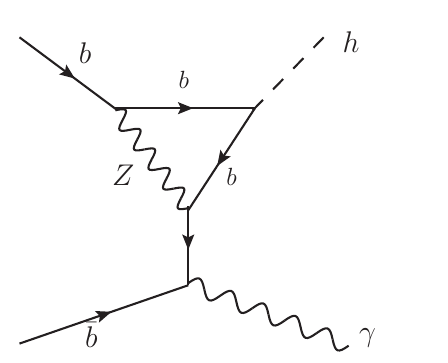}\\
		\vspace{0.6mm}
		(g) \hspace{35mm}(h) \hspace{40mm}(i)   \\
		\vspace{0.6mm}
		\includegraphics[width=3.9cm,height=3.0cm]{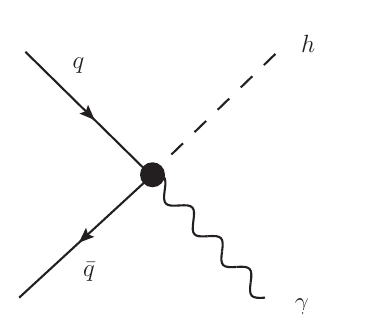}~
		\includegraphics[width=3.9cm,height=3.0cm]{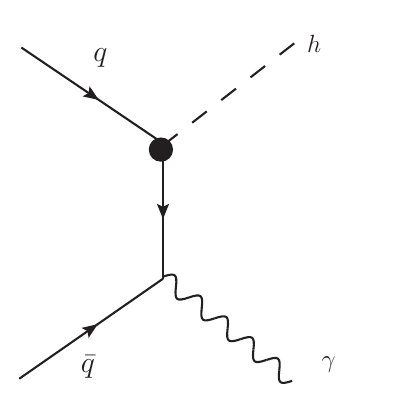}~
		\includegraphics[width=3.9cm,height=3.0cm]{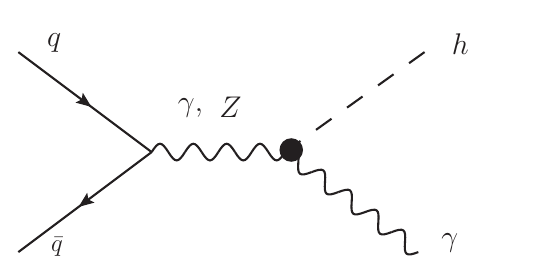}~
		\includegraphics[width=3.8cm,height=3.1cm]{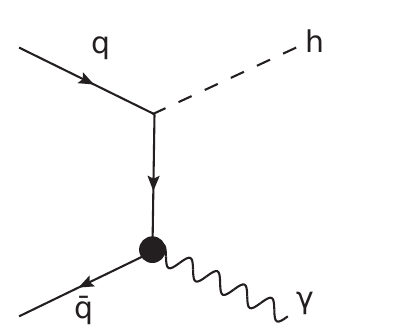}\\
		\vspace{0.6mm}
		\hspace{8mm}(j) \hspace{35mm}(k) \hspace{33mm}(l)  \hspace{32mm}(m)  \\
		\vspace{1.1mm}
		\caption{\label{tab:Feynmandiag} Representative Feynman diagrams showing production of a Higgs boson in association with a	 photon. Diagrams (a)-(i) correspond to the SM process, while  (j) involves the  contact interaction  denoted with a blob, between the Higgs, gauge boson and fermions (k),(l) processes involve the anomalous couplings of Higgs with gauge bosons and fermions respectively and (m) involves the anomalous $\gamma q\bar{q}$ anomalous coupling.}	
	\end{figure}

	In Fig.~\ref{tab:Feynmandiag}, we present a representative set of Feynman diagrams for $h \gamma$ production arising from the dimension-6 operator interactions and the SM $h \gamma$ production for illustration.
	\vspace{1mm}
	The $h\;\gamma$ production arises from heavy b and c quark initiated tree-level amplitudes  (see Fig.~\ref{tab:Feynmandiag} (a)). While evaluating such an amplitude, we have used the running heavy quark  masses in the Yukawa coupling, evaluated at scale $m_H = 125$ GeV with \cite{Gabrielli:2016mdd}: 
	\begin{eqnarray}
	m_t^{\overline{MS}}(m_H)= 173  {\;\rm GeV}, \; 
	m_b^{\overline{MS}}(m_H)&=& 2.765 {\;\rm GeV}, \;\; m_c^{\overline{MS}}(m_H)=0.616 {\;\rm GeV},
	\label{masses}
	\end{eqnarray}
	Amplitudes arising from Fig.~\ref{tab:Feynmandiag} (b-i) proceeding via one loop also contribute to $h \gamma$ rate. To the best of our knowledge, the previous studies on Higgs and photon production at the LHC \cite{Gabrielli:2016mdd,Khanpour:2017inb} only considered  the contribution from heavy quark-antiquark annihilation and one-loop contributions have been neglected. However, for completeness, we have taken into account all the contributions arising both from tree level and one loop diagrams.  We note that with the running heavy quark masses evaluated at $m_H$, the cross-section at  tree level is greater than the one-loop order process by $\mathcal{O}(10)\%$ at 14 TeV LHC and thus, the latter cannot be neglected. These SM one-loop amplitudes for triangle and box graphs can also be evaluated under the {\tt{MG5aMC\_@NLO}} \cite{Hirschi:2015iia} environment. However, we have used the expressions for these one-loop amplitudes in terms of Passarino and Veltman functions following Refs.~\cite{Abbasabadi:1997zr,Djouadi:1996ws,Passarino:1978jh} and incorporated these as one-loop form factors in the UFO model file obtained from FeynRules. This procedure has been followed to avoid the complications arising while calculating interference effects of SM loop amplitudes with dimension-6 operators. We have explicitly checked that in the $C_i \to 0$ limit, our procedure yields the same result with that evaluated by {\tt{MC@NLO}} formalism.
	
	In this work, we will investigate the effects of interactions such as $q\bar{q}h\gamma$ and $hV\gamma$ couplings in producing the Higgs in association with a photon at the LHC. We do not aim to probe the deviations of the Higgs couplings to first and second generation SM quarks via the higher dimensional operators (corrections to Yukawa interactions) as shown in Fig.~\ref{tab:Feynmandiag}(k). The inclusion of such dimension-6 operator modifies  the Higgs Yukawa couplings to the initial light quarks from their SM values by a factor of  $ \left(1+ \frac{\sqrt{2} C v^2}{ y_q(m_h) \Lambda^2}\right)$, where $y_q(m_h)$ is the corresponding light quark Yukawa coupling, evaluated at the scale of Higgs mass  and would only lead to an overall scaling on the rate of the process $q \bar{q} \to h \gamma$. We leave the details of how ones measures it and how it may need to be separated from other dynamics to another study. In our analysis, we include the interference with the SM and the quadratic terms ensuing from the dimension-6 operators for the signal process.\footnote{On including the contribution of higher-dimensional operators with the SM, the total cross-section can be expressed as a function of a Wilson-coefficient as given by $\sigma = \int \frac{1}{\rm initial~flux}|\mathcal{M}_{tot}(C_i)|^2 d\Pi$, where $d\Pi$ is the Lorentz invariant phase space factor and  $\mathcal{M}_{tot}(C_i)=\mathcal{M}_{SM}+\mathcal{M}_{D6}(C_i)$} It is to be noted that the operators $\mathcal{O}_{uB},\mathcal{O}_{uW},\mathcal{O}_{dB}$ and $\mathcal{O}_{dW}$  interfere with only Fig.\ref{tab:Feynmandiag} (a) SM process which connect fermions of different helicities. The bosonic operators, on the other hand, do not interfere with this tree level process. The EFT interference contribution, being proportional to the mass of the fermions is negligibly small. The bosonic type of operators interfere with processes in  Fig.\ref{tab:Feynmandiag} (b-i). However, we will see in the following that the SM contribution (both tree level and loop level) being very small, it is the quadratic contribution that dominates over the interference effect in the new physics signal.
	
\begin{figure}[t]
		\centering
		\subfloat{
			\begin{tabular}{cc}
				\includegraphics[width=7.5cm,height=6.5cm]{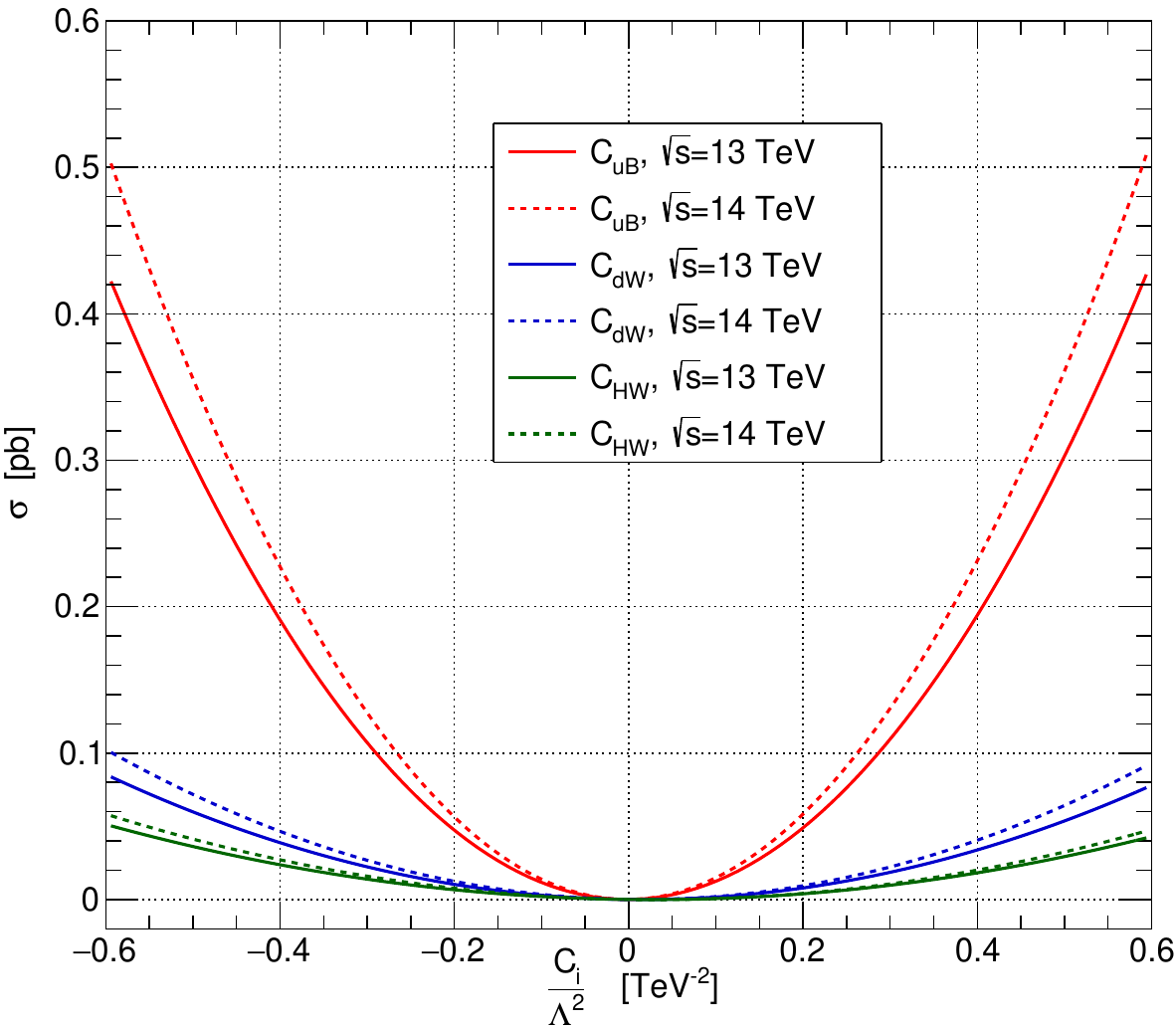}&
				\includegraphics[width=7.5cm,height=7.0cm]{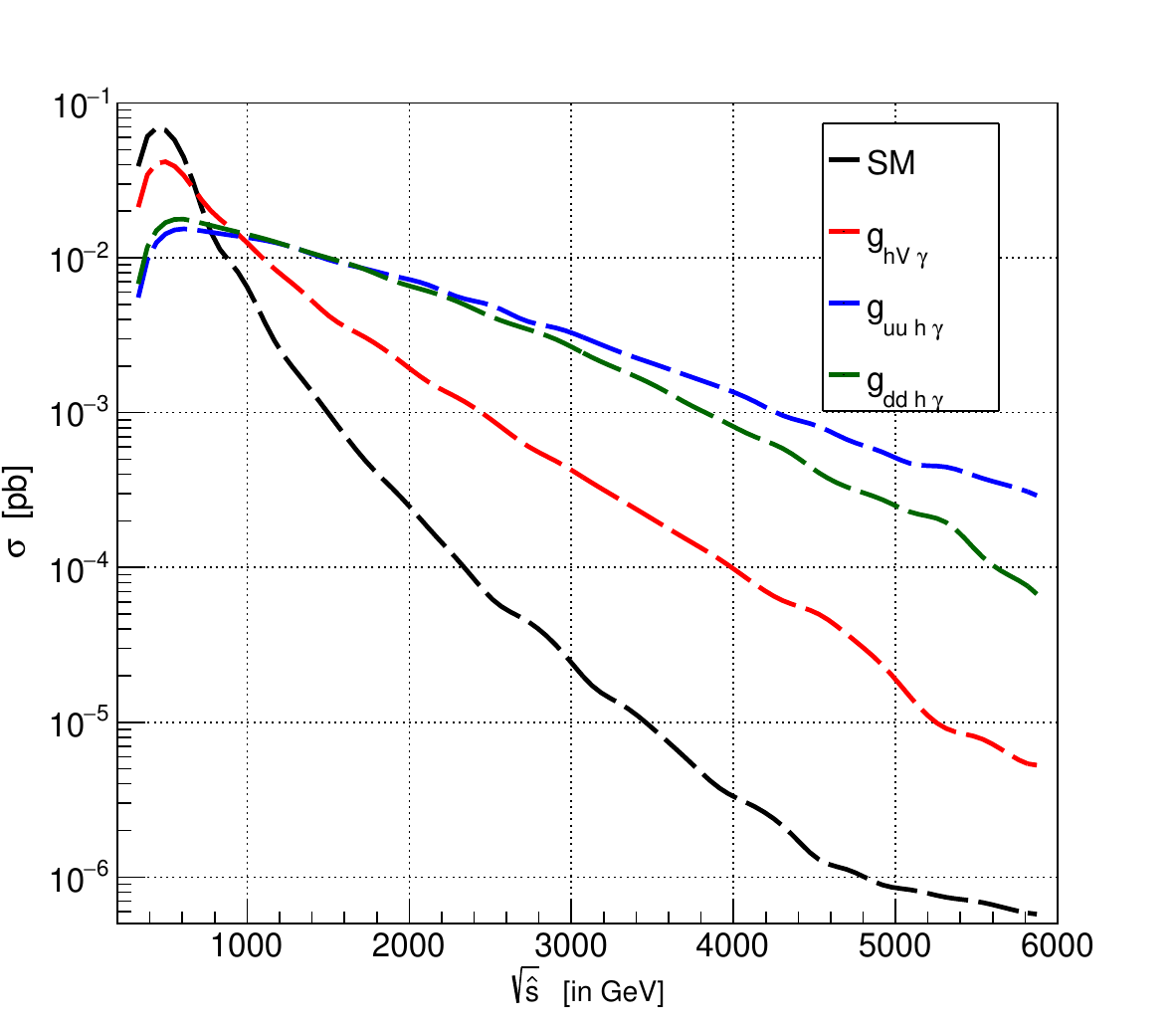}\\
				(a)&(b)
		\end{tabular}}
		\caption{(a) The cross-section variation for $pp \to h \gamma$ generated by different dimension-6 operators at the parton level (colored solid lines at 14 TeV and dashed lines at 13 TeV). (b) Variation of cross-sections with center of mass energies $\sqrt{\hat{s}}$ for the various Higgs anomalous couplings. Cross-section for SM background from $q \bar{q}$ annihilation shown as black dashed line.}
		\label{cxratio}
	\end{figure}

	
	The variation of  Higgs + photon  production  cross-section  for  different  values of  effective  couplings has been presented	in Fig.~\ref{cxratio} (left panel). One can see that the effect of fermionic operators is more pronounced than the bosonic operators. The  enhancement  of  cross-section  in  case  of  dimension-6  operators  involving  fermions  can  be  accounted by the absence of an extra propagator which is present in the SM-like $h \gamma$ production involving	the bosonic-type operators.  Most of the high-energy contribution due to these dimension-6 operators have an amplitude that is distinct from the SM contribution because of a quadratic growth  of energy.
	
	Cross-sections for $pp \to h \gamma$ as function of $\sqrt{\hat{ s}}$ have been presented for different dimension-6 operators of our interest along with the SM  in Fig.~\ref{cxratio} (right panel). A value of ${C_i\over \Lambda^2} = 0.3 ~\rm TeV^{-2}$ is used in this plot. The $h \gamma$ production rate, in presence of dimension-6 operators,  falls off at a rate slower than that in the SM. The reason behind this can be accounted as the following. In case of dipole-type operators, their interference with the SM (being proportional to the light quark mass) is extremely tiny. These operators generate four-point interactions which proceeds with absence of a propagator compared to the SM $q\bar{q} \to h \gamma $ process. Such an operator generates contribution which grows quadratically with energy of scattering.  On the other hand, $q\bar{q} \to h \gamma$ process driven by  the bosonic operators, having similar propagator structure with the SM, interfere with the SM amplitude. New physics contributions to $q\bar{q} \to h \gamma$ rate, proportional to $C\over \Lambda^2$ and $C^2 \over \Lambda^4$ become insensitive to the energy of scattering at high energy scales. These parton level cross-sections when folded with parton density functions result into physical cross-sections which fall with $\sqrt {\hat s}$, but the signatures of dimension-6 operators are captured in the slower rate of fall of cross-section when compared to the pure SM case (see Fig. 
	\ref{cxratio}(b)).  However, it must be borne in mind that the relative rate of fall  of the cross-section also depends on the value of $C$.  The aforementioned observation clearly tells that these operators can be  constrained from the considerations of unitarity.
	
	In this study, we consider the regime where the Higgs and the photon are highly boosted. The total hadronic decay products of Higgs can be contained in a large radius jet (fat-jet). We demand at least one Higgs jet in the signal sample by reconstructing its invariant mass using the jet substructure technique, along with a hard photon recoiling against this fat-jet in the transverse plane. The radius of the fat-jet can be approximated by the relation $R \sim \frac{2 m_h}{p_{T_{h}}}$. The following SM processes can mimic the $h \gamma$ signal.
	\begin{itemize}
		\item Continuum $\gamma j$ turns out to be the most dominant background. This background can be efficiently suppressed in the case of $h\gamma$ by applying a Higgs tagger to the jet. The efficiency of b-tagging and the misidentification
		rates of a $c$-jet or a light jet posing as a b-tagged jet are taken to be dependent on the transverse momentum of jets as given in Ref.~\cite{CMS:2012feb}.
		\item The $Z/W \gamma+jets$ process has almost similar topology to the signal, but is subdominant due to its low cross-section.
		\item The production of $t\bar{t}\gamma$ followed by hadronic decays of the top quarks will also contribute to the background.  However, demanding a high $p_T$ photon along with Higgs tagging would suppress this background. Similarly, single top production $tj\gamma, tb\gamma$ also contribute in the background.
		\item The $pp \to h \gamma$ associated production in the SM has a nominal rate either due to the fact that Higgs boson has negligibly small couplings with the partons or the process is driven by higher order one loop  contribution. The gluon-initiated contribution $gg \to h \gamma$ is vanishing due to Furry's theorem  \cite{Furry:1939,PeskinSchroeder}.
		\item Some other SM processes, such as QCD, $t\bar{t}$+jets, $W$+jets, $Z$+jets, despite having no direct sources of hard photons, may also contribute to the background owing to their large production cross-sections coupled with mistagging of jets or leptons leading to fake photons. However, the cumulative effect of hard transverse momentum of photon as well as the boosted sytem of the final state requirement and Higgs tagging renders these contribution negligible.
	\end{itemize}
	All the aforementioned background processes have been estimated at the LO and multiplied by appropriate K-factors to obtain the NLO cross-sections following Ref.~\cite{CMS:2018cno}. The cross-sections as used in present analysis are listed in Table \ref{tab:bkgxsec}.
	\begin{table}[b!]
		\begin{center}
			\renewcommand{\arraystretch}{1.5}
			\begin{tabular}{|c|c|c|c|}
				\hline\hline
				\multicolumn{2}{|c|}{Process} 			& $\sigma$ (fb) at 13 TeV 	& $\sigma$ (fb) at 14 TeV   \\ \hline\hline
				\multirow{2}{*}{$J \gamma $~\cite{Alwall:2014hca}}& $ b-tagged$  & $47.20 \times 10^3$& $54.19 \times 10^{3}$  \\ \cline{2-4} 
				& $ untagged $ 				& $188.9 \times 10^3$ & $212.92 \times 10^{3}$ \\ \hline
				\multirow{2}{*}{$V \gamma + jets$~\cite{Krause:2017nxq,Lombardi:2020wju}}   & $Z \gamma + jets$ & $140.80$	& $158.56$\\ \cline{2-4} 
				& $W \gamma + jets$  		& $541.89$ 	& $933.04$\\ \cline{2-4} \hline
				\multirow{2}{*}{Top $\gamma + jets$~\cite{Pagani:2021iwa}} & $t\bar{t} \gamma + jets$  & $6.69$ & $8.22$\\ \cline{2-4} 
				& $tj \gamma + jets$  		& $5.418$ & $6.978$\\ \cline{2-4} \hline
				Electroweak~\cite{Gabrielli:2016mdd} & $h \gamma$ & $0.01377$ & $0.01617$ \\ \hline
			\end{tabular}
			\caption{Cross-sections for the background processes considered in this analysis at the 13 TeV and 14 TeV LHC.  These processes except the electroweak $h \gamma$ production in the SM have been estimated at the LO and multiplied by appropriate K-factors to obtain the NLO cross-sections following the procedure in Ref~\cite{CMS:2018cno}. }
			\label{tab:bkgxsec}
		\end{center}
	\end{table}
	
\begin{figure}[t!]
		\centering
		\includegraphics[width=14.0cm,height=10.0cm]{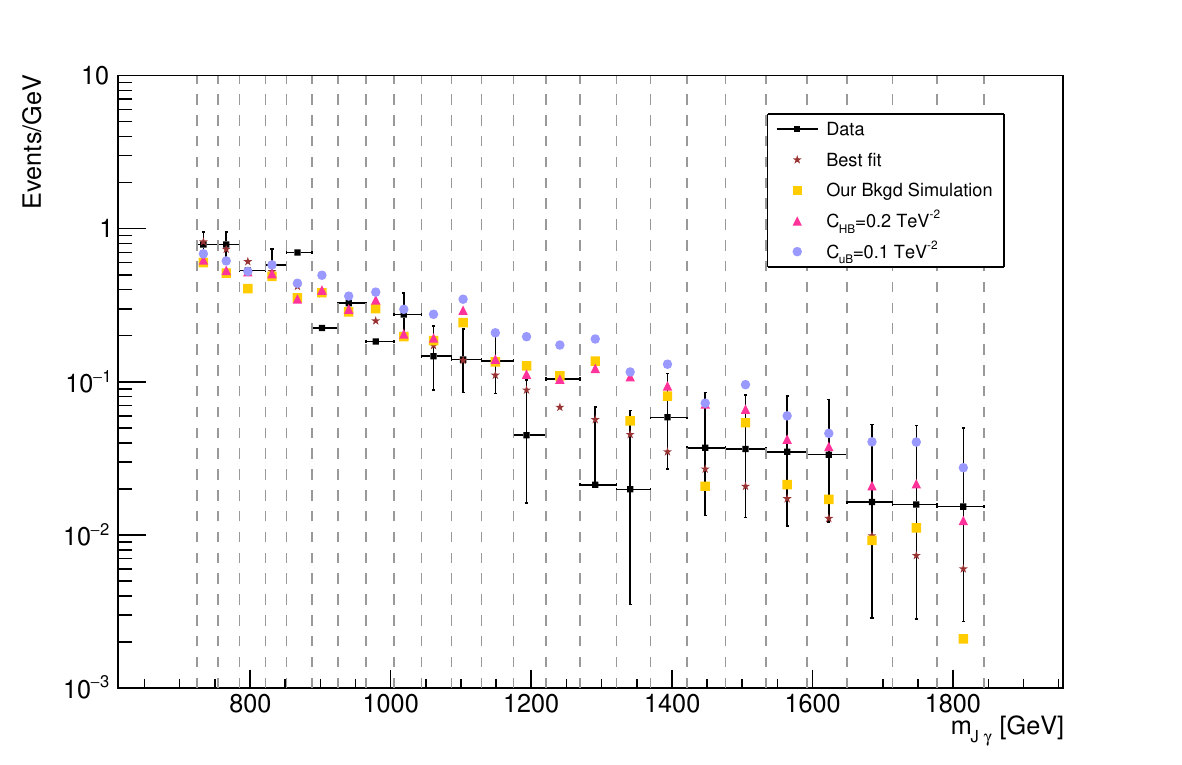}
		\caption{\label{limits} $J \gamma$ invariant mass distribution with $36~\rm fb^{-1}$ data at 13 TeV run of LHC. The SM prediction at the NLO level (yellow squares) are shown along with  predictions corresponding to Wilson coefficients $\frac{C_{uB}}{\Lambda^2}$ (purple circles) and $\frac{C_{HB}}{\Lambda^2}$  (pink triangles).  The data (black dots with error bars) and best fit points (brown stars) to data are also illustrated from  the analysis in Ref.~\cite{CMS:2018cno}.}
	\end{figure}

	The CMS \cite{CMS:2018cno} and the ATLAS \cite{ATLAS:2018sxj} Collaborations have looked the $h\gamma$ final state in search of a resonance decaying into Higgs and photon. No significant excess over the SM expectation has been reported. The CMS has collected 187  events with $\int \mathcal{L}dt = 36~\rm fb^{-1}$ in the signal region  $720~\rm GeV < m_{J \gamma} < 1.82~\rm TeV$. This corresponds to a cross-section of 5.2 fb including acceptance and efficiency of selection cuts. Our simulation of SM events yield 164 events with a scale uncertainty of 5\%. We have rescaled the LO cross-section 4.56 fb by the corresponding QCD-NLO factor. With this SM background expectation, we compare Fig. 2 (left panel) of Ref.~\cite{CMS:2018cno} as shown in Fig.~\ref{limits}.
	
	We now consider the process involving contributions from dimension-6 operators and perform a goodness-of-fit test with the nine SMEFT coefficients on the $J \gamma$ invariant mass distribution. We construct the $\chi^2$ function for the aforementioned EFT couplings as 
	\begin{equation}
	\chi^2(C) = \sum_{i}^{bins} \frac{(N_i(C)^{theo}-N_i^{exp})^2}{(\sigma_i^{exp})^2}
	\end{equation}
	where $N_i^{exp}$ denotes the observed number of events, $N_i(C)^{theo}$ estimates the expected number of events, assumed to be different from the SM due to the presence of non-zero EFT couplings. $\sigma_i^{exp}$ quantifies the error including both statistical and systematic uncertainties added in quadrature. We have assumed an overall systematic uncertainty which is $3\%$ of statistical uncertainty.
	\begin{equation}
	\sigma_i^{exp} = \sqrt{N_i^{exp} + \Delta_{sys}N_i^{exp}} \nonumber
	\end{equation}
	We minimise the $\chi^2$-function to derive the limits of the parameter space allowed by the experimental data at $2 \sigma$ level.	
	\begin{table}[h!]
		\centering
		\renewcommand{\arraystretch}{1.5}
		\begin{tabular}{|c|c|c|}
			\hline
			95 \% C.L. (TeV$^{-2}$) & Single Parameter & Profiled  \\ \hline \hline
			$\frac{C_{uW}}{\Lambda^2}$ & $ [-0.42,0.44] $&$ [-0.48,0.47] $\\ \hline
			$\frac{C_{uB}}{\Lambda^2}$ & $[-0.18,0.21] $& $ [-0.47,0.48] $ \\ \hline
			$\frac{C_{dW}}{\Lambda^2}$ & $[-0.38,0.39] $& $ [-0.47,0.53] $\\ \hline
			$\frac{C_{dB}}{\Lambda^2}$ & $[-0.28,0.30] $& $ [-0.48,0.43] $\\ \hline
			$\frac{C_{HW}}{\Lambda^2}$ & $[-1.03, 1.01] $& $ [-1.82,1.98] $\\ \hline
			$\frac{C_{HB}}{\Lambda^2}$ & $[-1.53, 1.42] $& $ [-1.87,1.95] $\\ \hline
			$\frac{C_{HWB}}{\Lambda^2}$ & $[-2.11,3.10]$ &  $ [-4.31,7.54] $  \\ \hline
			$\frac{C_{HD}}{\Lambda^2}$ & $[-22.88,17.62]$  &  $ [-29.85,34.57] $ \\ \hline
			$\frac{C_{H\Box}}{\Lambda^2}$ & $[-7.49,9.49]$ &$ [-17.45,20.13] $   \\ \hline
		\end{tabular}
		\caption{95 \% C.L. bounds for the fermionic and bosonic operator coefficients in the Warsaw basis. The bounds are given in $\rm TeV^{-2}$ units.}
		\label{bounds}
	\end{table}
	The numerical results obtained by fitting the $h \gamma$ production process using both profiling and single parameter are given in Table \ref{bounds}. In general, it can be observed that the single parameter limits are notably more restrictive than the profiled ones. This is expected as the profiling approach allows for the simultaneous variation of multiple parameters of interest in the $\chi^2$ function, capturing potential correlations and dependencies between them. Thus, by using the  profiling method, the complex and multidimensional nature of the fit method is demonstrated, compared to considering one operator  at a time that restricts the analysis to very specific BSM models that could be hard to realise. It is noteworthy that the limits on the dipole operators obtained here are more restrictive than our previous study on their effects on VBF Higgs production~\cite{Biswas:2021qaf}. We also note that the Wilson coefficients are more constrained from the higher energy bins. We will use these allowed values of the effective couplings to investigate the prospect of finding such a signal at 14 TeV run of the LHC with $\mathcal{L}=3000$ fb$^{-1}$ integrated luminosity in the following sections. 	
	
	\section{Collider Simulation}
	\label{sec:collider_analysis}
	\subsection{Cut-based Analysis}
	\label{sec:cut_based_analysis}
	We generated samples for the signal that lead to the $pp \to h~(\to \rm fat$-$\rm jet)\gamma$ final states where non-zero EFT couplings in dimension-6 are present along with the SM. The cross-sections are evaluated upto $\mathcal{O} \left( \frac{1}{\Lambda^4}\right)$ at 14 TeV $pp$ center-of-mass energy. We also generated the SM background events for the processes listed in Table~\ref{tab:bkgxsec}. We choose the following basic selection criteria to select events in the signal and background for further analysis:
	
	\begin{itemize}
		\item Photons are required to have $p_{T_{\gamma}} > 150~\rm GeV$ and $|\eta_{\gamma}|<2.47$ excluding $1.37<|\eta_\gamma|<1.52$.
		\item We veto the events if any lepton with $p_T > 10~\rm GeV$ lies in the central psuedorapidity range $|\eta| < 2.4$.
		\item We only select those events with at least one fat-jet of radius 0.8 and with   $p_T >  70$ GeV.
		\item To suppress any theoretical uncertainties and veto additional jets in the event sample, the fat-jet should satisfy the criteria of $|\Delta \phi(J, \gamma)| > 0.4$.
		\item To obtain statistically significant events, we select events with jet mass to be  $m_{J} > 50~\rm GeV$.
	\end{itemize}
	We now proceed to discuss the main kinematic characteristics of signal and the SM background.  To accentuate the distinguishability, we choose  a few benchmark points as listed in Table~\ref{BMpoints}. Operators $\mathcal{O}_{fW}$ and $\mathcal{O}_{fB}$ differ in the cross-section by a  factor of $\rm tan^{2} \theta_W$. BP3 closely mimics the SM interaction. Benchmark points $\{BP1,BP2,BP3\}$ have been chosen alongside the SM (Fig.\ref{tab:Feynmandiag} (a-i)) with only one non-zero EFT couplings  at a time. The values of such couplings are allowed by the  available  Higgs data from the LHC itself. We also choose benchmark couplings with two couplings non-zero at a time so as to emulate a scenario where a significant deviation from the SM is observed. These values are chosen based on the fact that certain linear combinations of couplings are more sensitive to the new physics interactions as expressed in Table~\ref{tab:rgb}. 
	\begin{table}[b!]
		\centering
		\renewcommand{\arraystretch}{1.5}
		\begin{tabular}{|c|c|}
			\hline
			Signal & $\frac{C_{i}}{\Lambda^2}$ (TeV$^{-2}$) \\ \hline \hline
			BP1 & \rm $\frac{C_{uB}}{\Lambda^2}= 0.1$ \\ \hline
			BP2 & \rm $\frac{C_{dW}}{\Lambda^2}= 0.2$ \\ \hline
			BP3 & \rm $\frac{C_{HW}}{\Lambda^2}= 0.2$ \\ \hline
			BP4 & \rm $\frac{C_{uW}}{\Lambda^2}= 0.08$,$\frac{C_{uB}}{\Lambda^2}= 0.08$ \\ \hline
			BP5 & \rm $\frac{C_{dW}}{\Lambda^2}= 0.15$,$\frac{C_{dB}}{\Lambda^2}= -0.15$\\ \hline
			BP6 & \rm $\frac{C_{HB}}{\Lambda^2}= 0.3 $,$\frac{C_{HW}}{\Lambda^2}= -0.3$\\ \hline
		\end{tabular}
		\caption{Listing of benchmark points used in our study}
		\label{BMpoints}
	\end{table}

\begin{figure}[t!]
		\centering
		\subfloat{
			\begin{tabular}{cc}
				\includegraphics[width=7.5cm,height=5.8cm]{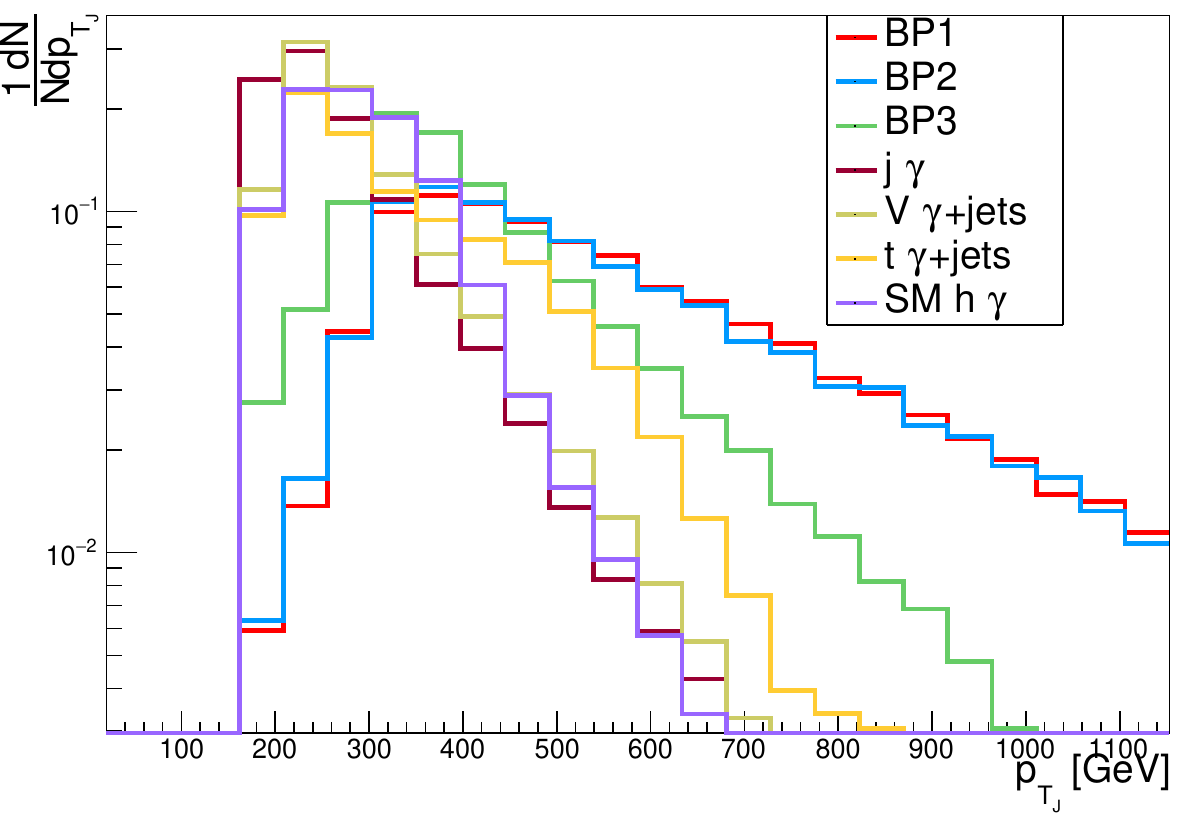}&
				\includegraphics[width=7.5cm,height=5.8cm]{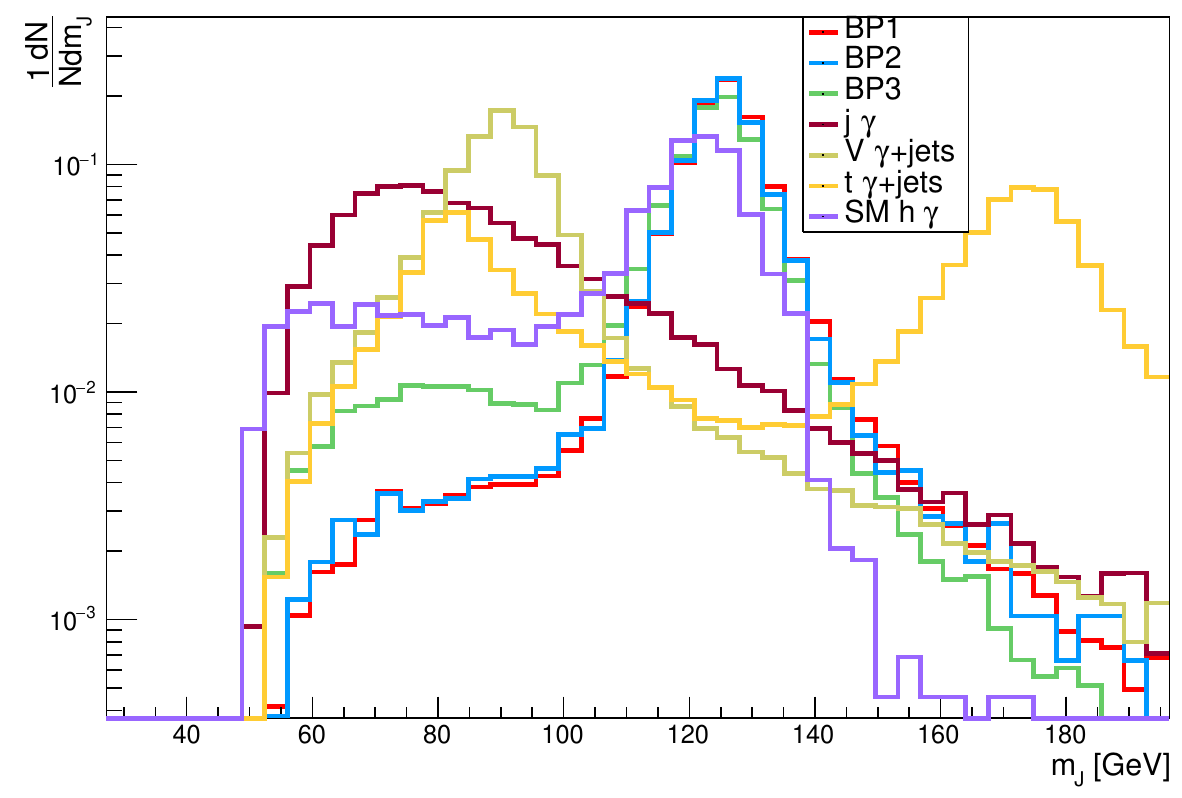}  \\
				\hspace{4mm}(a)&\hspace{8mm}(b)\\
				\includegraphics[width=7.5cm,height=5.8cm]{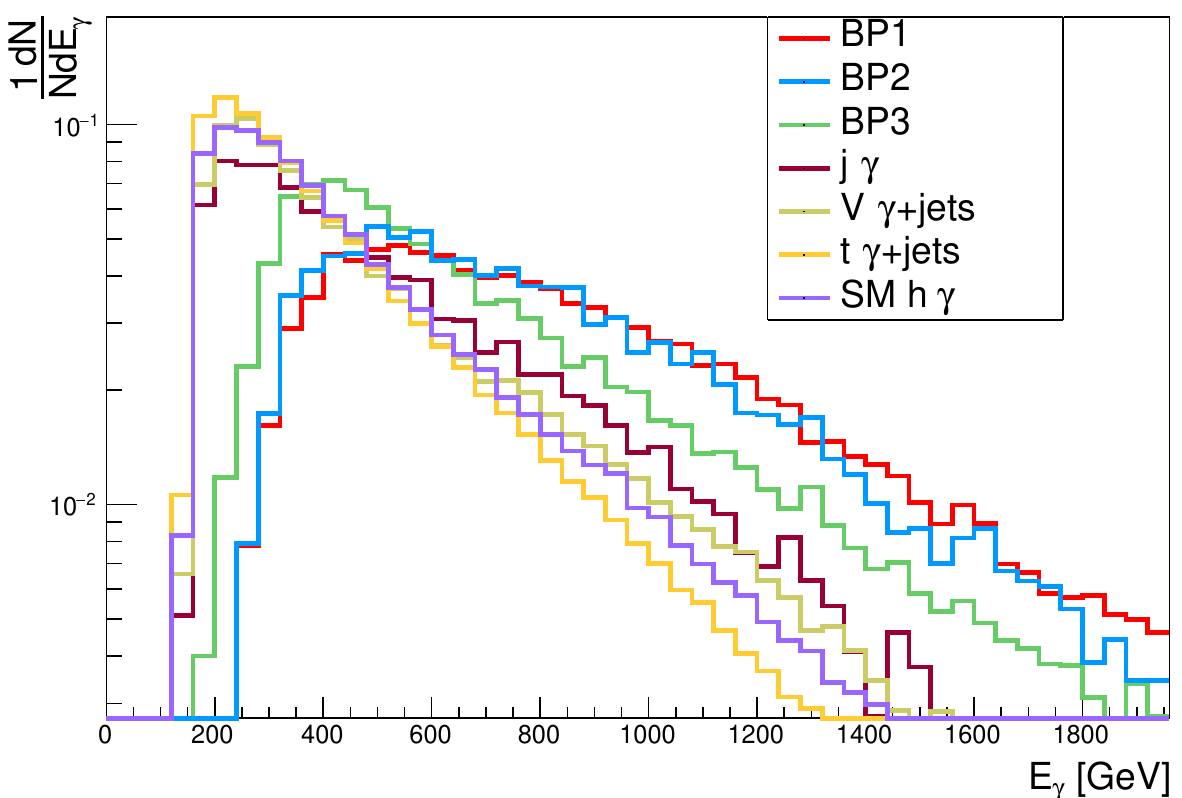}&
				\includegraphics[width=7.5cm,height=5.8cm]{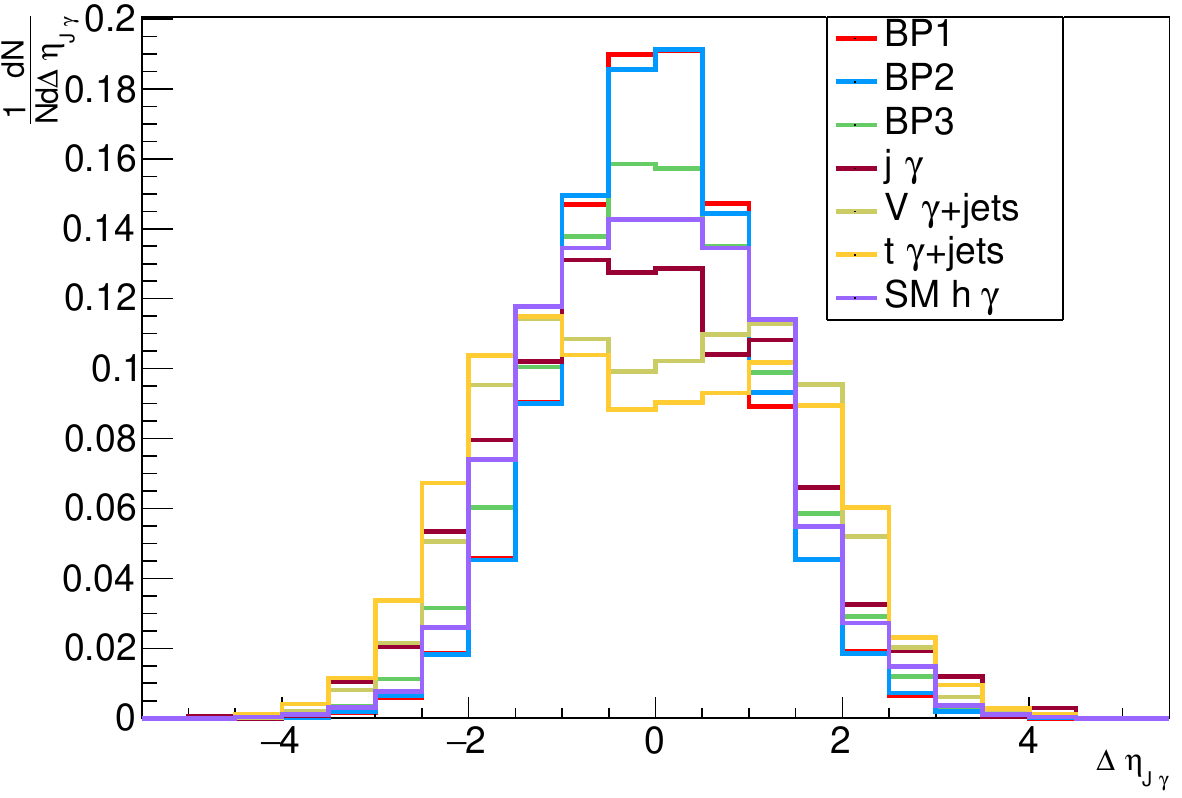} \\
				\hspace{4mm}(c)&\hspace{4mm}(d)
		\end{tabular}}
		\caption{Normalized distributions of (a) $p_{T_{J}}$, (b) $m_{J}$, (c) $E_{\gamma}$ and (d) $\Delta \eta_{J \gamma}$ for signal as well as SM backgrounds. The EFT contributions include the effects from the SM, the interference of SM and dimension-6 operator interaction and quadratic order dimension-6 operator contributions. For the EFT predictions, we have chosen BP1 = $\frac{C_{uB}}{\Lambda^2}= 0.1~\rm{TeV^{-2}}$ (red), BP2 = $\frac{C_{dW}}{\Lambda^2}= 0.2~\rm{TeV^{-2}}$ (blue) and BP3 = $\frac{C_{HW}}{\Lambda^2}= 0.2~\rm{TeV^{-2}}$ (green).}
		\label{kin_distribtn}
	\end{figure}

	Fig.~\ref{kin_distribtn} shows the normalized distributions of some of the kinematic variables for the three BPs and the SM background processes. The variables $p_{T_{J}}$ and $E_\gamma$ (Fig.~\ref{kin_distribtn}(a) and (c)) are quite efficient in distinguishing the new interactions from most of the SM backgrounds. The availability of larger parton center-of-mass energy in these interactions pushes the energy of photon ($E_\gamma$) and the transverse momentum of fat-jet ($p_{T_{J}}$) to higher values. The photons in the signal events exhibit a hard $p_T$ and mainly appear in the central region of the detector. Values of $E_\gamma$ and $p_{T_{J}}$ extend to 1.8 TeV and 1 TeV. We see that significant enhancement from the SM backgrounds can be seen due to the contribution of both the type of dimension-6 operators. In presence of dipole-type operators the cross-section grows faster at higher energies compared to the bosonic-type operators or in the case of SM.  This can be attributed  to the absence of an extra s-channel vector boson propagator in the amplitudes involving dipole operators.
	
	Instead of performing a standard resolved analysis, where two separate b-tagged jets are required, we demand a fat-jet with a considerably large cone radius. We employ the so-called BDRS approach~\cite{Butterworth:2008iy} with minor modifications to maximize the sensitivity. This technique helps in discriminating boosted electroweak-scale resonances from large QCD backgrounds.
	This approach recombines jets using the CA algorithm with a significantly large cone radius to contain all the decay products of the resonance (Higgs boson). The candidate jet J is broken into two subjets by unwinding the last stage of jet clustering. The two subjets $j_{1}$ and $j_{2}$ are labelled such that $m_{j_{1}} > m_{j_{2}}$. And a mass drop condition is applied such that $m_{j_{1}}<\mu m_J$ with $\mu=0.66$ ($m_{J}$ being the mass of the fat-jet) and further a symmetry criterion  $\frac{min(p_{T,j_1}^2,p_{T,j_2}^2)}{m_J^2}\Delta R^2_{j_1,j_2} > 0.09$ is applied between the two sub-jets. If the condition is not satisfied, the  softer subjet, $j_2$ is discarded and only the subjet $j_1$ forms the part of the remaining analysis and the process is reiterated. The algorithm is stopped with the final jet $J$ obtained if the mass drop condition is obtained. This criteria is fairly efficient in eliminating QCD jets, but still can vitiate the jet reconstruction at LHC due the underlying events, which is  also a problem at the enhanced energies and luminosities of LHC. To remove the remaining rare QCD events where a hard gluon is emitted and the underlying events, we need to filter the Higgs neighbourhood.  The constituents of $j_1$ and $j_2$ are again recombined using the CA algorithm with a cone radius $R_{filt} = min(0.2, R_{b\bar{b}}/2)$\footnote{We find that in our study, choosing $R_{filt} = min(0.2, R_{b\bar{b}}/2)$ facilitates in  removing the backgrounds.}. Only the hardest three filtered subjets are retained to reconstruct the resonance. We show the distribution of this filtered jet mass in  Fig.~\ref{kin_distribtn}(b) for the different signals and the dominant backgrounds. It is evident from these distributions that the peak around $115-140~\rm GeV$ reflect the Higgs peak for the signal processes whereas for most of the backgrounds, the peaks are below 50 GeV reflecting that the fat-jet mimicing from either single prong hard QCD jet or a peak around 90 GeV reflects $Z$ boson or peak about 165-185 GeV from a top.
	
	The most dominant backgrounds for this process produce events with larger angular separations, $\Delta \eta$ between the photon and the fat-jet. For the new physics signal, the separation of the final states is isotropic leading to narrow separations in $\eta$ as shown in Fig.~\ref{kin_distribtn}(d).
	
	Having discussed the kinematic observables, we proceed to find a suitable cut-flow, that will significantly discriminate between signal and background.
	\begin{table}[t!]
		\centering
		\scalebox{0.85}{%
			\begin{tabular}{||c||c|c|c|c|c|c||}
				\hline
				& \multicolumn{6}{c||}{Cross-sections at  $\sqrt{s}= 14 $ TeV (fb)}  \\ \cline{2-7}
				Cut flow   & \multicolumn{6}{c||}{EFT driven signals} \\
				\cline{1-7} 
				BPs & $BP1$ & $BP2$ & $BP3$ & $BP4$ & $BP5$  & $BP6$   \\
				\cline{1-7} 			
				Preselection& $14.41$ & $10.889$ & $8.849$ & $19.199$  & $18.660$ & $14.784$\\\hline
				
				Atleast 1 fat-jet with 2 B-meson tracks with $p_{T} > 200~\rm GeV $& $6.638$ & $4.984$ & $5.911$  & $8.798$  & $8.602$ & $9.876$\\\hline
				
				Atleast 1 isolated photon and lepton veto & $6.119$ &$4.507$& $5.576$ & $8.124$ & $7.799$ & $8.674$\\\hline
				
				Photon $p_{T} > 200$ GeV  & $5.879$ &$4.187$& $4.672$ & $7.796$ & $7.281$ & $8.081$\\\hline
				
				Atleast 1 fat-jet with two B-meson tracks with $p_T > 250$ GeV & $5.541$ &$3.733$& $2.945$ & $7.344$ & $6.602$ & $5.042$\\\hline
				
				2 mass drop subjets and $\ge 2$ filtered subjets & $2.504$ &$1.713$& $1.330$ & $3.319$ & $2.641$ & $2.136$\\\hline
				
				2 b-tagged subjets& $1.952$ & $0.762$& $0.385$  & $2.586$  & $1.096$ & $0.587$ \\\hline
				
				$110.0 < m_{H} < 140.0$ GeV & $0.966$ & $0.755$& $0.379$   & $1.281$  & $1.085$ & $0.335$\\\hline
				
				$\Delta R(\gamma, b_i) > 0.4, \slashed{E_T}<30 \rm GeV, |\eta_{h}|<2.5$  & $0.401$ &$0.302$& $0.181$  & $0.532$  & $0.435$& $0.193$\\\hline
				
				$\Delta R(\gamma, h) > 2.4$  & $0.393$ & $0.294$ & $0.178$ & $0.521$ & $0.392$ & $0.187$\\\hline
				
				$m_{J \gamma} > 600~ \rm GeV$& $0.370$ & $0.282$ & $0.171$ & $0.490$ & $0.353$ & $0.183$\\\hline
				
				$p_{T_{\gamma}} > 300~ \rm GeV$ & $0.338$ & $0.274$ & $0.168$& $0.448$ & $0.323$ & $0.179$\\\hline
				
		\end{tabular}}
		\caption{The cut-flow table for Higgs photon production from the dimension-6 operators.}
		\label{tab:bbgam_sig}
	\end{table}
	As evident from Table~\ref{tab:bbgam_sig}, cut efficiencies vary depending on the Wilson coefficients associated with the operators with different Lorentz structures involved.  
	\begin{table}[b!]
		\centering
		\scalebox{0.85}{%
			\begin{tabular}{||c||c|c|c|c||}
				\hline
				& \multicolumn{4}{c||}{Cross-sections at  $\sqrt{s}= 14 $ TeV (fb)} \\ \cline{2-5}
				Cut flow   & \multicolumn{4}{c||}{SM Backgrounds} \\
				\cline{1-5} 
				& jets + $\gamma$ & $W/Z + \gamma$ & $t\bar{t}\gamma+tj\gamma$ & $SM(h+\gamma)$ \\
				\cline{1-5}  			
				Preselection& $297407.2$ & $1148.86$ & $18.9696$ & $0.4731$  \\\hline
				
				Atleast 1 fat-jet with 2 B-meson tracks with $p_{T} > 200~\rm GeV $& $9932.125$ & $142.06$  & $16.483$  & $0.3763$  \\\hline
				
				Atleast 1 isolated photon and lepton veto& $7836.135$ & $119.946$ & $13.643$  & $0.3224$  \\\hline
				
				Photon $p_{T} > 200$ GeV & $3550.661$ & $66.034$  & $8.005$   & $0.1857$  \\\hline
				
				atleast 1 fat-jet with two B-meson tracks with $p_T > 250$ GeV& $2068.434$ & $48.855$  & $6.286$   & $0.0987$  \\\hline
				
				2 mass drop subjets and $\ge 2$ filtered subjets& $672.283$  & $23.737$  & $4.485$   & $0.0508$  \\\hline
				
				2 b-tagged subjets   & $115.134$  & $14.024$  & $0.824$   & $0.0263$  \\\hline
				
				$110.0 < m_{H} < 140.0$ GeV& $101.294$  & $7.541$   & $0.734$   & $0.0237$  \\\hline
				
				$\Delta R(\gamma, b_i) > 0.4, \slashed{E_T}<30 \rm GeV, |\eta_{h}|<2.5$& $38.532$   & $5.159$   & $0.279$   & $0.0115$  \\\hline
				
				$\Delta R(\gamma, h) > 2.4$& $38.066$   & $5.114$   & $0.276$   & $0.0108$  \\\hline
				
				$m_{J \gamma} > 600~ \rm GeV$& $20.063$   & $3.289$   & $0.179$   & $0.00892$ \\\hline
				
				$p_{T_{\gamma}} > 300~ \rm GeV$& $13.456$& $2.507$   & $0.147$   & $0.0060$ \\\hline
				
		\end{tabular}}
		\caption{The cut-flow table for jets + $\gamma $ production at $\sqrt{14}~\rm TeV$.}
		\label{tab:bbgam_bkg}
	\end{table}	
	It is worth pointing out here that with these choice of cuts, the SM background is reduced by 4 orders of magnitude.
	
\begin{figure}[t!]
		\centering
		\subfloat{	
			\begin{tabular}{cc}
				\includegraphics[width=11.5cm,height=7.5cm]{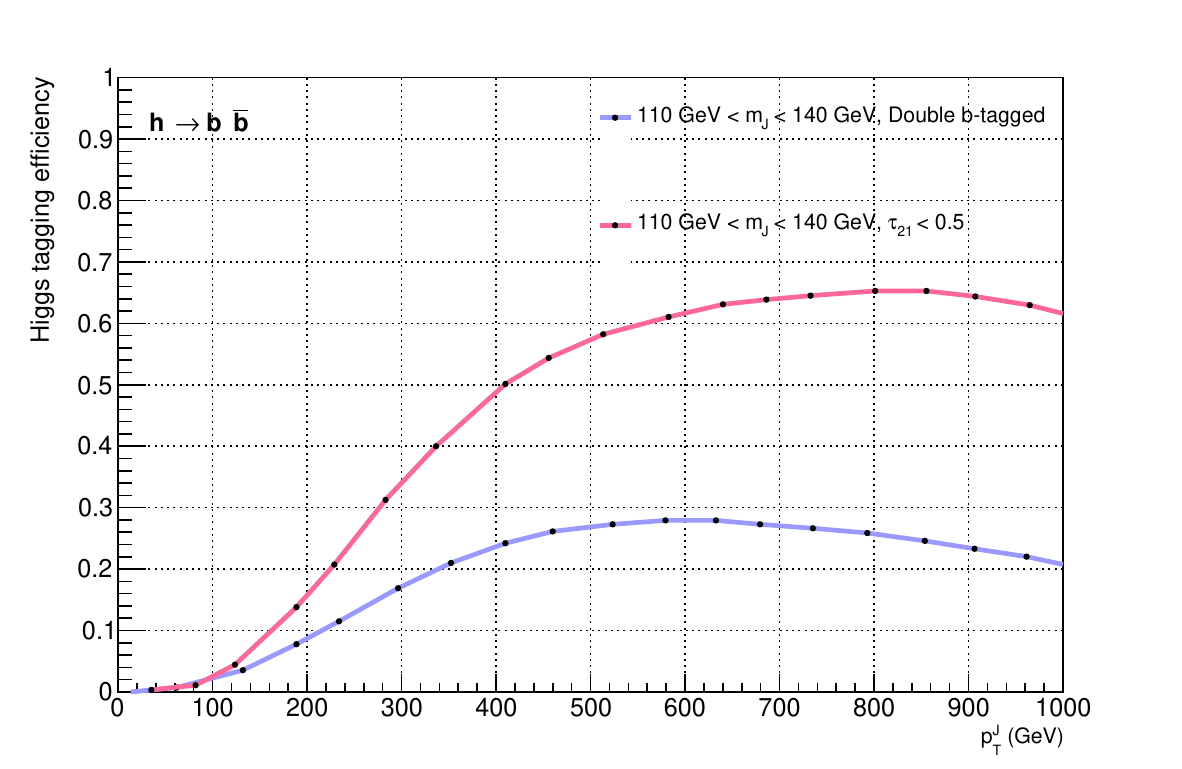} 
		\end{tabular}}
		\caption{Higgs tagging efficiency as a function of transverse momentum of large radius jet.}
		\label{higgstag}
	\end{figure}

	A comment is in order here. An immense progress has been made in  identifying the Higgs boson itself from its hadronic decay products despite the large QCD background at the LHC when the Higgs is being produced with sufficiently high $p_T$ or is in highly boosted regime \cite{Bhattacherjee:2012bu}. Using the properties of jet substructures also greatly helps in identifying signals  of double Higgs production via  various final states such as $b\bar{b}b\bar{b}$~\cite{FerreiradeLima:2014qkf}, $b \bar{b}\gamma\gamma$~\cite{Yao:2013ika} and $b\bar{b}WW^*$~\cite{Shao:2013bz}. Fig.~\ref{higgstag} shows the Higgs tagging efficiency, as a function of transverse momentum of Higgs boson for two cases.  The so-called  Higgs tagger algorithm described above is implemented in our analysis and its efficiency  i.e. the ratio of the number actually tagged as Higgs jets to the original number of Higgs produced is computed. In case one, the {\em{double b-tagged}} case,  both the harder sub-jets in the filtered fat-jet are required to be b-tagged. In the second case, a {\em{two-prong identifier}} is applied by requiring the fat-jet to have N-subjettiness, $\tau_{21}<0.5$ (see Section~\ref{sec:multivariate_analysis}).  This is also an effective way to identify such signal events with two-prong decays of Higgs~\cite{DiSipio:2019cfm}.  The Higgs tagging method by using 2-prong finder tagger is more efficient in finding the Higgs events than the b-tagging based technique. For instance, the method involving b-tagging, causes the event number to significantly fall to as low as $30 \%$ at $p_T=600~\rm GeV$ of Higgs. The latter can, however, be highly advantageous in removing events from the large QCD background untagged jets.
	
	It is to be noted that this figure is process independent and the tagging process can also serve to give information about the other heavy SM particles, distinct by their masses and the new particles that any UV theory predicts.
	
	As the process of our consideration is a hard $2 \to 2$ process, its amplitude can be completely specified by two variables, an energy variable and an angle.	We demonstrate the correlation between two such variables, the invariant mass of jet and photon, $m_{J \gamma}$ and center-of-mass opening angle, $cos~\theta^*$ in Fig.~\ref{2D_kinematics}, where it is found that the populated regions in the phase-space display a shift when the new physics effects due to higher dimensional operators are included. The following features emerge from Fig.~\ref{2D_kinematics}:
	\begin{figure}[h!]
		\centering
		\subfloat{
			\begin{tabular}{cc}
				\includegraphics[width=7.5cm,height=4.5cm]{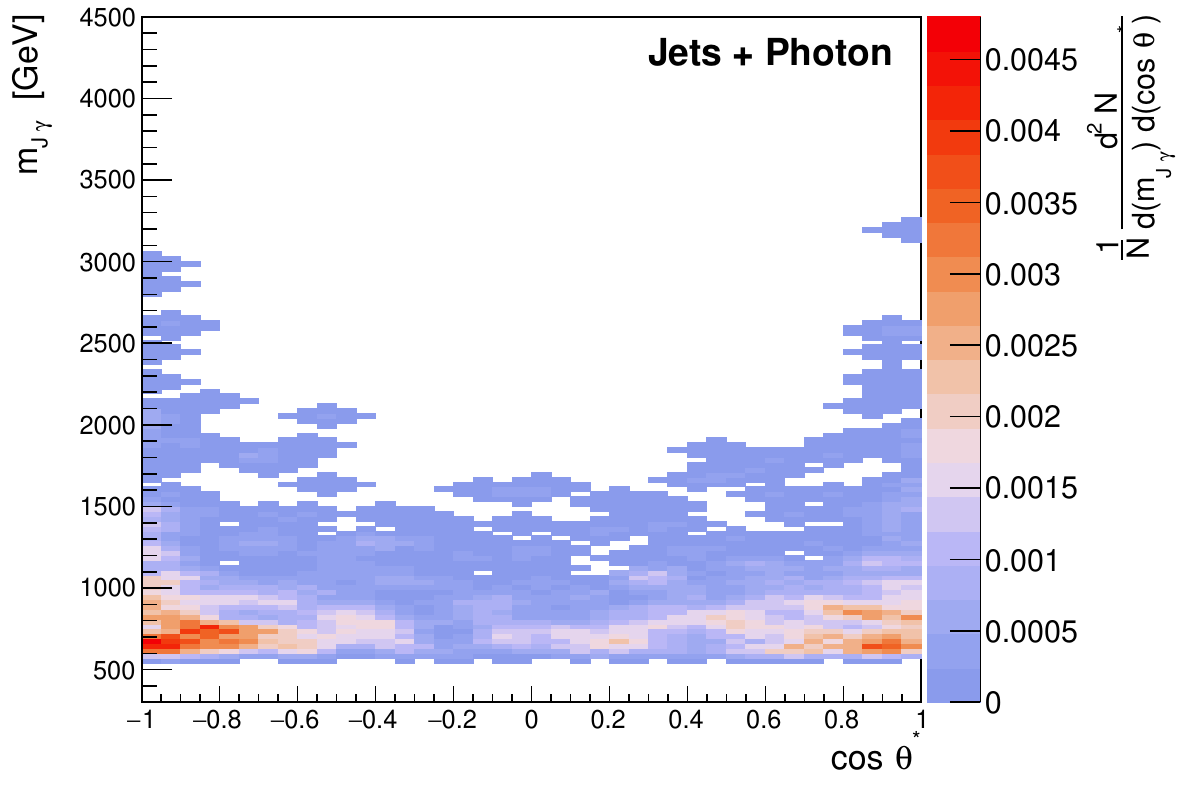}&
				\includegraphics[width=7.5cm,height=4.5cm]{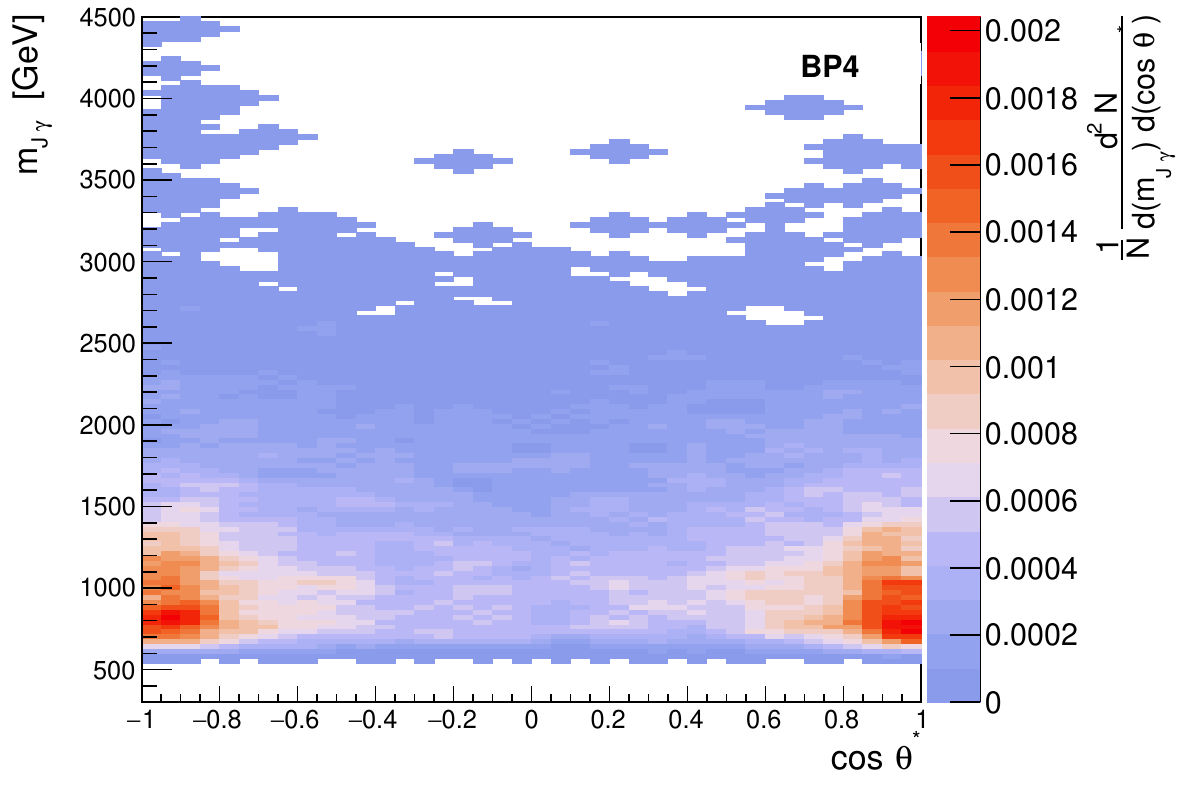}  \\
				\small (a)&(b)\\
				\includegraphics[width=7.5cm,height=4.5cm]{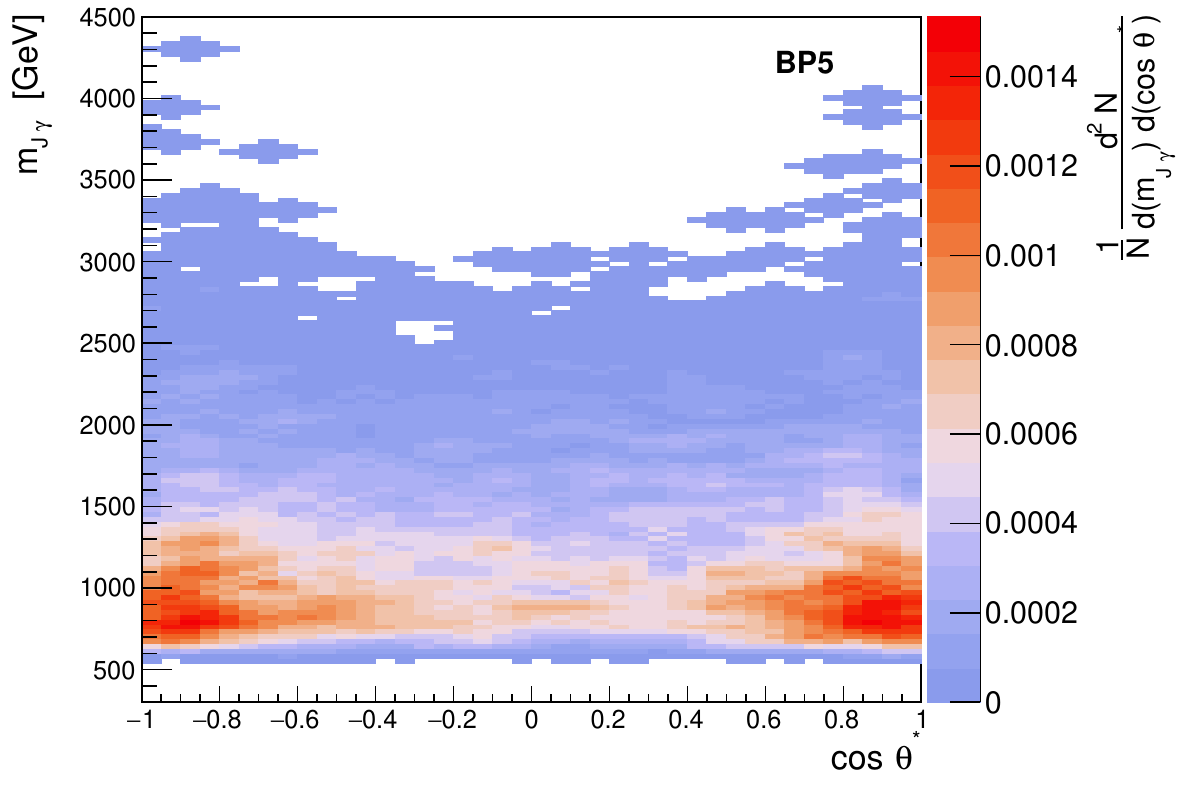}&
				\includegraphics[width=7.5cm,height=4.5cm]{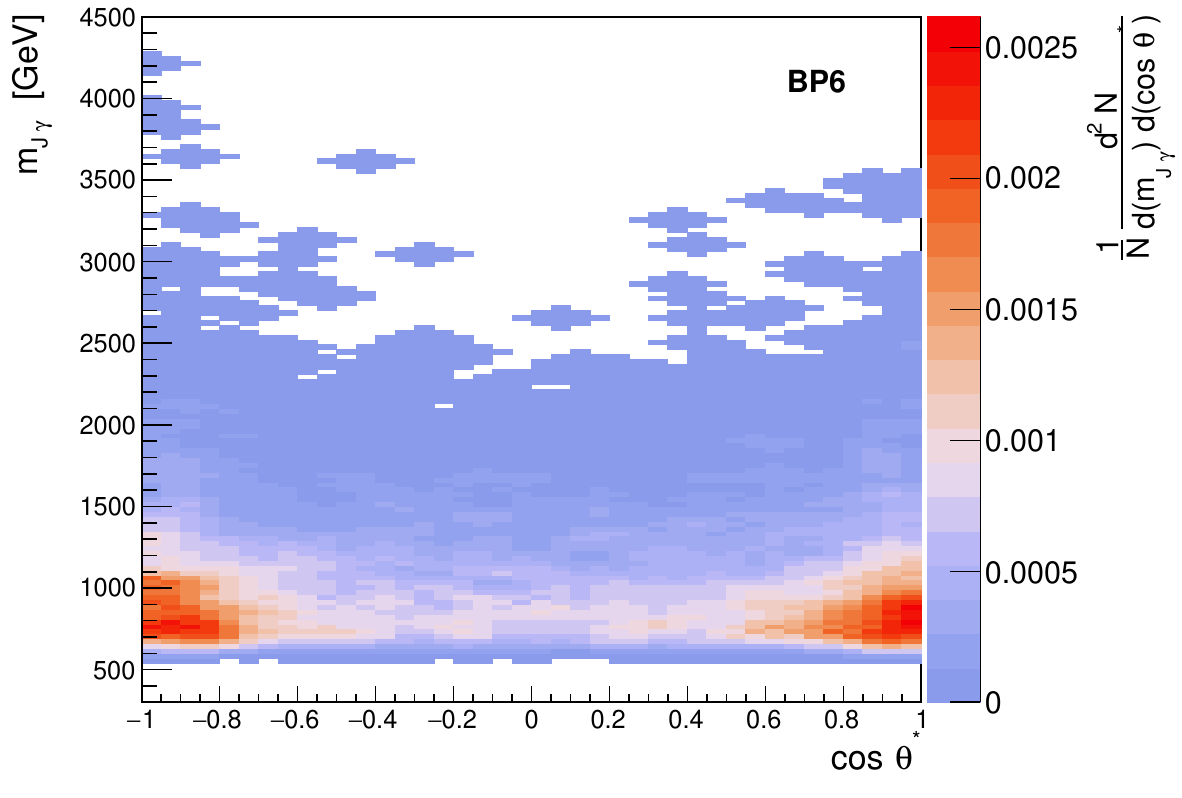} \\
				\small 	(c)&(d)
		\end{tabular}}
		\caption{Two dimensional histograms showing the correlation between
				invariant mass of jet and photon, $m_{J \gamma}$ and the scattering angle of photon in Higgs-photon rest frame, $cos \theta^*$. The $z$-axis indicates the normalized frequency of events, in arbitrary units. (a) represents the SM effect, comprising the dominant backgrounds, namely, jets+photon continuum from QCD processes, $V \gamma +$ jets, $t\gamma +$ jets and  SM $h \gamma$. The SM contributions in the absence of dimension-6 interactions have been subtracted in (b), (c) and (d), $i.e.$,they include the interference of SM and dimension-6 operator interaction and quadratic order dimension-6 operator contributions. ${|C_{uW}| \over {\Lambda^2}} = {|C_{uB}| \over {\Lambda^2}}= 0.08 ~\rm TeV^{-2}$ (BP4), ${|C_{dW}| \over {\Lambda^2}} = {|C_{dB}| \over {\Lambda^2}}= 0.15 ~\rm TeV^{-2}$ (BP5) and ${|C_{HW}| \over {\Lambda^2}} = {|C_{HB}| \over {\Lambda^2}} = 0.3 ~\rm TeV^{-2}$ (BP6) have been assumed in (b), (c) and (d) respectively.}
		\label{2D_kinematics}
	\end{figure}	
	\begin{itemize}
	
		\item For the background, the $m_{J\gamma}$ distribution is smoothly and rapidly falling while new physics contributions have rather slowly falling $m_{J\gamma}$ distributions. For the angular dependence analysis, it is to be noted that the main contributions to photon+jet final state background come from  a quark exchange leading to production of photon and jet, t-channel gluon exchange leading to QCD jet production along with a bremstrahlung radiation of a photon. In the center-of-mass frame, the angular distribution resulting from such background processes show a steeper increase as $|\rm cos \theta^*| \to 1$ than the new physics signal where either there is direct production of photon and jet from quarks or dominated by an s-channel propagator (red colour indicates larger larger number of events). A cut on the invariant mass of jet and photon of 1 TeV will characteristically feature the new phase space regions. Complementarily a cut $|\rm cos~\theta^*| \le 0.8$ for instance would remove a large sample of background events that peak in forward and backward directions.
		
		\item This implies that the new vertices tend to push the Higgs-like jet and the photon towards higher energy and the opening angle of the photon with respect to the beam line tends to be larger compared to the SM background. Such a correlation is most prominent for $C_{dB}$-$C_{dW}$ (Fig.~\ref{2D_kinematics}(d)) which has a very flat angular distribution as we look into regions with high $m_{J\gamma}$.  The effects of other two operators, too, extend to  $m_{J \gamma}$ values as high as 4 TeV.
	\end{itemize}
	\subsection{Projected sensitivities on the EFT couplings}
	The statistical significance $(\mathcal{S})$~\cite{cowan} of the signal $(s)$ over the total background $(b)$ has been calculated using
	\begin{center}
		\begin{equation}
		\mathcal{S}=\sqrt{2 \times \left[(s+b)ln(1+\frac{s}{b})-s\right]}
		\label{significance_eqn}
		\end{equation}
	\end{center}	
	We now focus on the sensitivity reach of the different benchmark points. We show the significance of the signal for all the benchmark points in the fat-jet$+\gamma$ channel in Table~\ref{significanceCBA} with the possible choice of integrated luminosity  (${\mathcal L}=300, 1000$ and $3000$ fb$^{-1}$). 
	\begin{table}[b!]
		\small
		\begin{center}
			\begin{tabular}{|c||c|c|c||c|c|}
				\cline{2-6}
				\multicolumn{1}{c||}{}&\multicolumn{3}{c||}{{Statistical significance}} & \multicolumn{2}{c|}{Required Luminosity} \\
				\multicolumn{1}{c||}{}&\multicolumn{3}{c||}{{(${\mathcal S}$)}} & \multicolumn{2}{c|}{(in $fb^{-1}$)} \\
				\hline
				Signal &   
				${\mathcal L}=300$ fb$^{-1}$  & 
				${\mathcal L}=1000$ fb$^{-1}$  & ${\mathcal L}=3000 $ fb$^{-1}$ & ${\mathcal S}=3{\sigma}$ & ${\mathcal S}=5{\sigma}$  \\ 
				\cline{1-6}
				BP1 & 1.410 & 2.574 & 4.458 & 1358.08 &  3772.44\\ 
				BP2 & 1.120 & 2.045 & 3.543 & 2150.91 & 5974.75  \\
				BP3 & 0.618 & 1.128 & 1.954 & 5494.64 & 15262.90 \\
				BP4 & 1.998 & 3.647 & 6.317 & 676.57 & 1879.36 \\ 
				BP5 & 1.334 & 2.437 & 4.221 & 1515.41 & 4209.48 \\ 
				BP6 & 0.628 & 1.147 & 1.987 & 5045.34 & 14014.80 \\ 
				\hline
			\end{tabular}
			\caption{Statistical significance of the signal for different benchmark points in the  Higgs $+\gamma$ analysis at 14 TeV LHC. The significance is estimated for three values of integrated luminosity (${\mathcal L}=300, 1000$ and $3000$ fb$^{-1}$). We also estimate the required integrated luminosity to achieve a $3\sigma$ and $5\sigma$ excess for each benchmark point at LHC with $\sqrt{s}=14$ TeV.}
			\label{significanceCBA}
		\end{center}
	\end{table}
	We find that BP4 is the most promising of all and should be observable in near future at the LHC with luminosity as low as  $ 700~\rm fb^{-1}$. BP3 and BP6 are the scenarios involving bosonic operators which would be difficult to observe in this final state.  In fact, an integrated luminosity of above $5000~\rm fb^{-1}$ would be required to observe a notable excess for the $h V \gamma$ coupling given by the above benchmark points.  This is however understandable as the corresponding process is SM-like  process, suppressed by a photon or $Z$ propagator. The rest of the benchmark points lead to $3 \sigma$ and $5 \sigma$ excess over the SM backgrounds with relatively nominal to higher integrated luminosities as shown in the last two columns of Table~\ref{significanceCBA}.
	
\begin{figure}[t!]
		\centering
		\subfloat{
			\begin{tabular}{cc}
				\includegraphics[width=7.6cm,height=6.8cm]{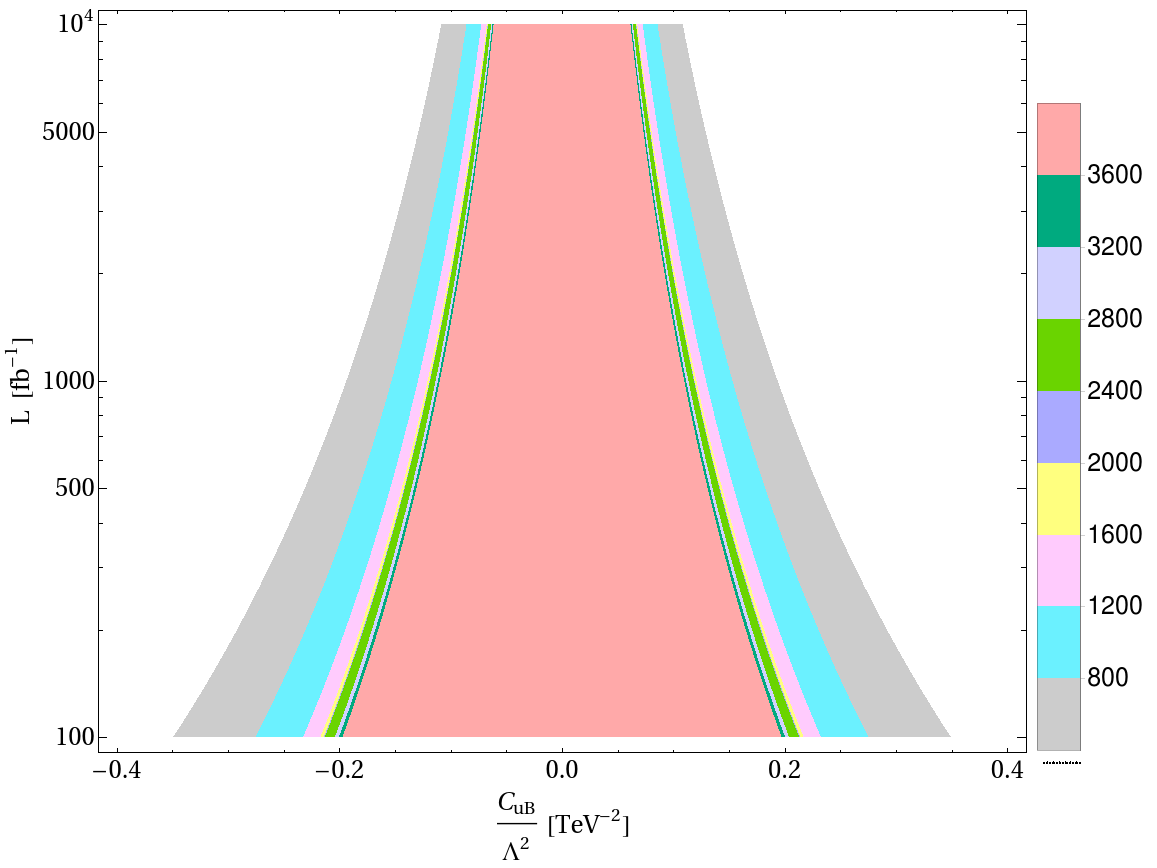} &
				\includegraphics[width=7.6cm,height=6.8cm]{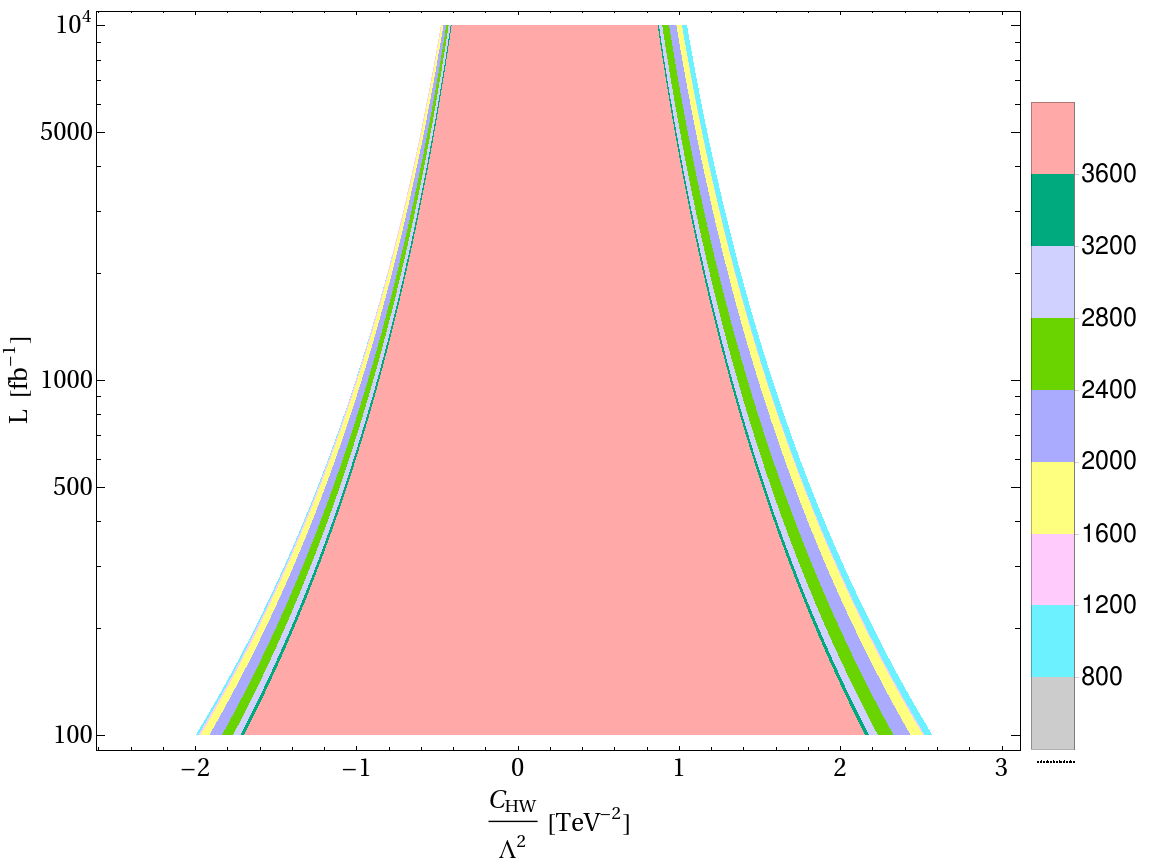}\\
				\hspace{4mm}(a) &\hspace{8mm}(b) 
		\end{tabular}}
		\caption{Required integrated luminosity at 14 TeV for achieving a 3$\sigma$ significance in bins of $m_{J \gamma}$ as a function of Wilson coefficients of operators (a)$\frac{C_{uB}}{\Lambda^2}$ and (b) $\frac{C_{HW}}{\Lambda^2}$. The different colored bands signify bins of $m_{J \gamma}$ each of width 400 GeV.}
		\label{reqdlumi}
	\end{figure}

	Using the distribution of $m_{J \gamma}$, we perform a binned analysis to project the reach of LHC in probing the higher dimensional operators. In this approach, we consider the range of  $m_{J \gamma}$  divided into bins of 400 GeV from $800$-$4000$ GeV. We study the required luminosity in each bin for achieving $3 \sigma$ significance as a function of the Wilson coefficients. This is shown for the operators $\mathcal{O}_{uB}$ and $\mathcal{O}_{HW}$ in Fig. \ref{reqdlumi}. Significance increases monotonically for both the operators with increasing bins of invariant mass of jet and photon. This is due to the quadratic growth of energy of the amplitude-square compared to that of the SM. With increasing energies, the photon and the Higgs get highly boosted allowing for a regime where background events are depleted. However, in such a region, the rate of cross-section of the EFT driven signal also decreases. It is the higher invariant mass bins that can give best signal to background sensitivity.
	
\begin{figure}[t!]
		\centering
		\subfloat{
			\begin{tabular}{ccc}
				\includegraphics[width=5.0cm,height=5.9cm]{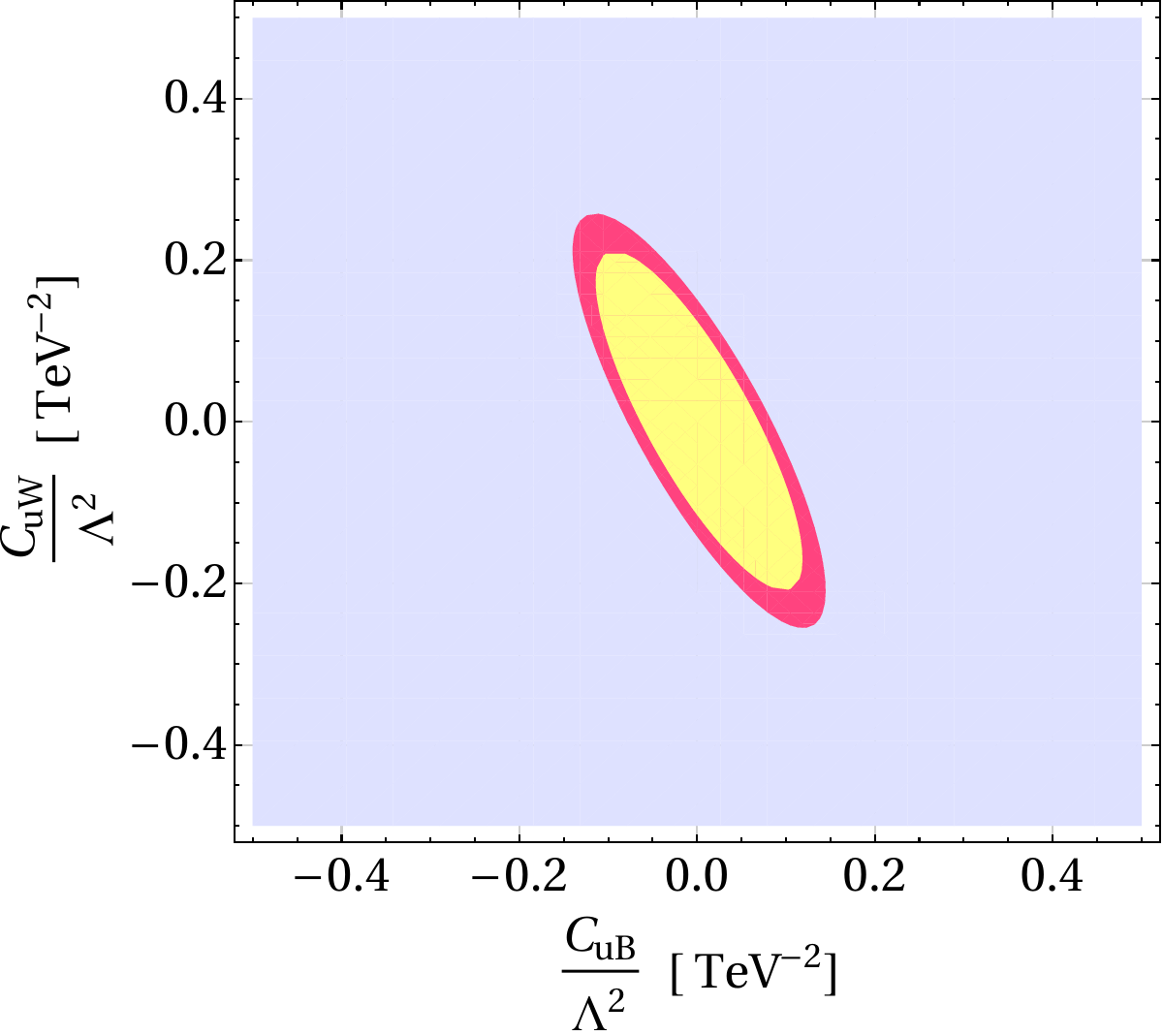}&
				\includegraphics[width=5.0cm,height=6.0cm]{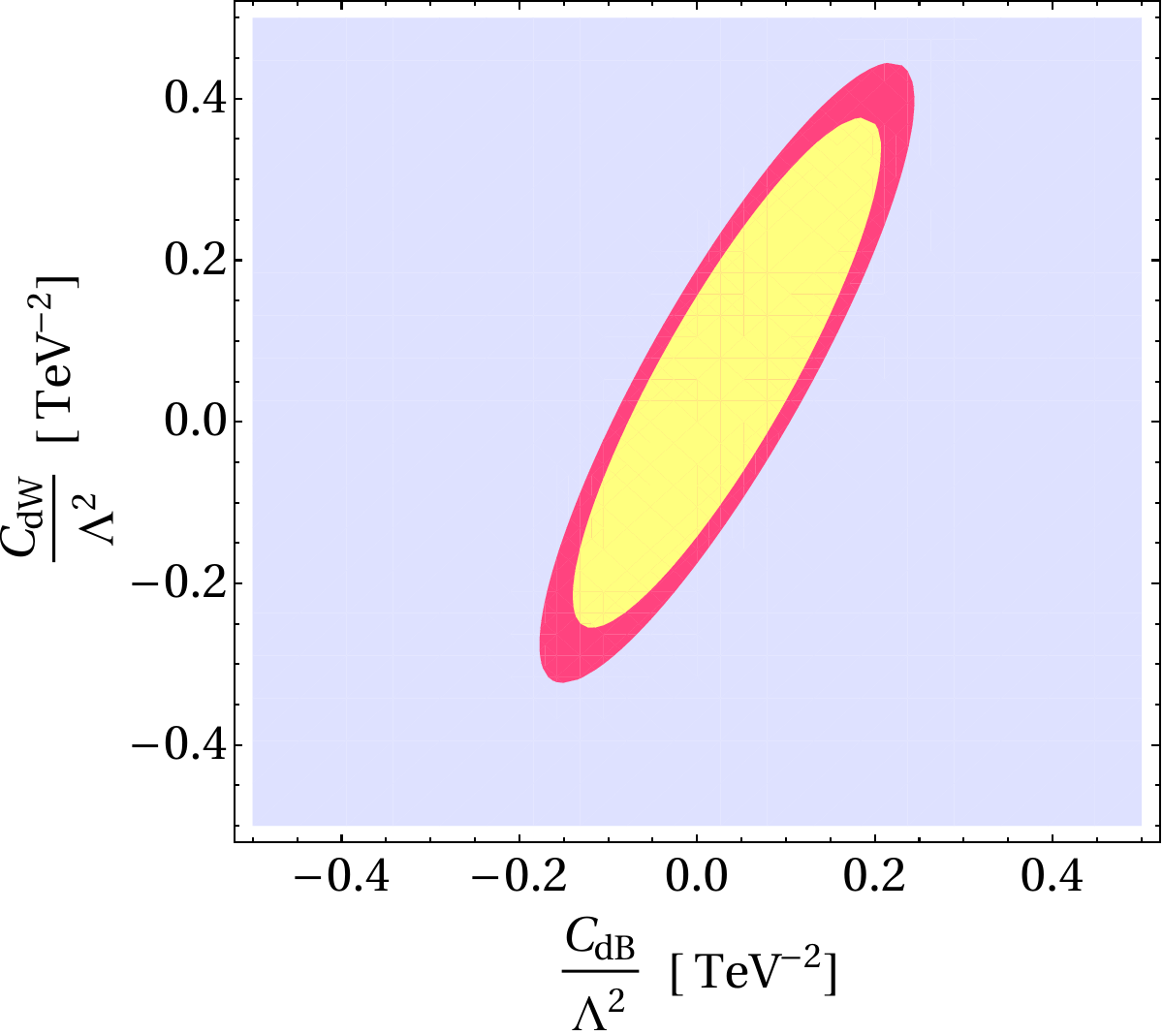}&
				\includegraphics[width=5.0cm,height=6.3cm]{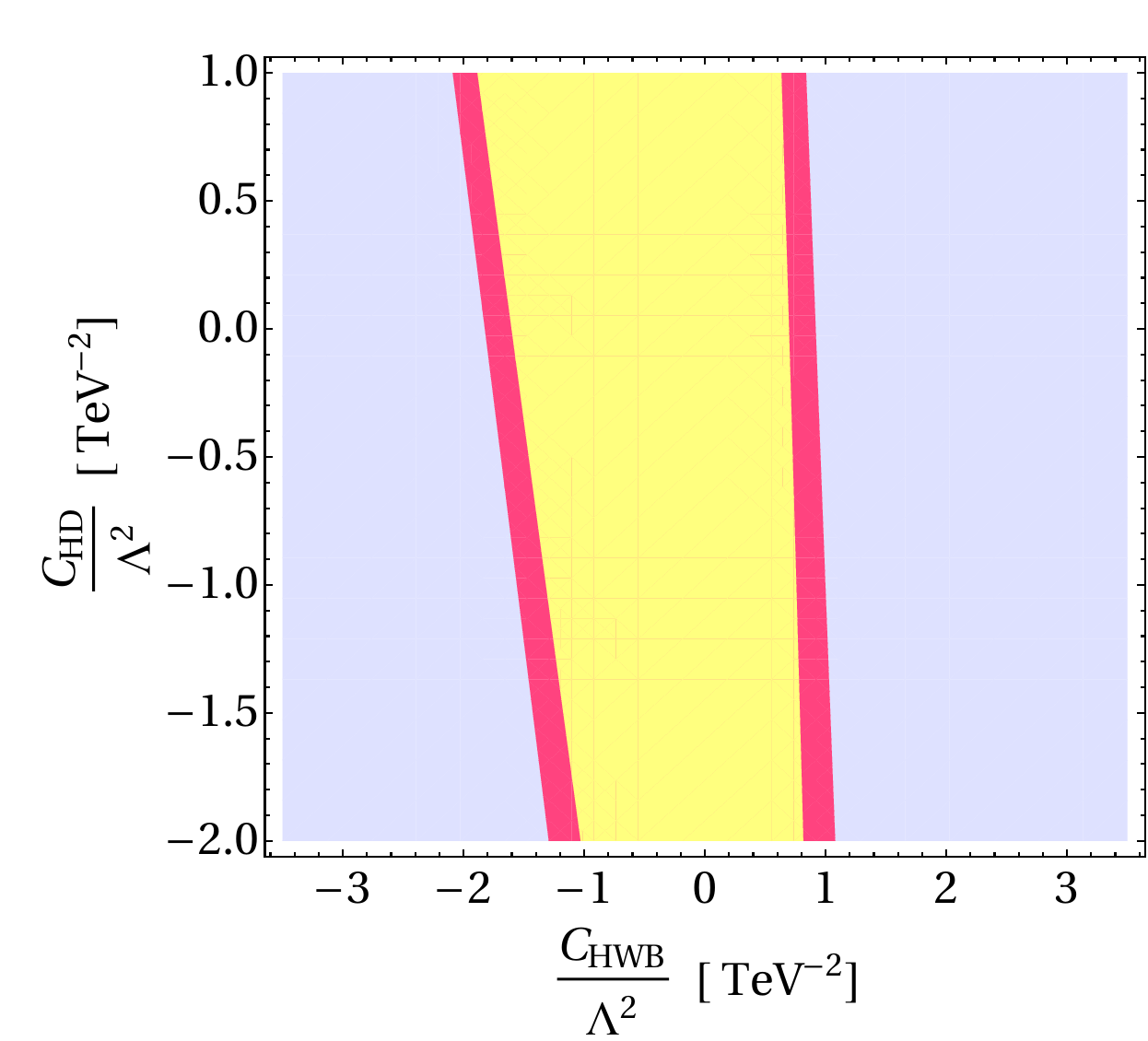}  \\
				(a)&(b)&(c)\\
				\includegraphics[width=5.0cm,height=5.9cm]{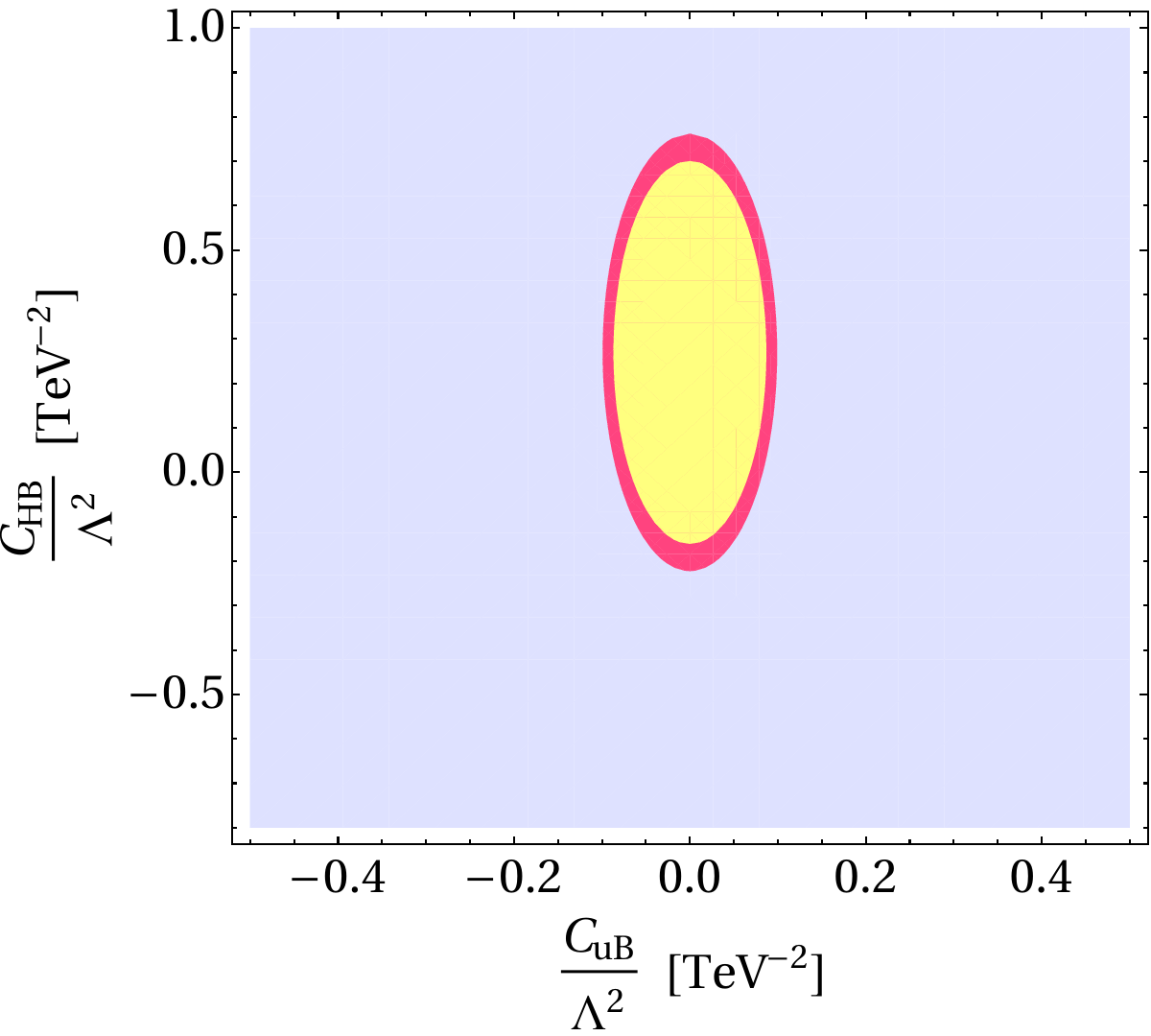}&
				\includegraphics[width=5.0cm,height=6.0cm]{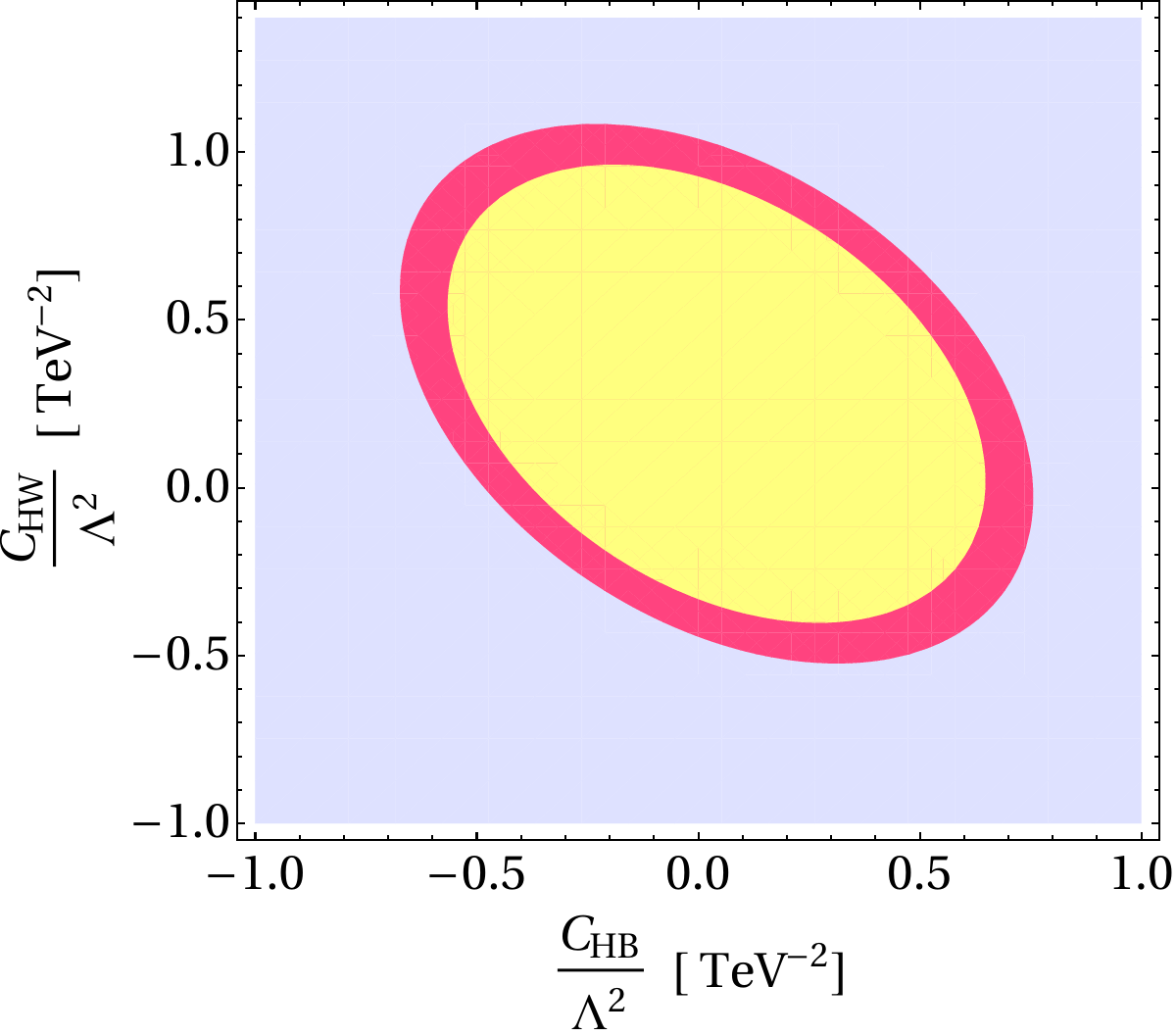}&
				\includegraphics[width=5.0cm,height=6.0cm]{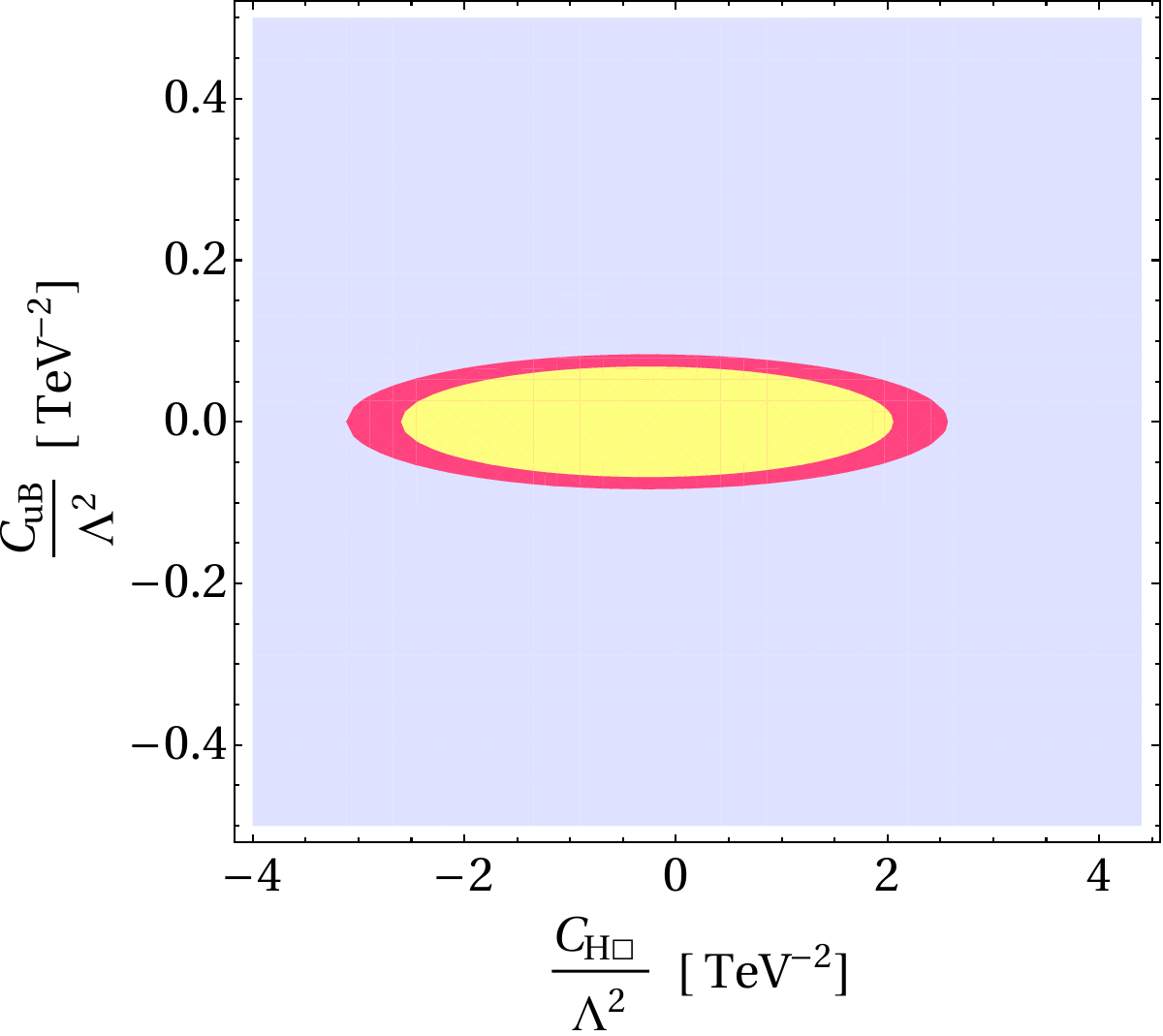} \\
			 	(d)&(e)&(f)
		\end{tabular}}
		\caption{Projections for a 3$\sigma$ signal significance in  blue shaded regions in the parameter space in  in the $C_{uB}-C_{uW}$ (top left) , $C_{dB}-C_{dW}$  (top middle), $C_{HD}-C_{HWB}$  (top right),  $C_{HB}-C_{uB}$ 	(bottom left), and $C_{HB}-C_{HW}$ (bottom middle) and $C_{uB}-C_{H\Box}$ (bottom right)	planes with all other EFT coefficients assumed to be zero with	3000 $fb^{-1}$ data at 14 TeV run of the LHC. The 2$\sigma$ level approximating the 95\% confidence level exclusion bounds are shown in magenta shaded regions.}
		\label{fig:2Dplots}
	\end{figure}

	Results for sensitivities of different benchmark points have been presented in Table~\ref{significanceCBA}. We now present the results for the sensitivity projections in six cases assuming two non-zero Wilson coefficients at a time,  quark-specific scenario ($C_{uB}$-$C_{uW}$ and $C_{dB}$-$C_{dW}$ planes), bosonic-specific scenario in $C_{HW}$-$C_{HB}$ and $C_{HD}$-$C_{HWB}$ planes and quark-bosonic interaction in $C_{uB}$-$C_{HB}$ and $C_{uB}$-$C_{H\Box}$ planes (Fig.~\ref{fig:2Dplots}).
	
	In the limit of vanishing fermion masses,  operators $\mathcal{O}_{uB}$ and $\mathcal{O}_{uW}$ do not interfere with the SM amplitude, however they interfere constructively among themselves which is clearly evident in Fig.~\ref{fig:2Dplots}(a).  The blue shaded region marks the bounds that can be probed at $3\sigma$ significance level. Similar interference effect prevails in the case  of  non-zero $C_{dB}$ and $C_{dW}$ (shown in  Fig.~\ref{fig:2Dplots}(b)).  These couplings in pairs of $(C_{uB},C_{uW})$ and $(C_{dB},C_{dW})$ are only slightly correlated with each other due to a smaller interference effect in the $h \gamma$ production. In addition, the correlations between $(C_{uB},C_{uW})$ and $(C_{dB},C_{dW})$ are opposite in nature. A more optimistic sensitivity on the Wilson coefficients of the operators with $u$-type quarks are obtained than that due to the $d$-type fermionic operators. In Fig.~\ref{fig:2Dplots}(c), the projection in $C_{HD}$-$C_{HWB}$ plane is presented for $pp \to h\gamma$ process. We note that among the aforementioned Wilson coefficients, $C_{HD}$ and $C_{HWB}$ contribute to $T$ and $S$ parameters,  respectively, at tree level. However, bounds coming from $S$ and $T$ on $C_{HD}$-$C_{HWB}$ parameter space is more constraining than the constraints derived in the present analysis. Operator $\mathcal{O}_{uB}$ neither interferes with the SM amplitude  and nor with the bosonic-type operator and thus, the leading BSM effect arises at $\mathcal{O}\sim \frac{1}{\Lambda^4}$. This can be seen from the $C_{uB} \to -C_{uB}$ symmetry in the  Fig.~\ref{fig:2Dplots}(d) whereas the $C_{HB}$ interferes with the SM. Similar is the case in Fig.~\ref{fig:2Dplots}(f) for $C_{uB}$ and $C_{HWB}$ plane. The yellow shaded region in Fig.~\ref{fig:2Dplots}(e) shows the $2\sigma$ significance region in the $C_{HW}$-$C_{HB}$ parameter space. In this plane, there are two couplings contributory. One, the coupling proportional to $g_{h\gamma \gamma}$  whose contribution to the Higgs-diphoton interaction depends on $s_W^2 C_{HW}+ c_W^2 C_{HB}$ combination. The other coupling due to the presence of Z propagator $g_{hZ\gamma}$ receives contribution via the combination  $ s_W c_W(C_{HW}-C_{HB})$, thus indicating a region in the same place when the latter decreases the rate for positive $C_{HB}$. In all, their total contribution requires for larger values of the Wilson coefficients for probing the $h\gamma$ production at $3 \sigma$ significance.

	\subsection{Multivariate Analysis}
	\label{sec:multivariate_analysis}
	In the previous sub-section, we presented the sensitivity of the new interactions at the end of a cut-based analysis. We now explore the possibility of any improvement in our analysis using a multivariate analysis (MVA).  We use the adaptive Boosted Decision Tree (BDT) algorithm in the TMVA framework~\cite{Hocker:2007ht}. The multivariate analysis complements the cut-based analysis in the following way. A cut-based analysis can select out the phase space by applying a series of grids in succession (bounded by either one  sided or both sided boundaries). However, many of the events may be characterised by kinematic variables that may be correlated in a complicated way. The decision trees proceed by recursively partitioning the phase space into hyperspace regions with edges aligned along the axes of the phase space. Each of these regions in the hyperspace is then classified for signal and background segregation. A non-linear boundary in the multi-dimensional phase space can, thus, help identifying the signal region more accurately.
	
	We have used slightly looser cuts for the MVA compared to those used in the previous section (in Table~\ref{tab:bbgam_bkg}). In addition to the preselection cuts, a minimum transverse momentum of $175$ GeV requirement has been applied on the leading fat-jet and  the photon. The reconstructed fat-jet mass is required to be at least 60 GeV. We ensured that the cuts are very effective in reducing the large background but not the signal. This is because the MVA does not perform well if we use events just with the baseline cuts as the signal is very small compared to the large background. Also, the strong cuts of Table~\ref{tab:bbgam_bkg} do not give an improved sensitivity while performing MVA. Thus, the cuts in MVA are chosen in such a way that they are neither the same nor too relaxed cuts compared to the cut-based analysis. 
	
\begin{figure}[h!]		
		\centering
		\subfloat{	
			\begin{tabular}{cc}
				\includegraphics[width=7.0cm,height=4.22cm]{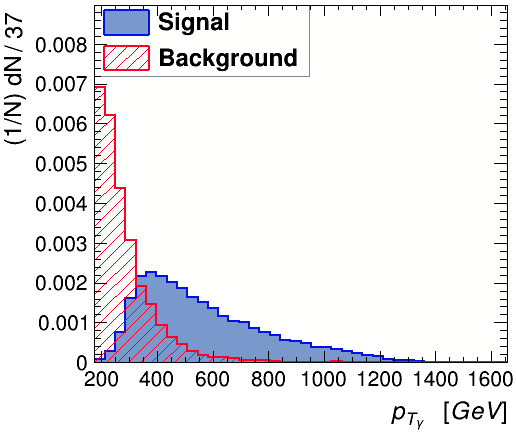} &
				\includegraphics[width=7.0cm,height=4.22cm]{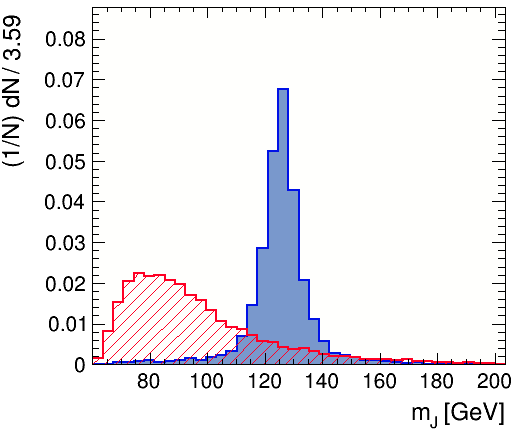}\\
				\hspace{4mm}(a) &\hspace{8mm}(b) \\
				\includegraphics[width=7.0cm,height=4.22cm]{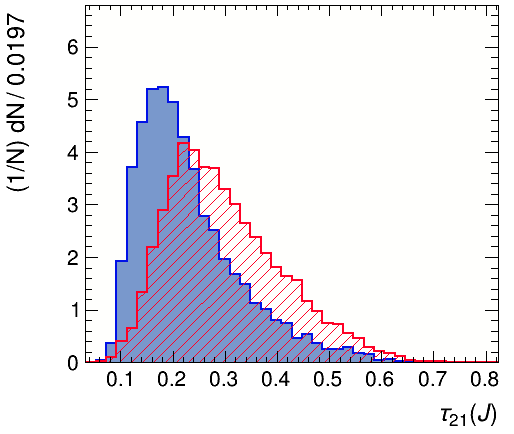}&
				\includegraphics[width=7.0cm,height=4.22cm]{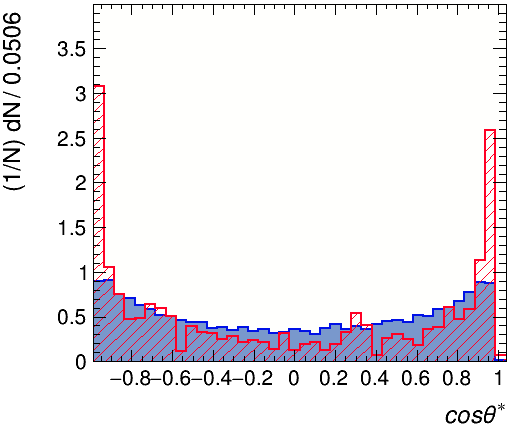}\\
				\hspace{4mm}(c) &\hspace{8mm}(d) \\
				\includegraphics[width=7.0cm,height=4.22cm]{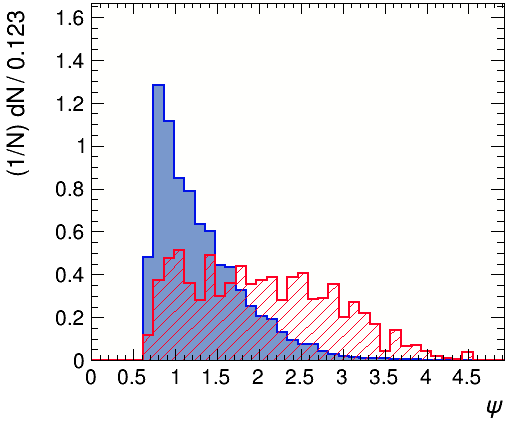}&
				\includegraphics[width=7.0cm,height=4.22cm]{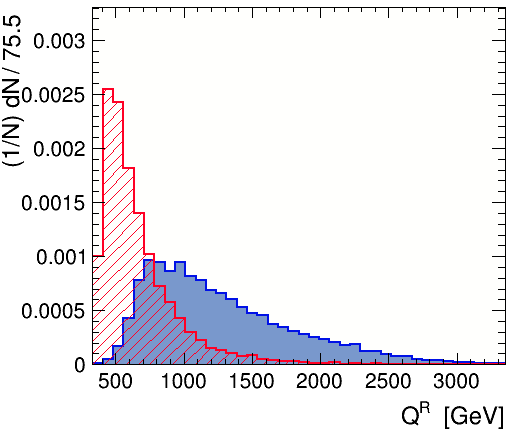}\\
				\hspace{4mm}(e) &\hspace{8mm}(f) \\
				\includegraphics[width=7.0cm,height=4.22cm]{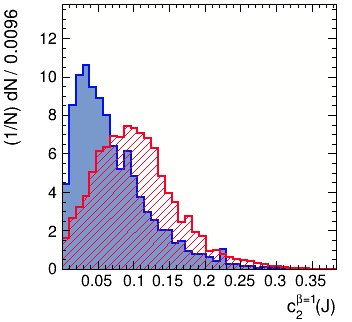}&
				\includegraphics[width=7.0cm,height=4.22cm]{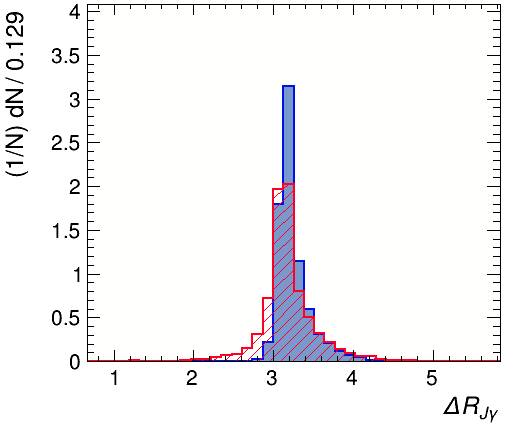}\\
				\hspace{4mm}(g) &\hspace{8mm}(h)
		\end{tabular}}
		\caption{Normalized distributions for the signal (blue) and the background (red) of the input variables at the LHC (at $\sqrt{s}=14$ TeV) used in the MVA. Signal distributions are obtained for BP4 and the background includes all the dominant backgrounds discussed in Sec.~\ref{sec:eft_framework}}.
		\label{fig:inputvarBP4}
	\end{figure}

	We identify the various observables with maximal discerning potency which are chosen to be the input variables to BDT. The normalized distributions of these input variables are shown in Fig. \ref{fig:inputvarBP4}.
	\begin{itemize}
		\item $p_{T_{\gamma}}$, the transverse momentum of photon.
		\item $m_J$, the mass of leading jet.			
		\item The fat-jet appearing from the $h \to b \bar{b}$ potentially retains some information of its two prong structure. This is revealed by the jet-shape variable, N-subjettiness~\cite{Thaler:2010tr,Thaler:2011gf}, defined as:
		\begin{eqnarray}
		\tau_N^{(\beta)} = \frac{1}{\mathcal{N}_0} \sum\limits_i p_{i,T} \, \min \left\lbrace \Delta R _{i1}^\beta, \Delta R _{i2}^\beta, \cdots, \Delta R _{iN}^\beta \right\rbrace
		\label{eq:nsub_N}
		\end{eqnarray}
		Here, $N$ is the axis of the subjet assumed within the fat-jet, $i$ runs over the constituents of the jet, $p_{i,T}$ are their respective transverse momenta, $\Delta R_{ij}=\sqrt{(\Delta \eta)^{2} + (\Delta \phi)^{2}}$ is the distance in $\eta-\phi$ plane, between candidate subjet $j$ and constituent particle $i$ and $\mathcal{N}_{0}=\sum_{i}p_{i,T} R_{0}$ is the normalization factor. $R_0$ is the radius parameter of the fat-jet. In particular, ratio $\tau_N/\tau_{N-1}$ discriminates between jets, predicably having $N$ internal energy clusters and those with ($N$-$1$) clusters.  In our case, the jet coming from the Higgs tends to have lower values for $\tau_{21}$ as compared to QCD jets as shown in Fig~\ref{fig:inputvarBP4}(c).
		\item $cos \theta^*$, the scattering angle of photon in $h \gamma$ rest-frame. 
		\item A variable, $\psi=\rm log(1+\rm exp|\eta_\gamma-\eta_J|)$ renders the dominant QCD background distributions  flat, and  the distribution we expect from a new physics signal are peaked at low $\rm{exp|\eta_\gamma-\eta_J|}$.
		\item We introduce few other new observables namely $Q^R$, $Q^T$ and $R_2$. They are defined as follows. 
		\begin{eqnarray}
		Q^R &=& \sqrt{(|\vec{p}_{J}| + |\vec{p}_{\gamma}|)^2 - ({p}_{Z_{J}} + {p}_{Z_{\gamma}})^2}  \\
		Q^T &=& \sqrt{{p}_{T_{J}}{p}_{T_{\gamma}}(1-\rm cos(\Delta \phi_{J \gamma}))/2.}\\
		R_2 &=& \left(\frac{{Q^T}}{{Q^R}}\right)^2
		\end{eqnarray}  
		The $Q^R$ variable contains both longitudinal and transverse information of the final-state particles while the $Q^T$ variable is obtained only from the transverse momenta of the final particles. The ratio $R_2$ of $Q^R$ and $Q^T$ encapsulates information about the flow of energy along the plane perpendicular to the beam and separating the final states.
		\item At the LHC, boosted objects such as the gauge bosons, the Higgs boson and the top quark decaying into quarks can be studied by identifying the jet substructures. These substructures are characterised by energy correlation functions~\cite{Larkoski:2014gra}, discriminating against the QCD jets, defined as:
		\begin{eqnarray}\label{C2}
		C_2^{\beta} = \frac{ e_3^{\beta}}{ (e_2^{\beta})^2}, \cr
		e_2^{\beta} &=& \frac{1}{p^2_{T,\jet}} \sum_{i,j \in \jet, i<j} p_{T,i}p_{T,j} R^{\beta}_{ij},\cr
		e_3^{\beta} &=& \frac{1}{p^3_{T,\jet}} \sum_{i,j,k \in \jet, i < j <k} p_{T,i}p_{T,j} p_{T,k}R^{\beta}_{ij} 
		R^{\beta}_{jk} R^{\beta}_{ki},  
		\end{eqnarray}
		where, $e_2^{\beta}$ and $e_3^{\beta}$ are 2-point and 3-point energy correlation functions respectively. The $\beta$ represents the angular exponent. If the jet has a two-prong like substructure, $C_2$ is much less than 1. The distribution for $C_2$, distinguishing the Higgs jet from the backgrounds is shown in Fig~\ref{fig:inputvarBP4}(g).
		\item Angular distance between the jet and the photon, $\Delta R_{J \gamma}$.
		\item Subjet variable $\Delta R_{b_i,b_j}$ given by the angular separation between two b-tagged subjets.	
	\end{itemize}
	It is to be noted that although $m_{J \gamma}$ is one of the most important distinguishing feature between signal and background, to minimize the bias  we do not use it as an input variable to the BDT. 	

\begin{figure}[t!]
		\subfloat{
			\begin{tabular}{cc}
				\includegraphics[width=8.2cm,height=8.0cm]{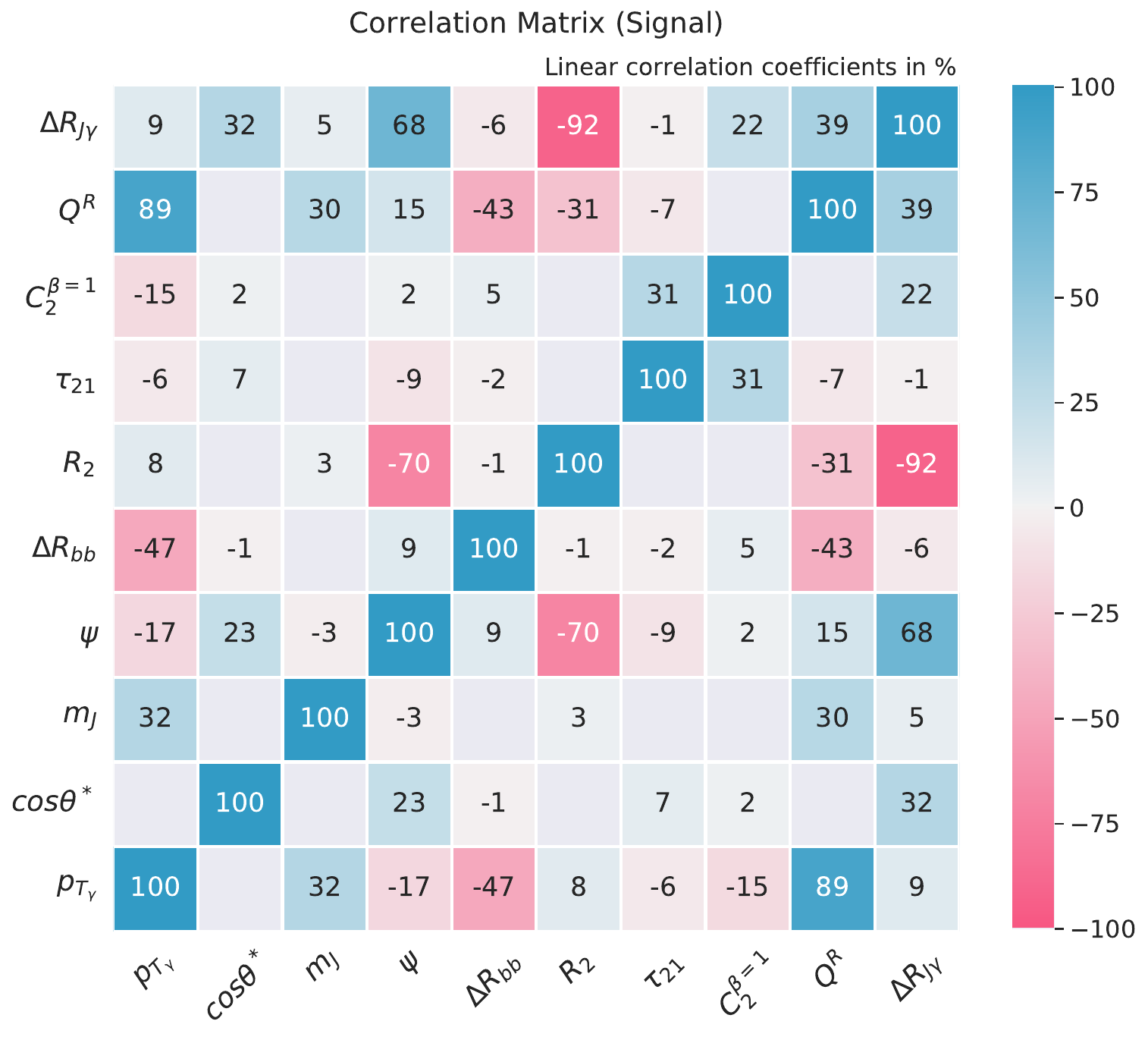}&
				\includegraphics[width=8.2cm,height=8.0cm]{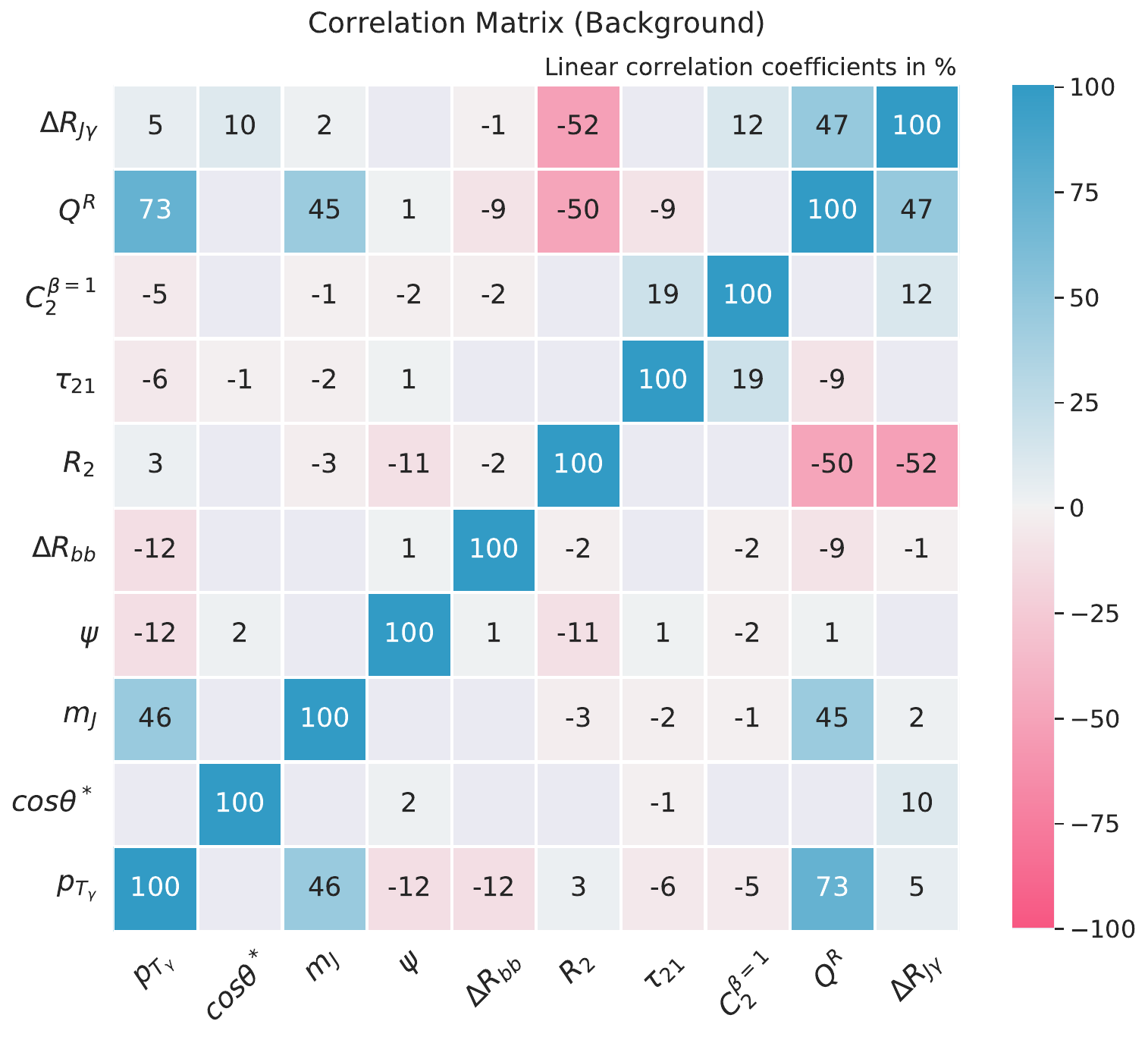}\\
				(a)&(b)
		\end{tabular}}
		\caption{The linear correlations coefficients in \% between the input variables that are used for the MVA   for (a) signal (with benchmark point BP4) and (b) background. Two variables are positively correlated and negatively-correlated if the coefficients have positive and negative signs respectively.}
		\label{fig:correlation}
	\end{figure}

\begin{figure}[h!]
		\centering
		\includegraphics[width=10.0cm,height=7.0cm]{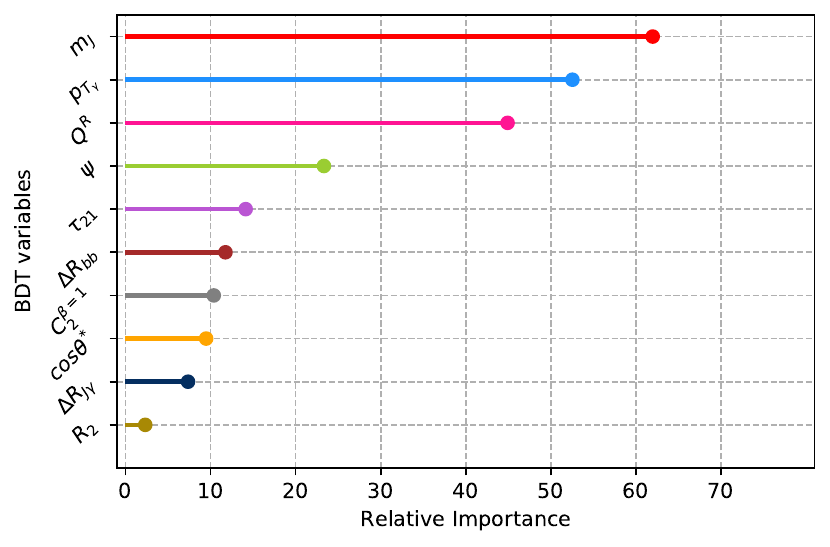}
		\caption{Kinematic variables used as input for our MVA and their relative importance. We obtain these using numbers for the benchmark point BP4.}
		\label{fig:Importance}
	\end{figure}

	The linear correlation between two variables $x$ and $y$ is computed with the correlation coefficient, $\rho$, defined by the following equation:
	\begin{equation}
	\rho=\frac{cov(x,y)}{\sigma(x) \sigma(y)} = \frac{E(xy)-E(x)E(y)}{\sigma(x) \sigma(y)}
	\end{equation}
	where $E(x),E(y)$ and $E(xy)$ are the expectation value of the variables $x, y$ and $xy$ respectively. Here, $\sigma(x), \sigma(y)$ represent the standard deviation of $x$ and $y$ respectively. The value of linear correlation coefficient, $\rho$ lies between -1 and 1 and its sign indicates the direction of linear relationship among the variables. The linear correlation coefficient matrix for signal and background is illustrated in Fig. \ref{fig:correlation}. Here, most of the variables are independent of each other. The energy variables such as $p_{T_{\gamma}}$ and $Q^R$ and angular variables $\psi$ and $\Delta R_{J \gamma}$ show slightly strong linear correlation in case of signal events while are less correlated in background events. The choice of kinematic variables used for the BDT network relies on factors that they should less correlation among themselves and more discriminating power between the signal and the background. The analysis always has the scope of improvement, by choosing a better set of variables. However, the variables that we have used are good discriminators as demonstrated in the following.
	Based on the separation between signal and background, we show the method unspecific ranking (relative importance) for each
	observable in Fig.~\ref{fig:Importance}. The separation power in  terms of an observable $w$ is defined using the integral:
	\begin{equation}
	<S^2> = \int \frac{(\hat{w}_s(w)-\hat{w}_b(w))^2}{\hat{w}_s(w)+\hat{w}_b(w)} dw
	\end{equation}
	where $\hat{w}_s$ and $\hat{w}_b$ are the probability distribution functions for signal and background for a given observable $w$ respectively. The limits of integration correspond to the allowed range of $w$. Here $<S^2>$ quantifies the discrimination performance of an observable $w$. The separation $<S^2>$ lies within the range $[0,1]$. If $<S^2> = 0$ implies $\hat{w}_s(w) = \hat{w}_b(w)$, which means identical signal and background distributions for the observable $w$ and $<S^2>  = 1 $ corresponds to distributions with no overlap. Among the variables used, the five most important variables to distinguish the signal from the backgrounds are : $m_J, p_{T_{\gamma}}, Q^R, \psi~\rm{and}~\tau_{21}$.
	
\begin{figure}[h!]
		\centering	
		\subfloat[]{\label{fig:Train_test}\includegraphics[width=8.0cm,height=6.0cm]{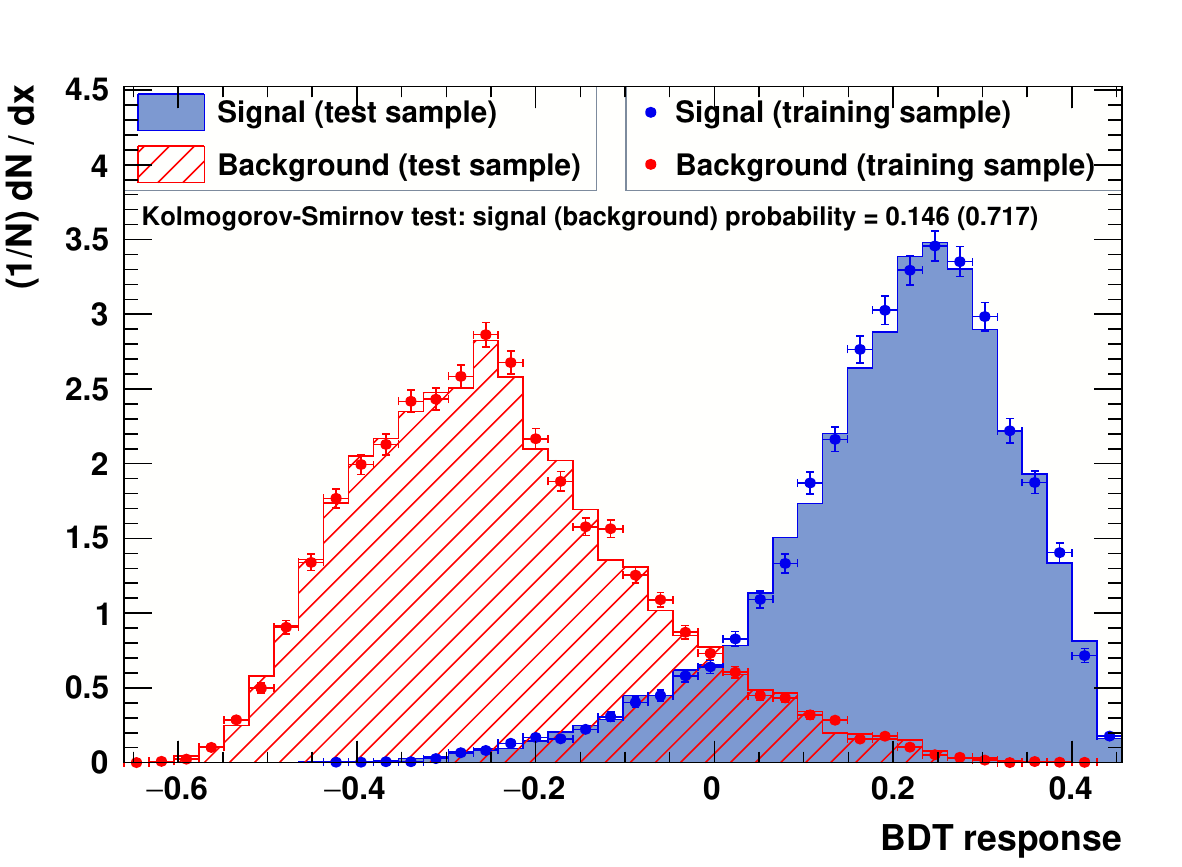}}~
		\subfloat[]{\label{fig:Significance}\includegraphics[width=8.0cm,height=6.0cm]{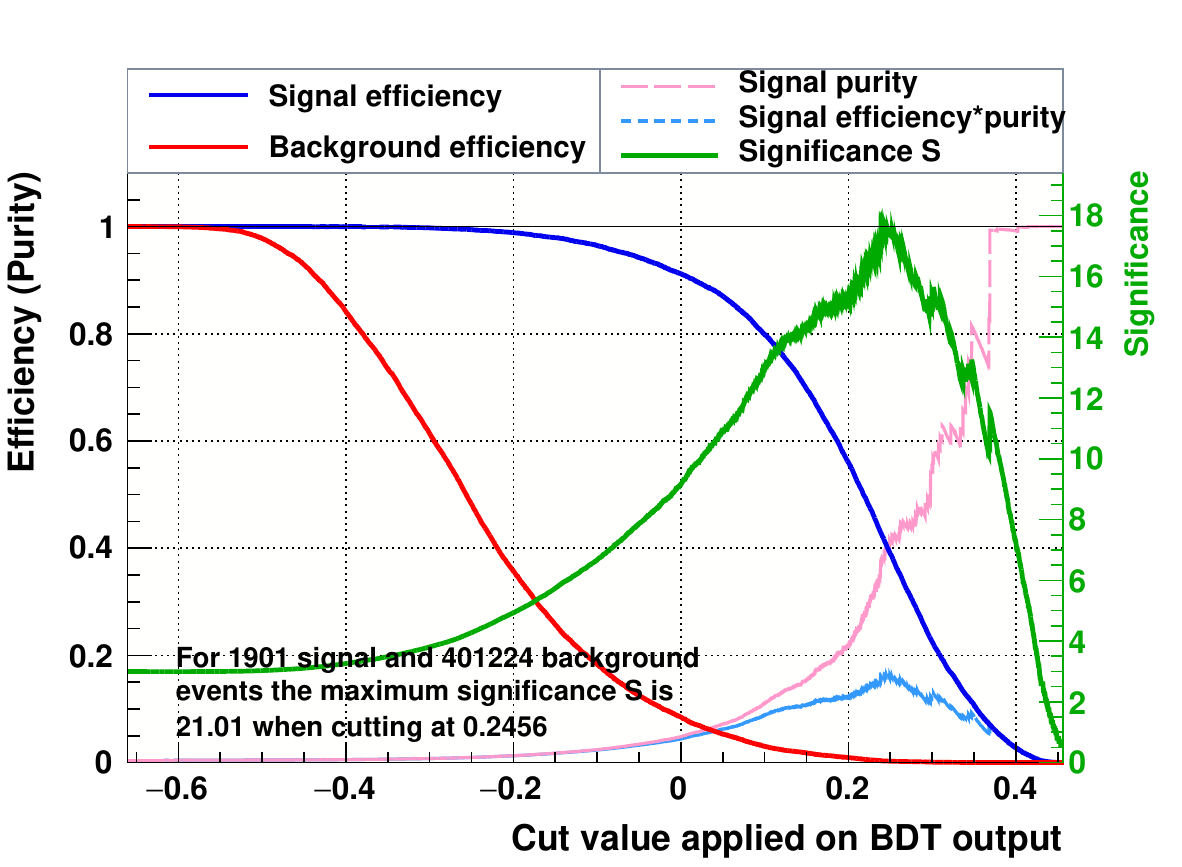}}
		\caption{(a) The normalized BDT response distributions for the signal and the background and 
			(b) the cut efficiencies as functions of BDT cut values for the benchmark point  BP4, using integrated luminosity of $3000$ fb$^{-1}$ at the 14 TeV LHC.}
		\label{fig:BDTresponse}
	\end{figure}

	We then employ the BDT algorithm.  The datasets consisting of the statistically independent event samples for the signal and the background have been split randomly in two equal parts. One part of the dataset is used to train the BDT algorithm and the other part is used for validation for both the signal and the background. The parameters used to train the BDT algorithm are listed in Table~\ref{tab:BDT_parameter}.
	\begin{table}[t!]
		\centering
		\small
		\renewcommand{\arraystretch}{1.5}
		\begin{tabular}{|c|c|c|}
			\hline
			NTrees& Number of trees in the forest  & 400 \\
			\hline
			MaxDepth& Max depth of the decision tree allowed& 2 \\ 
			\hline
			MinNodeSize & Minimum \% 
			of training events required in a leaf node & $5.6 \%$\\ 
			\hline
			BoostType&  Boosting type for the trees in the forest & AdaBoost \\
			\hline
			AdaBoostBeta& Learning rate for AdaBoost algorithm	& 0.5 \\
			\hline
			nCuts& Number of grid points in variable  & 20\\
			
			& range used in finding optimal cut in
			node splitting& \\
			\hline
		\end{tabular}
		\caption{The list of BDT parameters (definition and values) used}
		\label{tab:BDT_parameter}
	\end{table}
	For optimal performance of the BDT, we also need to minimise the possibility of overtraining. The case of overtraining the signal/background usually arises during the training of the algorithm due to insufficient statistics to train or without the proper choices of the algorithm specific tuning parameters. Here, the distinction between signal and background becomes extremely good for the training sample while for the test sample, it fails to achieve the same level of difference. We have explicitly checked that our algorithm is not over-trained by ensuring that  Kolmogorov-Smirnov probability is atleast of the order $\sim 0.1$ during training and that the area under the Receiver Operator Charateristic (ROC) curve (that estimates the degree of rejecting the backgrounds with respect to the signal) remains almost same for training and testing sample. We train the BDT network for each benchmark point separately. Kolmogorov-Smirnov probability for training and testing samples are shown in Fig.~\ref{fig:BDTresponse} (a), indicating that both signal and background samples are not overtrained.
	We apply an appropriate cut on BDT response and obtain the corresponding number of signal $\mathcal{N}_{S}$ and background $\mathcal{N}_{B}$. The cut value of BDT response is $\mathrm{BDT}_{opt}$, where the maximum significance is achieved. We calculate the statistical significance using formula for  $\mathcal{S}$ defined in Eqn.~\ref{significance_eqn}.  Results from BDT analysis considering one sample benchmark point (BP4) is shown in Fig.~\ref{fig:BDTresponse} (b). The results for all benchmark points are displayed in Table  \ref{tab:BDT}.	
	\begin{table}[h!]
		\centering
		\begin{tabular}{|c||c|c|c|c|c|c|}
			\hline \hline
			BPs & $\mathcal{N}_S^{bc}$ & $\mathrm{BDT}_{opt}$ & $\mathcal{N}_{S}(\epsilon_S)$ & $\mathcal{N}_{B}(\epsilon_B \times 10^{-2})$ &  $\mathcal{S}$ & $\mathcal{L}^{req}_{(5\sigma)} (fb^{-1})$ \\ 
			\hline 
			BP1 & 2665  & 0.2442 & 1304(0.4893) & 2285(0.5695) &25.15 & 118.49 \\ 
			\hline
			BP2 & 2048  & 0.2559 & 574(0.2803) & 491(0.1223)  &22.39& 149.63\\ 
			\hline
			BP3 &  497  & 0.3694 & 9(0.0181)  &$1(2.492\times 10^{-4})$   & 5.29 & 2673.63  \\ 
			\hline
			BP4 & 1901  & 0.2456 & 765(0.4024) &1093(0.2724) & 21.01& 169.83\\ 
			\hline
			BP5 & 4112    &  0.2307 &  1891(0.4598)&  2305(0.5744)  &35.28 & 79.21  \\ 
			\hline
			BP6 & 1430  & 0.2337 & 649(0.4538) & 6270(1.5627)  & 8.06 & 1154.34 \\ 
			\hline \hline 
			$\mathcal{N}_{\textrm{Bkg}}$ & 401224 & - & - & - &-& \\ 
			\hline \hline
		\end{tabular} 
		\caption{Number of signal events $\mathcal{N}_S^{bc}$ and background events $\mathcal{N}_{\textrm{Bkg}}$ before  imposing the optimum BDT criteria ($\mathrm{BDT}_{opt}$) for an integrated luminosity of $3000$ fb$^{-1}$ at the 14 TeV LHC. $\mathcal{N}_S$ and $\mathcal{N}_{B}$ are the number of signal and background events respectively after applying the $\mathrm{BDT}_{opt}$ cut. $\epsilon_S$ and $\epsilon_B$ represent the signal acceptance and background acceptance efficiency at the $\mathrm{BDT}_{opt}$ cut value. Finally, $\mathcal{S}$  and  $\mathcal{L}^{req}_{(5\sigma)}$ denote the statistical significance for an integrated luminosity of $3000$ fb$^{-1}$ and required luminosity for a $5\sigma$ discovery in case of each BP, respectively.}
		\label{tab:BDT}
	\end{table}
	\begin{figure}[h!]
		\centering
		\includegraphics[height=7cm,width=10cm]{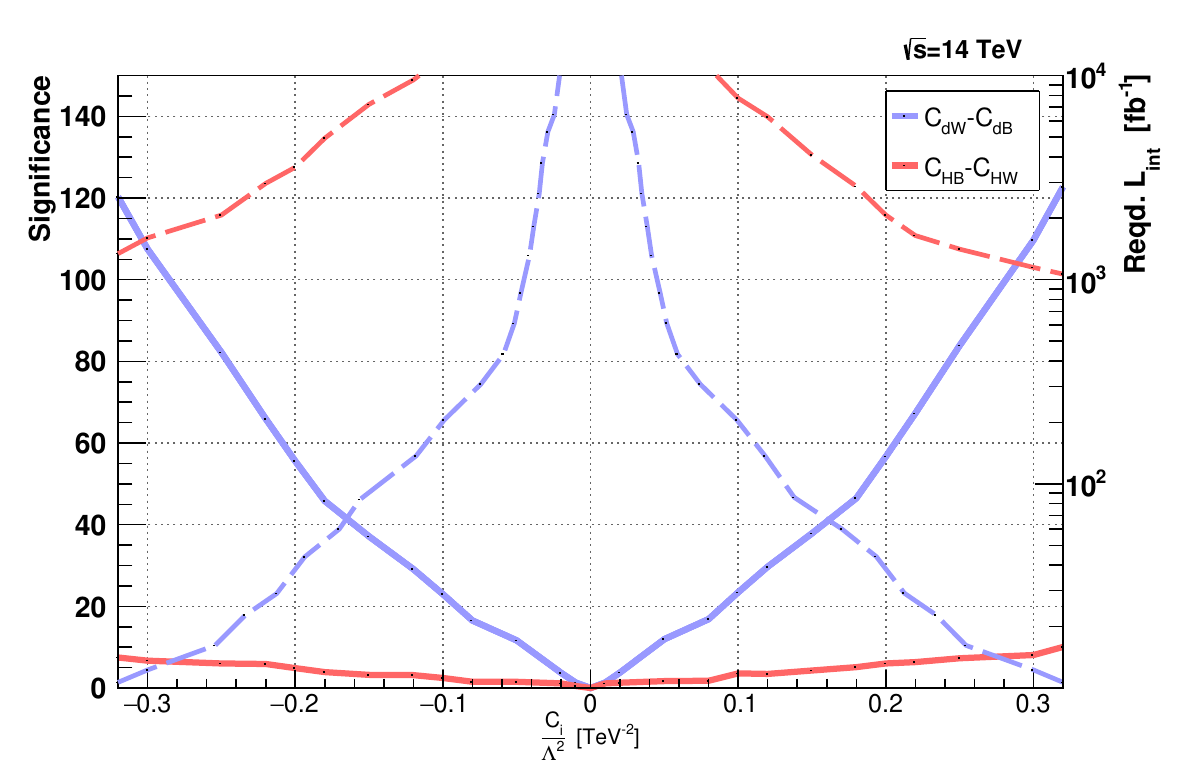}
		\caption{Significance as a function of Wilson coefficient $\frac{C_i}{\Lambda^2}$ (solid lines) at the 14 TeV LHC with 3000 fb$^{-1}$ integrated luminosity. We also present the required luminosity for $5 \sigma$ discovery (dashed lines) as a function of the Wilson coefficient $\frac{C_i}{\Lambda^2}$.}
		\label{fig:signi}
	\end{figure}
	We also show the signal significance as a function of $\frac{C_i}{\Lambda^2}$ at the 14 TeV LHC with 3000 fb$^{-1}$ integrated luminosity in Fig.~\ref{fig:signi}. While presenting these results, we have assumed two non-zero couplings of equal strength but with opposite signs at a time. The blue curve represents ($\mathcal{C}_{HW},\mathcal{C}_{HB}$) combination while the pink curve represents the ($\mathcal{C}_{dW},\mathcal{C}_{dB}$) combination. For ($\mathcal{C}_{HW}$,$\mathcal{C}_{HB}$) coupling, the interference with the SM explains the slightly asymmetric nature of significance as a function of $\frac{C_i}{\Lambda ^2}$.  It is evident from Fig.~\ref{fig:signi} that with the BDT analysis, for ${|C_i|\over \Lambda ^2} = 0.2~\rm TeV^{-2}$, an integrated luminosity of  $200~\rm fb^{-1}$  for the contact $ff\gamma h$ coupling ($C_{dW},C_{dB}$) and  $2000~\rm fb^{-1}$ for the effective $hV\gamma$ coupling ($C_{HW},C_{HB}$) is needed to achieve a 5$\sigma$ significance. We observe a clear improvement of the statistical significance of the signal resulting from the BDT analysis over the cut-based analysis for all benchmark points (also see Tables~\ref{significanceCBA} and \ref{tab:BDT}).
	
	\subsection{Correlation of $h\gamma$ signal with other Higgs signal}
	We now turn our attention to any possible correlation between the Higgs-photon production and other  Higgs signals which are driven by the same dimension-6 operators in which we are interested. With this in mind, we define the following quantities as ratio of event rates:
	\begin{equation}
	R_{h\gamma} = \frac{\sigma_{pp\to h\gamma}(C_i) \times BR_{h \to b \bar{b}}}{\sigma_{pp\to h\gamma} \times BR_{h \to b \bar{b}}};~
	R_{\gamma\gamma} = \frac{\sigma_{pp \to h} \times BR_{h\to \gamma\gamma}(C_i)}{\sigma_{pp \to h} \times BR_{h\to \gamma\gamma}};~
	R_{Z\gamma} = \frac{\sigma_{pp \to h} \times BR_{h\to Z\gamma}(C_i)}{\sigma_{pp \to h} \times BR_{h\to Z\gamma}} 
	\end{equation}
	\begin{equation} 
	R_{Vh} = \frac{\sigma_{pp\to h V}(C_i)\times BR_{h \to b\bar{b}}}{\sigma_{pp\to h V} \times BR_{h \to b\bar{b}}};~~~
	R_{VBF} = \frac{\sigma_{pp\to hjj}(C_i)\times BR_{h \to b\bar{b}}}{\sigma_{pp\to hjj} \times BR_{h \to b\bar{b}}} 
	\end{equation}	
	
\begin{figure}[t!]
		\centering
		\includegraphics[width=7.5cm,height=6.5cm]{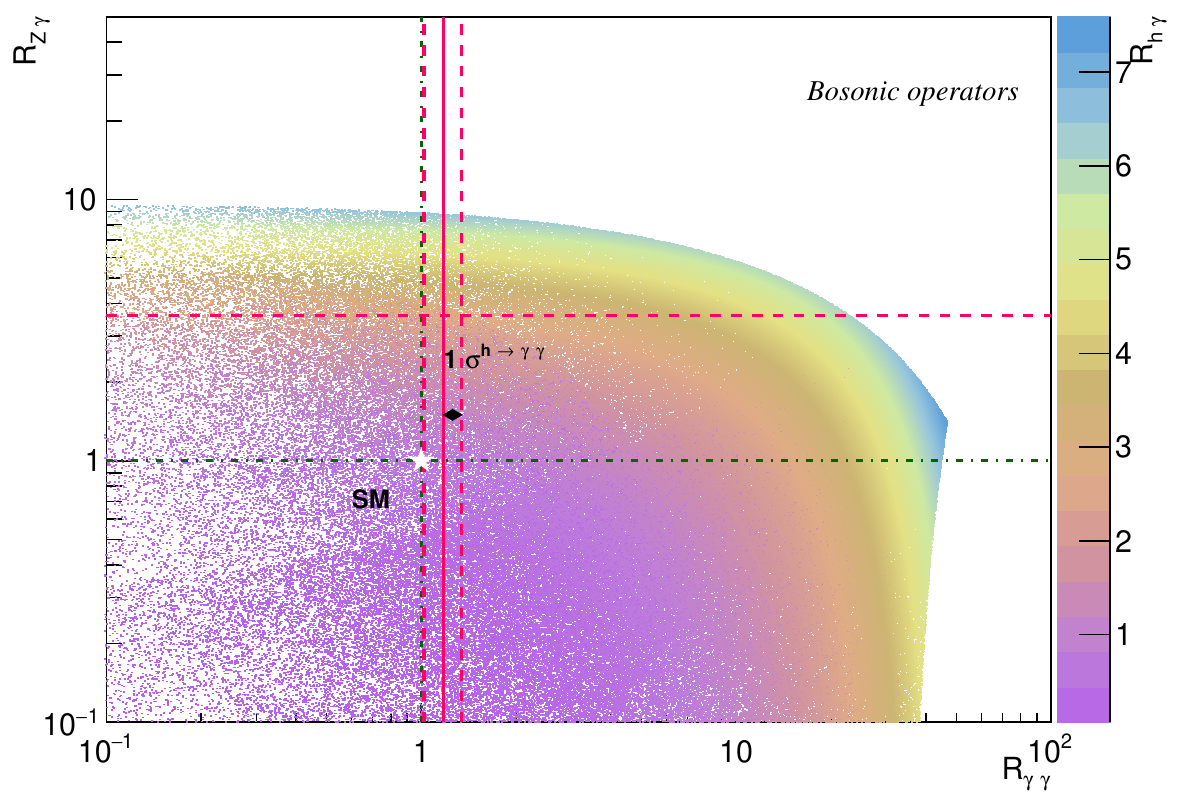}
		\includegraphics[width=7.5cm,height=6.5cm]{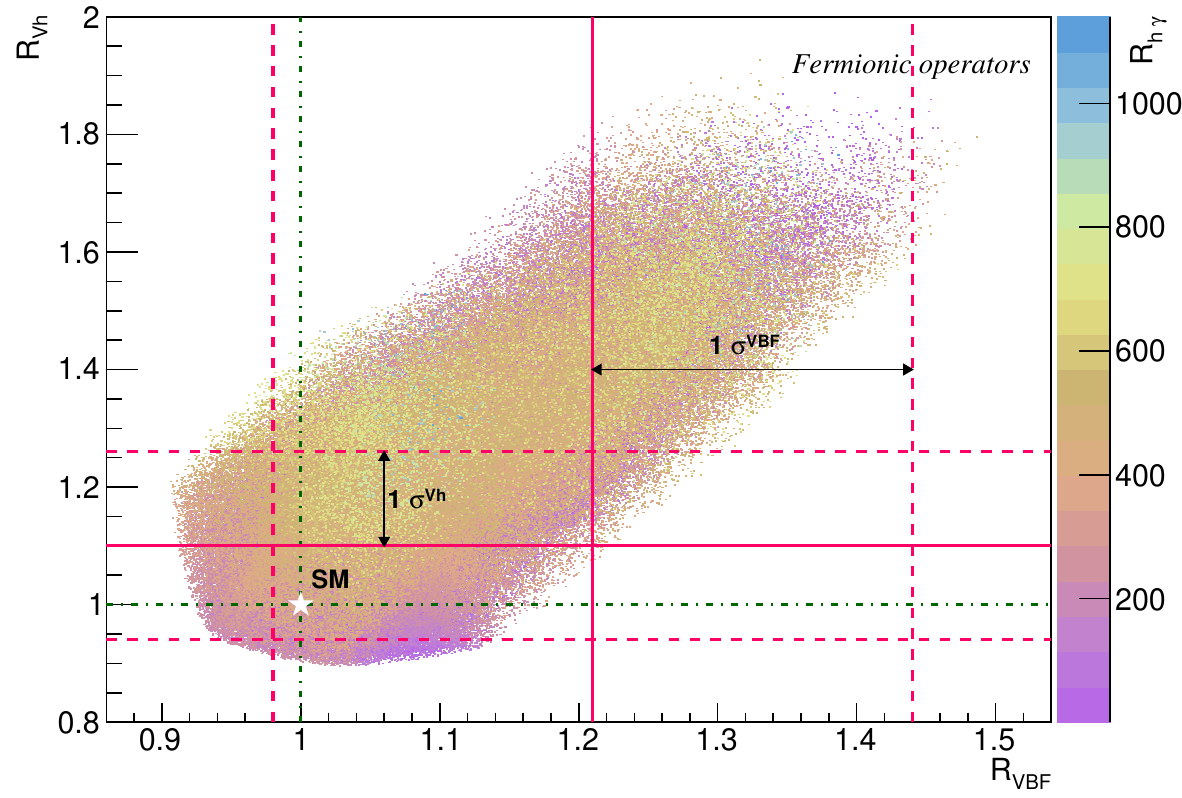}
		\vspace*{-0.5cm}
		\caption{Left: The correlations of $R_{h\gamma}$  with $R_{\gamma\gamma}$ and $R_{Z\gamma}$ and Right: correlations of $R_{h\gamma}$  with $R_{VBF}$ and $R_{Vh}$ at the LHC. The solid lines correspond to experimental values of signal strengths measured and the dashed lines are $1\sigma$ uncertainties in the measurements.}\label{ratio_correlation}
	\end{figure}

	Note that, in the aforementioned ratios, any dependence on the PDF and scale choice are absent. 
	In Fig.~\ref{ratio_correlation} (left panel), we study the correlation between $R_{h\gamma}$ ,  $R_{\gamma\gamma}$ and $R_{Z\gamma}$ evaluated at the 13 TeV run of LHC. These ratios depend on two particular combinations of bosonic Wilson coefficients $C_{HB}$, $C_{HW}$ and $C_{HWB}$ as listed in Table~\ref{tab:rgb}. However, $C_{HWB}$  is constrained to be small from the measurement of $sin^2\theta_W$. In the following analysis, we set $C_{HWB}=0$. The other two couplings have been varied in the range allowed by $h \gamma$ measurement and listed in Table~\ref{bounds}. The $h \to \gamma \gamma$ rate with respect to the SM is anticorrelated to $h \to Z \gamma$. But $pp \to h \gamma$ is positively correlated to both the aforesaid processes. This is accounted by the fact that $R_{\gamma\gamma}$ and $R_{Z\gamma}$  are proportional to $s_W^2C_{HW}+c_W^2C_{HB}$  and $s_W c_W(C_{HW}-C_{HB})$ respectively while both the above combinations of couplings contribute to $h \gamma$ production. The $h \to \gamma \gamma$ signal strength measured by ATLAS collaboration at 13 TeV  is $1.18^{+0.17}_{-0.14}$~\cite{CMS:2018piu} while we have only upper limits on $R_{Z \gamma}$ so far which is a factor of 3.6 ~\cite{ATLAS:2020qcv}. We have shown the $1\sigma$ uncertainty in the signal strength with the dashed lines. It is evident from the plot that the $R_{h\gamma}$ can be as large as 3 in the region of parameter space allowed by both $R_{Z\gamma}$ and $R_{\gamma \gamma}$. This is a consequence of more stringent limits of $C_{HB}$ and $C_{HW}$  imposed from $R_{\gamma \gamma}$ and $R_{Z\gamma}$ measurement at the LHC.

	Now let us see how the fermionic operators could correlate some of the Higgs observables. To illustrate this, we have shown the correlation of $R_{h\gamma}$  with $R_{Vh}$ and $R_{VBF}$ in Fig.~\ref{ratio_correlation} (right panel). A positive correlation between $R_{Vh}$ and $R_{VBF}$ is observed from this plot. One can see that the rate of  $h \gamma$ production is largely enhanced compared to $Vh$ production and $VBF$ Higgs production over the allowed range of couplings for $g_{uu h\gamma}$  and $g_{ddh\gamma}$ (see Table~\ref{tab:rgb}). For instance, a particular region occurs where the rate gets enhanced by 100-200 times the SM rate of $h \gamma$ and yet  in allowed band of signal strength $Vh~(1.1^{+0.16}_{-0.15})$ and $VBF~(1.21^{+0.24}_{-0.22})$~\cite{CMS:2021kom}. This result clearly shows the efficacy of $h \gamma$ channel over $VBF$ and $Vh$ channels atleast in case of fermionic operators.	
	
	At this end, we would like to comment that the $h \gamma$ production via non-standard interactions followed by $h \to b \bar{b}$ decay via the SM would produce  $b \bar{b} \gamma$ final state, with $b \bar{b}$ mass peaked around Higgs mass $m_h$. This final state also arises when Higgs is being produced resonantly via SM interactions and decaying to $b \bar{b} \gamma$ via the same non-standard interaction. These two processes could easily be separated by looking at the invariant mass of $b \bar{b} \gamma$ system. The ratio of the above two rates defined below can be expressed as a function of  Wilson coefficients $C_{dB}$ and $C_{dW}$ and can be used as a discriminator of new physics from the SM. 
	
	\begin{equation}
	r_b (C_i)=\frac{\sigma_{h \gamma}(C_i) \times BR_{h \to b \bar{b}}}{\sigma_{ggF} \times BR_{h \to b \bar{b} \gamma}(C_i)}
	\end{equation}
	
\begin{figure}[t!]
		\begin{center}
			\includegraphics[width=9.0cm, height=6.5cm]{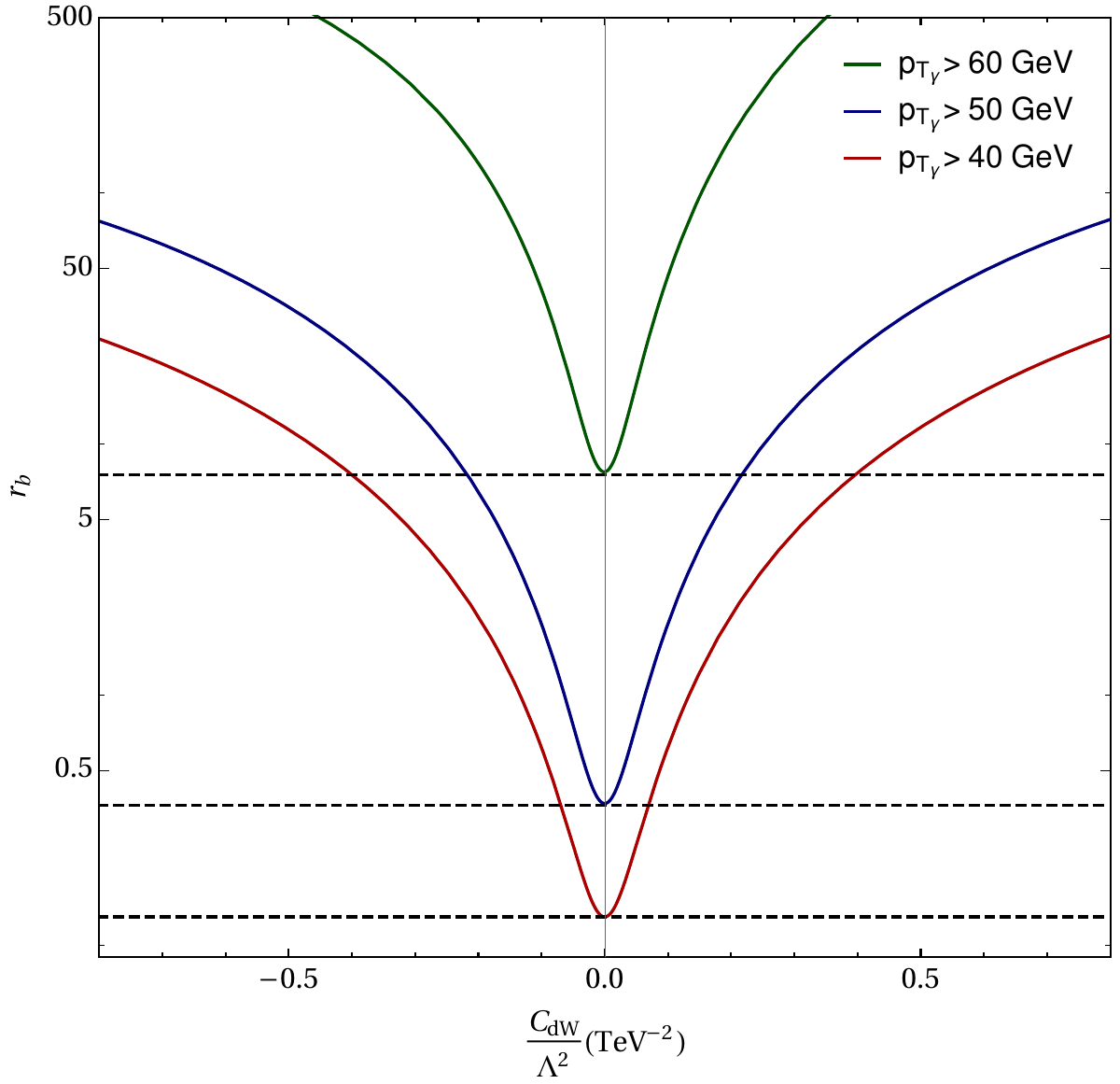}
			\caption{The variation of ratio $r_b$ with $\frac{C_{dW}}{\Lambda^2}$.}
			\label{ratio_observable}
		\end{center}
	\end{figure}

	This ratio also crucially depends on the $p_T$ of the photon. In Fig.~\ref{ratio_observable}, we have plotted the variation of the ratio $r_b$ with $\frac{C_{dW}}{\Lambda^2}$ for three different choices of minimum value of photon $p_T$. The values of $r_b$ for vanishing  $\frac{C_{dW}}{\Lambda^2}$  correspond to its SM values.
	
	In the case of $b \bar{b} \gamma$ coming from the decay of the Higgs, all the final state particles tend to have a small transverse momentum and small mutual angular separation. Consequently, the method of Higgs tagging will not be effective in this scenario. Moreover, the photon in the background process arising from final-state radiation, tends to be close  to one of the jets and have a small $p_T$. So, we would like to distinguish the signal from background in case when the effective operator controls the Higgs decay and produces the $b \bar{b} \gamma$ final state. We have specifically checked that distributions of (i) $E_\gamma$ (ii) the larger ($\Delta \eta_{j \gamma}^{max}$) and the smaller ($\Delta \eta_{j \gamma}^{min}$) one of the rapidity separations between the b-jets and photon, (iii) the larger ($m_{j \gamma}^{max}$) and the smaller ($m_{j \gamma}^{min}$) one of the invariant mass of the leading jets and photon, (iv) transverse momentum of the three final states from $h \to b \bar{b} \gamma$, (v) angular separation between the two jets and (vi) the larger $p_{Tj}^{max}$ and the smaller $p_{Tj}^{min}$ transverse momentum of the two leading jets, when given as input variables to the BDT network serve as distinguishing features between signal and background characteristics leading to a signal significance of 2.83 with $C_{dW} = 0.2~\rm{TeV^{-2}}$ at 3000 fb$^{-1}$ 14 TeV LHC. While the inclusive $pp \to h (\to b \bar{b} \gamma)$ process alone may not be a promising channel to probe these new contact interactions, however, the ratio $r_b$ can serve as a better discriminator between the SM and the effective interactions.
	\section{Validity of EFT and interpretation of the results}
	\label{sec:further_interpretation}
	\subsection{EFT Applicability}
	The SMEFT framework is a good approximation to describe the effects of new physics at low energy but cannot be used above a certain energy scale. As the SMEFT Lagrangian is expanded in power series of $\frac{\rm energy^2}{\Lambda^2}$, the steadily growing cross-section with the partonic center-of-mass energy  $\sqrt{\hat{s}}$ indicates the breakdown of the EFT expansion when $\sqrt{\hat{s}}$ is comparable with the masses of the heavy particles which have been integrated out. We need to estimate the validity of the EFT expansion which is particularly important for the collider bounds, since a wide range of energies are naturally probed at the hadron collider experiments. We thereby discuss the viable range of Wilson coefficients that admit a valid EFT interpretation of cross-sections. While estimating the cross-sections, we have kept the interference $(\mathcal{O}(\frac{1}{\Lambda^2}))$ of dimension-6 operators with the SM along with the square of the purely dimension-6 operator contributions $(\mathcal{O}(\frac{1}{\Lambda^4}))$. We find that for the operators of our interest, their effects in $h \gamma$ production is such that the dominant contribution comes from the term proportional to the square of $\frac{C_i}{\Lambda^2}$ than that of the interference term, which may raise questions about the validity of the EFT nature of interaction. However, such a doubt is dispelled so long as the dimension-8 effects interfering with the SM remains smaller than the dimension-6 quadratic terms. The validity of EFT expansion is ensured provided that their contribution dominates over the SM in the energy range $\Lambda > E > \frac{g_{SM}}{g_x} \Lambda$, where $g_x$ is the dimensionless coupling in the UV completion~\cite{Contino:2016jqw}. The dimension-8 operator interfering with the SM is indeed suppressed by a factor of $g_{SM}^2$ with respect to the dimension-6 squared term, although both contributing at $\mathcal{O}\left(\frac{1}{\Lambda^4}\right)$ and thus, dimension-8 effect can be neglected.
	
\begin{figure}[t!]
		\centering
		\subfloat{\begin{tabular}{cc}	
				\includegraphics[width=7.5cm,height=6.2cm]{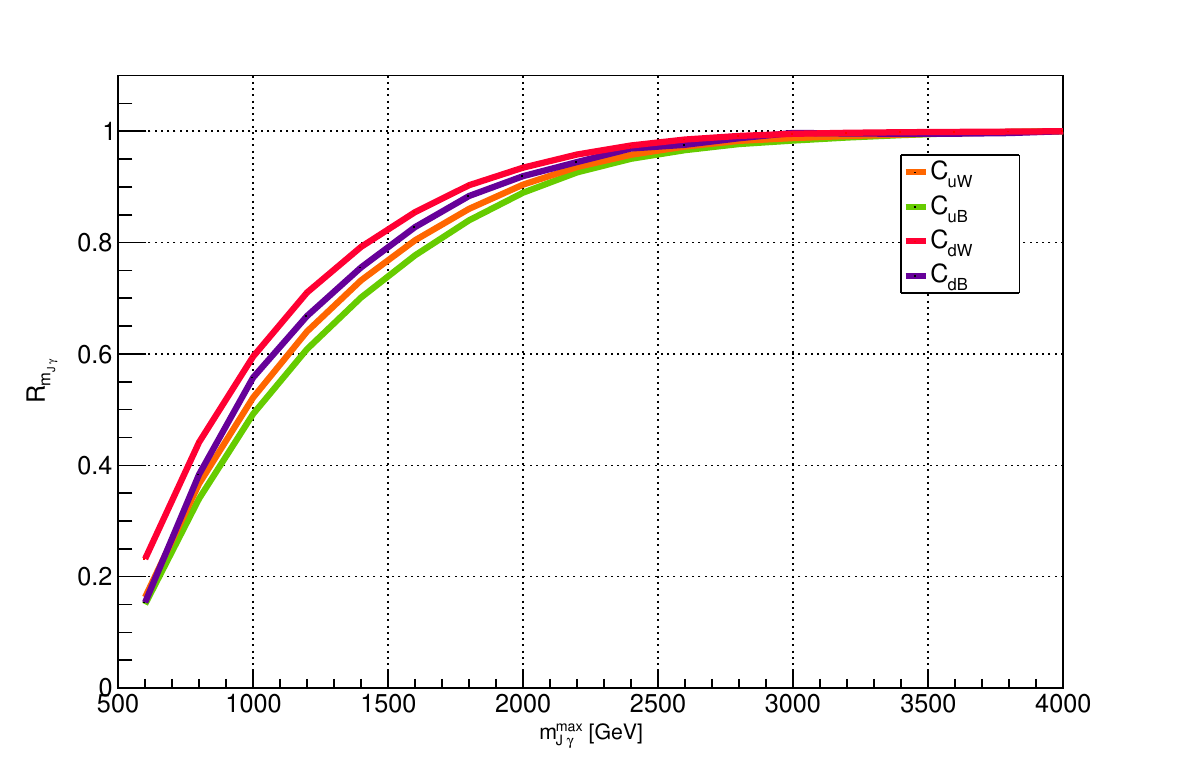}&
				\includegraphics[width=7.5cm,height=6.2cm]{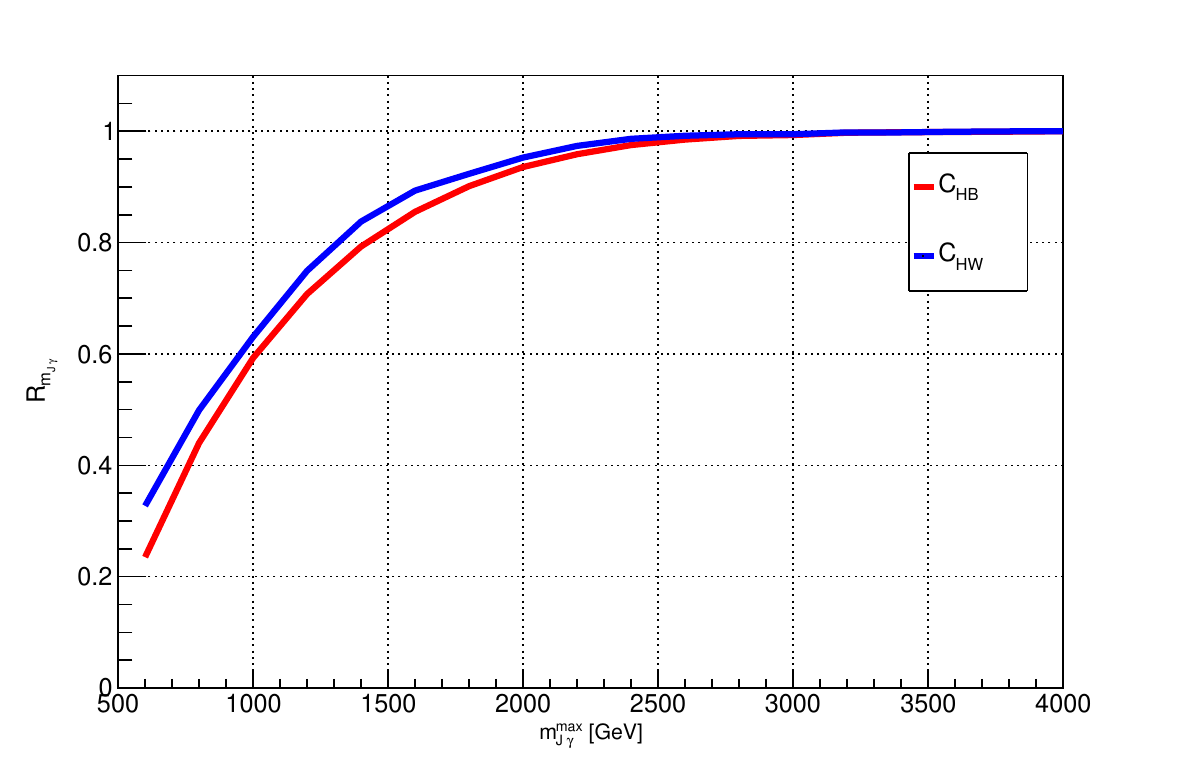}\\
			(a)&(b)
\end{tabular}}
		\caption{The ratio $R_{m_{J\gamma}}$ variation as a function of maximum $m_{J \gamma}$ in the $h \gamma$ production process with (a) the dipole type operators and (b) the bosonic operators. }
		\label{tab:eft_validity}	
	\end{figure}

	The invariant mass of the jet-photon system ($m_{J \gamma}$) determines the energy scale being probed in a collision.  The validity of our analysis can be ensured by restricting $m_{J \gamma}$ to be less than the cut-off scale ($\Lambda$). However, the cutoff scale for the EFT cannot be defined precisely in a model independent approach, without knowing the details of the underlying UV complete theory.  To have some idea about the cut-off scale we  use a method following Ref.~\cite{Busoni:2013lha,Bhattacharya:2015vja}. When $m_{J \gamma}$
	remains smaller than $\Lambda$ for most of the collisions taking place, the ratio $R_{m_{J \gamma}}$, defined in the following, would tend to 1.
	\begin{equation}
	R_{m_{J \gamma}} \equiv \frac { \int^{m_{J \gamma} < {m ^{max} _{J \gamma}}} \frac{d \sigma} {d m_{J \gamma}}\; d m_{J \gamma} }  { \int \frac{d \sigma} {d m_{J \gamma}}\; d m_{J \gamma} }
	\end{equation}
	We illustrate this effect  in Fig.~\ref{tab:eft_validity}  with different non-zero EFT couplings at $\sqrt{s}=$14 TeV. Values of  $R_{m_{J \gamma}}$ close to unity signifies that transfer of energy in the process is much smaller than ${m ^{max} _{J \gamma}}$, the maximum $m_{J \gamma}$ value. Such values of $m_{J \gamma}^{max}$ could be identified as the cut-off scale $\Lambda$ of the EFT.  In fact, for both the fermionic and bosonic  operators, $R$ tends to unity as  $m^{max}_{J \gamma} > 2.5~\rm TeV$. Minimum values of Wilson coefficient for which $3 \sigma$ significance sensitivity can be obtained, have been assumed while estimating the above ratio. 
	However, higher values of the Wilson coefficients  would not lead to better sensitivity on $m^{max}_{J \gamma}$.

	
	\subsection{Reinterpretation in UV theory}
\begin{figure}[b!]
		\centering
		\subfloat{
			\begin{tabular}{ccccc}	
				\includegraphics[width=3.5cm,height=2.8cm]{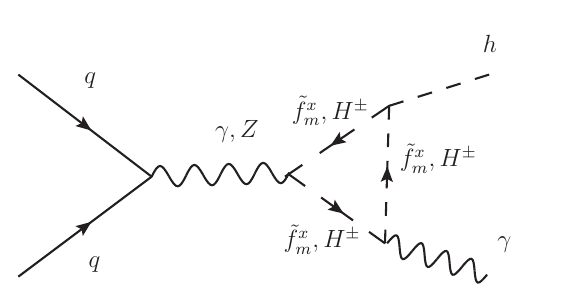}&
				\includegraphics[width=3.5cm,height=2.8cm]{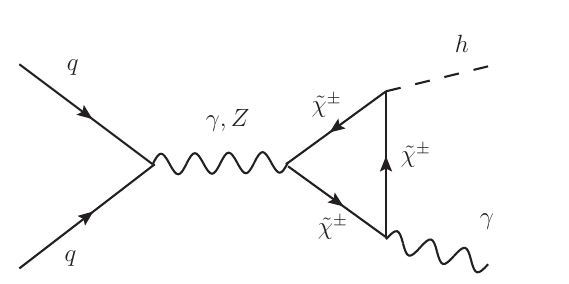}\\
				(a)&(b)
			\end{tabular}
		}
		\caption{Additional triangle-type diagrams contributing to the process $q \bar{q} \to h \gamma$ at one-loop level that can be parameterised with $\mathcal{O}_{HW}$, $\mathcal{O}_{HB}$ and $\mathcal{O}_{HWB}$.}
		\label{tab:mssm_feynmandiag_bosonic}	
	\end{figure}

\begin{figure}[t!]
		\centering
		
		\begin{tabular}{cccc}	
			\includegraphics[width=2.8cm,height=2.5cm]{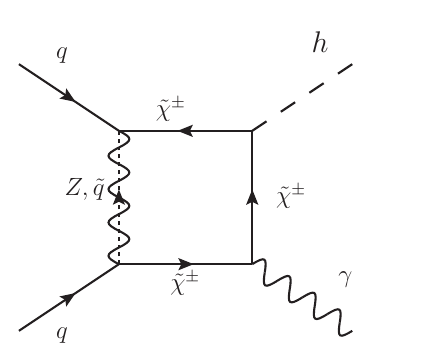}&
			\includegraphics[width=2.8cm,height=2.5cm]{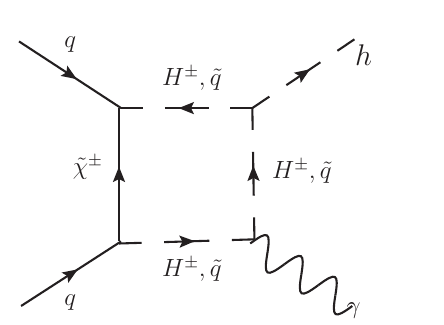}&
			\includegraphics[width=2.8cm,height=2.5cm]{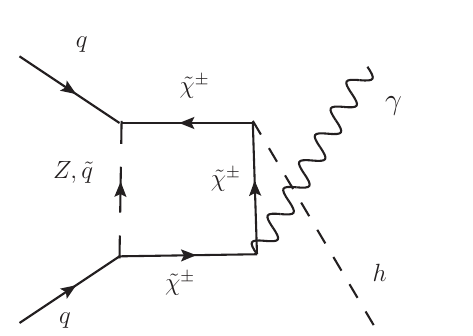}&
			\includegraphics[width=2.8cm,height=2.5cm]{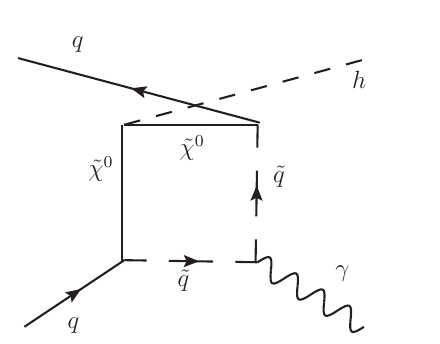}\\
			(a)&(b)&(c)&(d)\\
			\includegraphics[width=2.8cm,height=2.5cm]{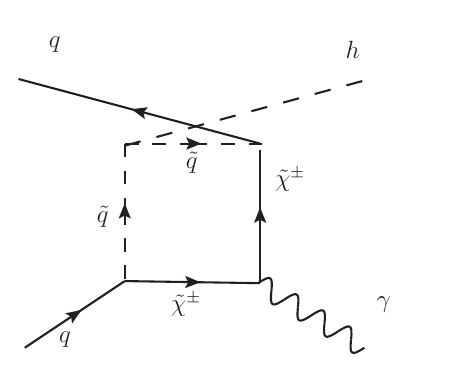}&
			\includegraphics[width=2.8cm,height=2.5cm]{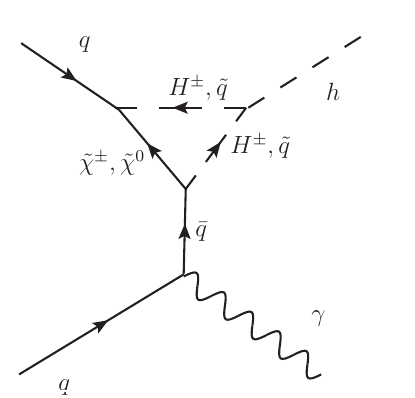}&
			\includegraphics[width=2.8cm,height=2.5cm]{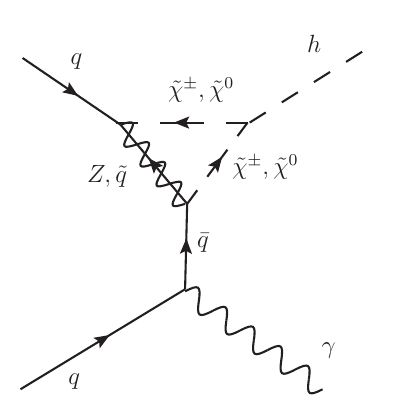}&
			\includegraphics[width=2.8cm,height=2.5cm]{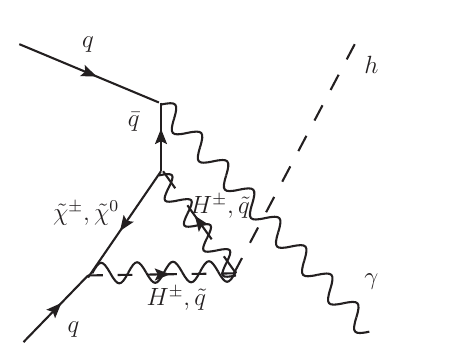}\\
			(e)&(f)&(g)&(h)
		\end{tabular}
		\caption{Additional box-type and triangle-type diagrams contributing to the process $q \bar{q} \to h \gamma$ at one-loop level that can be parameterised with $\mathcal{O}_{fB}$ and $\mathcal{O}_{fW}$.}
		\label{tab:mssm_feynmandiag_dipole}	
	\end{figure}

	It is of interest to assimilate how the SMEFT operators parametrise a scenario with ultra-violet completion. In a top-down approach, the SMEFT provides an interface between the different  UV models and the \emph{low energy} theory. Thus, SMEFT approach gives a good approximation for both the cases where new physics has effects at tree level as well as loop level. However, the sensitivity to the scale of new physics varies in each case. The Wilson coefficients of the higher dimensional operators contributing to a process like  $pp \rightarrow h \gamma$   have particular relationships with the  parameters of the UV completion. Thus, unlike in the model-independent SMEFT case, where the operator couplings are independent of each other, when translated into regions of parameter space of specific models, can be related to each other. The matching of dimension-6 operators to the UV models such as extensions in the Higgs sector of the SM have been discussed in recent literature \cite{Henning:2014wua,Dawson:2020oco,DasBakshi:2018vni}. For example, one of such popular models, the minimal supersymmetric standard Model (MSSM), that predicts several new particles such as the additional Higgs bosons and the  superpartners of SM particles, can give additional contributions to the loop-induced production of Higgs and  photon. Figs.~\ref{tab:mssm_feynmandiag_bosonic} and~\ref{tab:mssm_feynmandiag_dipole} illustrate few representative one-loop processes where the effects can be parametrised by dimension-6 operators.
	
	The triangle and box contributions in Fig.~\ref{tab:mssm_feynmandiag_dipole}  can lead to the dipole-type operators (see Eqn.~\ref{eq:op2}) without a requiring a chirality flip of the SM fermion lines. Fig.~\ref{tab:mssm_feynmandiag_bosonic} lead to bosonic-type operators (see Eqn.~\ref{eq:op3}) that can contribute to the loop-induced $h \gamma$ production as well as decays of $h \to \gamma \gamma$ and $h \to Z \gamma$. 
	
	As  an  illustration, we show how the limits of bosonic operators obtained in our analysis can provide interesting constraints on masses and mixings of stop squarks in the framework of MSSM in the limit when the electroweak gauginos are much heavier than the stop-quarks. In presenting our results, we follow \cite{Henning:2014gca}, assuming a common  diagonal mass term $m_{\tilde{t}}$ for right- and left- stop squarks and  $X_{t}=A_t-\mu~\rm cot~\beta$ that governs the $\tilde{t}_L$-$\tilde{t}_R$ mixing and the splitting between the two stop mass eigenstates. It has been already pointed out that $C_{HB}$ and $C_{HW}$ are  severely constrained from $R_{\gamma \gamma}$. The contribution of stop squarks to the  $h \gamma \gamma$ vertex at one loop  \cite{Dermisek:2007fi}, is given by:
	
	\begin{align}
	C^{MSSM} &=\frac{N^{\tilde{t}}}{D^{\tilde{t}}}\,, {\rm{here}}~N^{\tilde{t}}~{\rm{and}}~D^{\tilde{t}}~{\rm{are~given~by~the~ following~equations:}}\nonumber\\
	N^{\tilde{t}}&=2{m_t^4}+2{m_t^2}m_{\tilde{t}}^2-{m_t^2} {X_t^2}+\frac{1}{8}{m_Z^4} {c_{2\beta}^2} + {m_t^2} {m_Z^2} {c_{2\beta}}+\frac{1}{2}{m_Z^2} c_{2\beta}m_{\tilde{t}}^2\notag\\
	&-\frac{1}{144}	{m_Z^2} c_{2\beta}(-3+8s_W^2)c_{2\theta_t} (144 m_{\tilde{t}}^4 (1+r)^2-288 m_{\tilde{t}}^4 (1-r^2)+144 m_{\tilde{t}}^2(1-r)^2\notag\\
	&+18 m_Z^4 - 96 m_Z^4s_W^2 + 128 m_Z^4 s_W^4 + 576 m_t^2 X_t^2 +288 m_Z^2 c_{2\beta}m_{\tilde{t}}^2-768 m_Z^2 s_W^2 c_{2\beta}m_{\tilde{t}}^2 \notag\\
	&+m_Z^4 (18-96 s_W^2 + 128 s_W^4) c_{4 \beta} )^{1/2}  \notag\\
	D^{\tilde{t}}&={m_t^4}+m_{\tilde{t}}^4(1-r^2)+2 m_t^2m_{\tilde{t}}^2+\frac{4}{3}s_W^2m_Z^2 m_{\tilde{t}}^2c_{2\beta}r+\frac{1}{2}m_Z^2 m_{\tilde{t}}^2 c_{2\beta}(1-r) + \frac{1}{2}m_Z^2 m_t^2 c_{2\beta}  \notag\\
	&+\frac{1}{3}m_Z^4 c_{2\beta} s_W^2-\frac{4}{9}m_Z^4 s_W^4 c_{2\beta}^2 - m_t^2 X_t^2 
	\label{mssm_eft_loop}
	\end{align}	
	where $m_{t}$ denotes the top mass and $\tan \beta$ is equal to the ratio of vacuum expectation values of the two Higgses: $\tan \beta = \frac{\langle H_{u} \rangle}{\langle H_{d} \rangle}$ and  $m_{\tilde{t}}^2 = \frac{m_{\tilde{t}L}^2+m_{\tilde{t}_R}^2}{2}$ and $r=\frac{m_{\tilde{t}_L}^2-m_{\tilde{t}_R}^2}{m_{\tilde{t}_L}^2+m_{\tilde{t}_R}^2}$. While deriving Eqn.~\ref{mssm_eft_loop}, it has been assumed that the electroweak gauginos are much heavier than the stop quarks.
	
\begin{figure}[b!]
		\centering
		\subfloat{
			\begin{tabular}{cc}	
				\includegraphics[width=7.5cm,height=6.6cm]{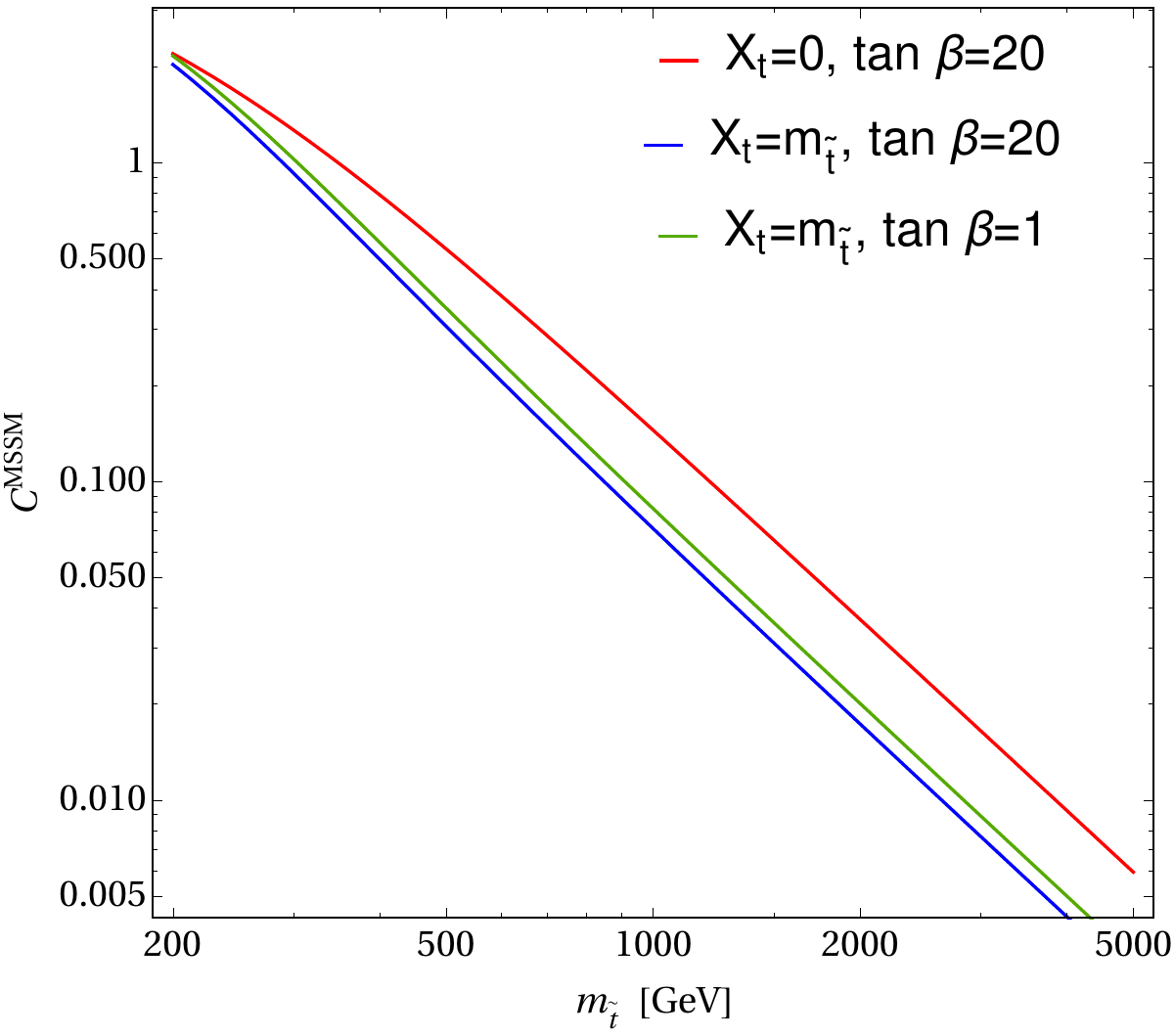}&
				\includegraphics[width=7.5cm,height=6.7cm]{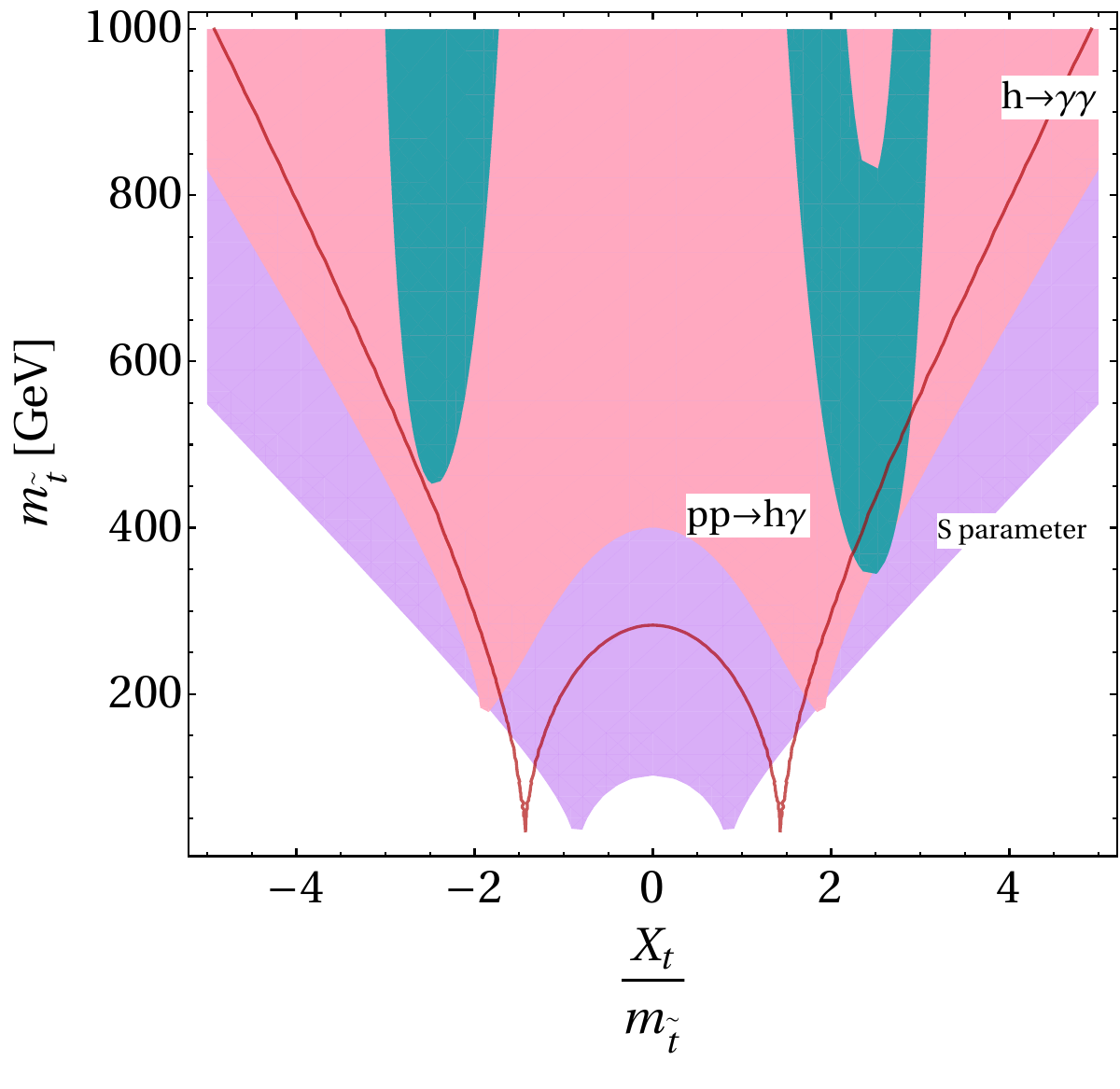}\\
				(a)&(b)
		\end{tabular}}
		\caption{(a) One loop-level MSSM contribution to the $h \gamma \gamma$ coupling, for $\tan\beta = 1$ and $\tan\beta = 20$ as a function of $m_{\tilde{t}}$   and (b) The allowed  parameter space: $(\frac{X_{t}}{m_{\tilde{t}}}, m_{\tilde{t}})$ plane with $\tan\beta = 20$ obtained from the $C_{HW},C_{HB}~\rm{and}~C_{HWB}$  projections from $pp \to h\gamma$ process. The purple shaded region corresponds to the $2$-$\sigma$ allowed region from the measurement of S parameter. The brown curve represents the $2$-$\sigma$ sensitivity contour from modification to $h \to \gamma \gamma$ decay. The green shaded region corresponds to  $m_h \in (122,\,128)~\textnormal{GeV}$. For both the above plots, degeneracy of left and right stop masses has been assumed.}	
		\label{fig:MSSM}
	\end{figure}

	Thus, the loop contribution arising from MSSM scales as  inverse square of the stop mass in the limit when stop is heavier than all other SM particles. This is also evident from Fig.~\ref{fig:MSSM}(a).  For large stop masses, this $C^{MSSM}$ can be interpreted from the constraints on anomalous coupling $g_{h\gamma\gamma}$. This coupling can be expressed as the following combination of effective operators,  $C^{EFT} = c_W^2 C_{HB}+s_W^2 C_{HW}  - c_W s_W C_{HWB} $ (see Table~\ref{tab:rgb}). Thus, $g_{h\gamma\gamma}$ obtained from the upper limits of  parameters $C_{HB}$, $C_{HW}$ and $C_{HWB}$ can be translated to stop parameters $m_{\tilde{t}}$ and $X_{t}$  through the same relation \cite{J_Ellis_Tevlong_Y}.
	
	We note that the EFT driven $pp \to h \gamma$ rate involves both the anomalous $g_{h \gamma \gamma}$ and $g_{h Z \gamma}$ interactions. Both these effective vertices depend on $C_{HB},C_{HW}$ and $C_{HWB}$. All these Wilson coefficients can be expressed in terms of $m_{\tilde{t}}$ and $X_t$. Consequently, these Wilson coefficients, which are independent in SMEFT framework, 
	now become related to each other. 
	So, we can use the $3 \sigma$ projected values of  $C_{HB}, C_{HW}$ and $C_{HWB}$, all non-zero at a time, to  map them in $X_{t} - m_{\tilde{t}}$ plane. One can interpret the region in $X_{t} - m_{\tilde{t}}$ plane, sensitive to the search of $h \gamma$ signal at the 14 TeV LHC, following this algorithm.
	
	 We show (in right panel of Fig.~\ref{fig:MSSM}) the allowed region in  $(\frac{X_{t}}{m_{\tilde{t}}}, m_{\tilde{t}})$ plane : shaded in pink from our analysis of $q \bar{q} \to h \gamma$ and the region shaded in purple corresponds to $2$-$\sigma$ allowed constraints from the measurements of $S$ parameter~\cite{Henning:2014gca}. We additionally present the deviation (with respect to the SM)  to $h \to \gamma \gamma$ width (Eqn.~\ref{mssm_eft_loop}) arising from $C_{HB}, C_{HW}$ and $C_{HWB}$.  The brown curve represents such a deviation at $2 \sigma$ level. However, the entire of pink and purple region is not allowed from the consideration of Higgs mass\footnote{To have the lightest CP-even Higgs boson in the MSSM to be identified with the 125 GeV scalar discovered at the LHC, one requires large stop masses and large stop mixing.}. One can easily see that the green shaded region with $m_h \in (122,\,128)~\textnormal{GeV}$ runs almost parallel to $m_{\tilde{t}}$ axis implying a loose constraint on $m_{\tilde{t}}$ resulting from our analysis. So, the green region which is the overlap of all three aforementioned regions allowed by different considerations is of interest. When the stop mixing parameter $X_t=0$, the stop mass $m_{\tilde{t}}\sim \mathcal{O}(400)$ GeV. Limits on stop masses extend upto 1 TeV for large stop mixings. These limits also depend on the values  of lightest neutralino masses, vanishing entirely for $m_{\tilde{\chi}_1^0} \le 400~\rm GeV$~\cite{ATLAS:2021twp}. Thus, these Wilson coefficient constraints, generated at one-loop level provide complementary information to those from direct searches for gluinos and squarks production. Also, a global SMEFT fit of data resulting from  top production, Higgs production and decay  along with di-boson production  at the LHC resulted into tighter constraint on stop parameters as discussed in Ref.~\cite{Ellisglobalfit2021}.
	
	Future lepton colliders also provide a viable prospect of precise and model independent measurement of Higgs and photon production. Predictions for the model independent parameters  would not only allow to test the $h\gamma$ interactions but also will be helpful to distinguish the production of a possible new scalar in association with a photon. A study including $h \gamma$ production at the lepton colliders and taking into account the additional Higgs decay modes such as $h \to WW^*, h \to \tau^+ \tau^-$  will be reported in a follow-up study. The production of Higgs and photon followed by the Higgs decays into diphoton would lead to a distinctive $3 \gamma$ peak, that would provide more precise limits on the couplings and masses of the UV dynamics.
	
	\section{Conclusions }
	\label{sec:discussions}
	
	We have presented a study of the production of Higgs boson in association with a high $p_T$ photon at the LHC, as the probe of new physics beyond the Standard Model. The $pp \to h \gamma$  rate being highly suppressed in the framework of SM, it can serve as a highly sensitive probe of new interactions. Instead of confining ourselves to a particular BSM scenario, we stick to the SMEFT framework in which operators involving SM fields of dimension-6 are constructed keeping various local and global symmetries of the SM.  The coefficients of these operators encapsulate the effects of any high scale dynamics that may  originate from a particular BSM scenario or even simply from the higher order processes in the SM. We have been particularly interested in the {\em boosted} kinematic regime where a very high $p_T$ photon recoils against a fat-jet originating from a pair of $b$-quarks resulting from the decay  of the Higgs boson. Dimension-6  SMEFT operators which  contribute in  the $h \gamma$ production rate arise in a couple of varieties, namely the {\em bosonic-type} involving  a Higgs boson and two gauge bosons and {\em fermionic-type}, involving two fermions of opposite helicities along with a Higgs boson and a gauge boson. These interactions have energy dependencies in the cross-sections that essentially grow with the partonic center-of-mass energy.
	
	To begin with, we have briefly reviewed the constraints on the dimension-6 operators of our interest, arising from the unitarity, electroweak precision tests and the available Higgs data. Finally, we compare the mass distribution of boosted $h \gamma$ system  estimated in SMEFT framework with the same measurement done  at the Run II phase of 13 TeV LHC by the CMS collaboration. This exercise helps us in validating our simulation as well as to constrain the relevant Wilson coefficients from $h \gamma$ data in particular. We choose six representative benchmark points from the  allowed parameter space and explore their detection possibility at the 14 TeV LHC with 3000 $fb^{-1}$ integrated luminosity. Different Lorentz structures appearing in the dimension-6 SMEFT operators leave their signatures in the different kinematic distributions which can be exploited to distinguish new physics signal from the  SM backgrounds. We have  constructed new observables like $\psi, Q^R, Q^T, R_2$ to  enhance the characteristic features of the tensor structures of $hV\gamma$ coupling and the $h\gamma ff$ contact interactions and discriminate against the SM backgrounds. The fact that  hadronic decay products of the Higgs boson can form boosted fat-jet,  opens up an intriguing new possibility that can be beneficial, contrary to {\em cleaner} final states from the $ZZ^\ast$ (or $WW^\ast$) decays of the Higgs.  The reconstructed fat-jet still carries the characteristics of the parent particle in its mass and substructures, thus enabling one to identify the Higgs boson itself with {\em significant} tagging efficiency. We have exploited such properties to counter the large SM background coming from the QCD jets of light flavour. The projected 3$\sigma$ upper limits derived from our cut-based analysis for integrated luminosity of 3000 $\rm fb^{-1}$  seem to be more stringent than the bounds coming from precision electroweak observables.
	
	To explore any possible improvement of the sensitivity for the new interactions at the LHC, we further undertook a {\em multivariate analysis} where  a total of 10 kinematic variables have been used as input to the boosted decision tree network which provide the optimum separation between the signal and the background. Unlike the rectangular cut-based analysis, the multivariate analysis can use the potential of various kinematic variables at one go. The above analysis clearly established that the LHC sensitivity can be greatly improved  for most of the benchmark points used by us. Benchmark points involving the  {\em fermionic} operators can be probed with more than  5$\sigma$ significance with 200 $\rm fb^{-1}$ data only while more accumulated data is needed to probe {\em bosonic} operators at the same level. 
	
	Finally, we have briefly discussed the connection between the model-independent coefficients of dimension-6 operators with an ultra-violet complete model, namely the MSSM. After qualitatively discussing how the bosonic and fermionic operators may arise from MSSM,  constraints have been derived on the MSSM parameters, namely stop masses and mixing,\footnote{We have assumed that electroweak gauginos are heavy} responsible for generating $h \gamma \gamma$ coupling at one loop.  As we are headed towards the new phase of LHC run with the higher energy and luminosity, it is important to push for the rare processes like $pp \to h \gamma$ production which are very sensitive to the physics beyond the standard paradigm.
	
	\section*{Acknowledgements}
	We thank Satyaki Bhattacharya and Amit Chakraborty for useful discussions at various stages of the work. TB acknowledges the support from Council of Scientific and Industrial Research, Government of India. TB is thankful to Somenath Pal for providing computational facilities at various stages of the work.
	
	\providecommand{\href}[2]{#2}\begingroup\raggedright
	\endgroup
\end{document}